\documentclass[aps,prd,floatfix,superscriptaddress,showpacs,nofootinbib,twocolumn]{revtex4}
\usepackage[latin1]{inputenc}     
\usepackage{bbm,slashed}
\usepackage{mathrsfs} 
\usepackage{graphicx,epsfig}  
\graphicspath{{plots/}}
\usepackage{amsmath,amsfonts,amssymb} 
\usepackage{booktabs}
 
\usepackage{epstopdf}

\usepackage{url}   
\usepackage{dsfont} 
\usepackage{bm,bbm,xcolor}    

\usepackage{color}

\newcommand{\eq}{\eqref}
\def\s0#1#2{\mbox{\small{$ \frac{#1}{#2} $}}} 
 
\newcommand{\fdi}{\slashed{\partial}}
   
\newcommand{\cD}{\mathcal{D}}

 \DeclareMathAlphabet{\boldmathe}{T1}{cmr}{bx}{it}

\begin{document}
\title{Phases of  supersymmetric $O(N)$ theories}

\author{M.  Heilmann}
\affiliation{ Theoretisch-Physikalisches Institut, Friedrich-Schiller-Universit{\"a}t
Jena,
Max-Wien-Platz 1, D-07743 Jena, Germany}
\author{D. F. Litim}
\affiliation{Department of Physics and Astronomy, University of Sussex, BN1 9QH, Brighton, UK.}

\author{F. Synatschke-Czerwonka}
\affiliation{ Theoretisch-Physikalisches Institut, Friedrich-Schiller-Universit{\"a}t
Jena,
Max-Wien-Platz 1, D-07743 Jena, Germany}
\author{A. Wipf}
\affiliation{ Theoretisch-Physikalisches Institut, Friedrich-Schiller-Universit{\"a}t
Jena,
Max-Wien-Platz 1, D-07743 Jena, Germany}

\begin{abstract}
We 
perform a global renormalization group 
study of  $O(N)$ symmetric Wess-Zumino theories and their phases
in three euclidean dimensions. 
At infinite $N$  the theory  is solved exactly. 
The phases and phase transitions are worked out for finite and infinite short-distance cutoffs.   
A distinctive new feature arises at strong coupling, where the effective
superfield potential becomes multi-valued, signalled by 
divergences in the fermion-boson interaction.
Our findings  resolve the long-standing 
puzzle about the 
occurrence of degenerate $O(N)$ symmetric phases.
At finite $N$, we find
a strongly-coupled fixed point 
in the local potential approximation and explain its impact on the phase transition. We   also examine the possibility for a supersymmetric Bardeen-Moshe-Bander phenomenon,
and relate  our findings with the spontaneous breaking of 
supersymmetry in other models.
\end{abstract}
\pacs{05.10.Cc,12.60.Jv,11.30.Pb,11.30.Qc} 
\maketitle

\section{Introduction}

Supersymmetry, the symmetry which links bosonic with fermionic degrees of freedom, is an intriguing concept with many applications in quantum field theory and statistical physics. It  plays a prominent role for open challenges  in the Standard Model of Particle Physics such as the hierarchy problem, and continues to inspire
the construction of models for new physics. In statistical physics, supersymmetry also appears as a technical symmetry in the exploitation of systems which otherwise are too difficult to handle. It is thus of great interest to further the understanding of interacting supersymmetric theories, and to clarify the impact of supersymmetry on the phase structure and the critical behavior at lowest and highest energies.

This work is devoted to the supersymmetric extension of $O(N)$ symmetric scalar
theories in three euclidean dimensions, continuing a line of research 
initiated in  \cite{Litim:2011bf}.
Without supersymmetry, the bosonic theory with a microscopic $(\phi^2)^3$ potential is described by three renormalized parameters permitting
first-order phase transitions at strong coupling as well as second order phase transitions with Ising-type critical behavior \cite{ZinnJustin}. In the limit of infinitely many scalars, the analytically solvable spherical model also admits an ultraviolet fixed point with broken scale invariance, the Bardeen-Moshe-Bander (BMB) phenomenon, allowing for  a non-trivial continuum limit  \cite{Bardeen:1983rv,David:1984we, David:1985zz}. 
With supersymmetry, additional fermionic degrees of freedom are present and their fluctuations modify the quantum effective theory. 
The $O(N)$ symmetric Wess-Zumino model with a microscopic $(\Phi^2)^2$ superpotential
is determined by only two renormalized parameters and critical and tricritical theories are the same. 
Its phase structure  has attracted   some attention in the past \cite{Bardeen:1984dx, Moshe:2003xn, Dawson:2005uw,Suzuki:1985uk,Suzuki:1985pw, Gudmundsdottir:1984yk, Feinberg:2005nx,Matsubara:1987iz}.  
 In the limit of infinitely many superfields, four different phases have been
observed \cite{Bardeen:1984dx,Moshe:2003xn}, including peculiar degenerate $O(N)$ symmetric phases with several mass scales.
Similar to the scalar case, a supersymmetric version of the BMB fixed point has equally been found at a critical coupling
where the bosons and fermions  become massive
while a Goldstone-boson (dilaton) and a Goldstone-fermion (dilatino) 
are dynamically generated. The supersymmetric $O(N)$ model has also
been discussed in  the $1/N$ expansion  \cite{Gudmundsdottir:1984yk}, where the authors found a non-trivial UV fixed-point and a stable dilaton phase. At next-to-leading order the 
dilaton acquires a mass of order $1/N$ showing that a phase with 
spontaneously broken scale invariance only exists in the limit of infinitely many superfields \cite{Matsubara:1987iz}. 

A method of choice in the study of phases and phase  transitions  is Wilson's renormalisation group (RG)  \cite{Wetterich:1992yh}. It is based on a path-integral representation of the theory, where the continuous  integrating-out of momentum modes
permits a smooth and controlled interpolation
between the microscopic and the full quantum effective theory  \cite{Berges:2000ew}. 
Physically-motivated approximations schemes together with analytic versions of the RG 
 \cite{Litim:2000ci,Litim:2001fd,Litim:2001up}  allow for a global analysis and a classification of phase transitions and critical exponents even at strong coupling.
The method has been successfully applied to phase transitions in Ising-type universality classes  \cite{Tetradis:1992xd,Tetradis:1995br,Berges:2000ew,Bagnuls:2000ae}
including high-precision computations of its critical exponents with increasing levels of sophistication \cite{Litim:2002cf,Litim:2003kf,Canet:2003qd,Bervillier:2007rc,Benitez:2009xg,Litim:2010tt}.
The extension of the functional RG towards supersymmetric theories  
\cite{Vian:1998kv,Bonini:1998ec,Synatschke:2008pv,Gies:2009az,
Synatschke:2009nm,Synatschke:2010jn,Synatschke:2010ub,Synatschke:2009da,
Falkenberg:1998bg,Rosten:2008ih,Sonoda:2009df,Sonoda:2008dz,Litim:2011bf} 
 therefore bears the promise for deeper insights into the phases and the critical behavior of supersymmetric $O(N)$ theories.

This paper is organized as follows: 
We recall the main features of  supersymmetric $O(N)$ models including a
supersymmetric version of Wilson's RG (Sec.~\ref{sec:SuSyFlow}), followed by a  discussion of its exact analytical solution in the large-$N$ limit (Sec.~\ref{sec:EffPot}). We then give a detailed account of  the phase diagram and phase transitions in the renormalized theory, and examine the  appearance of a multi-valued effective potential, also in comparison with earlier findings (Sec.~\ref{sec:RenormalizedTheory}). We repeat this exercise with a finite short-distance cutoff including a thermodynamical derivation of scaling exponents (Sec.~\ref{sec:EffectiveTheory}), and  examine the supersymmetric BMB phenomenon (Sec.~\ref{sec:ScaleInvariance}). At finite $N$, we derive an exact fixed point to leading order in a gradient expansion and evaluate its impact on the phase transition, and on the fate of the BMB mechanism (Sec.~\ref{sec:FiniteN}). We close with a brief summary and some conclusions (Sec.~\ref{Conclusions}).

\section{Supersymmetric RG flow}
\label{sec:SuSyFlow}

In this section we sketch the features of supersymmetric $O(N)$ models 
and recall the supersymmetric renormalization group flow in the local potential approximation. 
For a  detailed discussion and derivation see \cite{Litim:2011bf}.

\subsection{Action}
The three-dimensional supersymmetric $O(N)$ models are
built from $N$ real superfields 
\begin{equation}
\Phi_{i}(x,
\theta)=\phi_i(x)+\bar{\theta}\psi_i(x)+\frac{1}{2}\bar{\theta}\theta F_i(x)
\label{eq:superfield}
\end{equation}
containing scalar fields $\phi_i$, Majorana fermions $\psi_i$ and auxiliary 
fields $F_i$ as components and a two-component anticommuting Majorana spinor $\theta$.
The invariant action
\begin{equation}
S = \int d^3x\,
\left.
\left(-\frac{1}{2}\Phi\,\bar{\cD}\,\cD\,\Phi + 2W(R)\right)
\right|_{\bar\theta\theta}\,,
\label{eq:action}
\end{equation}
wherein we suppress the internal summation index $i$,
contains the $O(N)$-invariant composite superfield
\begin{equation}
R=\frac{1}{2}\Phi^2=
\bar\rho+(\bar{\theta}\psi)\phi+\frac{1}{2}\,\bar\theta\theta\left(\phi
F-\frac{1}{2}\bar{\psi}\psi\right)\,,
\end{equation}
where $\bar\rho\equiv \phi^2/2$. The supercovariant derivatives
\begin{equation}
\cD = \frac{\partial}{\partial {\bar\theta}} + i\fdi\theta
\quad \mbox{and} \quad \mathcal{\bar{D}} = -\frac{\partial}{\partial \theta}
- i \bar\theta\fdi
\label{eq:derivatives}
\end{equation}
obey $\{\cD_k,\bar{\cD}_l\}=-2i(\gamma^{\mu})_{kl}\partial_{\mu}$. 
An expansion in component fields
yields the off-shell Lagrangian density
\begin{eqnarray}
\mathcal{L}_{\rm off} &=& \s0{1}{2}\left(-\phi\Box\phi -i\bar{\psi}\fdi
\psi +F^2\right) +W'(\bar\rho)\,\phi F \nonumber\\
&-&
\s012W'(\bar\rho)\,\bar\psi\psi
-
\s012 W''(\bar\rho)\,(\bar{\psi}\phi)\,(\psi\phi)\,.
\label{eq:offshell}
\end{eqnarray} 
By eliminating the auxiliary fields $F$ through their algebraic equation of motion,
$F=-W'(\bar\rho)\,\phi$, we obtain the on-shell density.
The field-dependent fermion mass $m_\psi$, the bosonic potential $V$, 
and the field-dependent Yukawa-type coupling $\lambda_Y$ all follow 
from the superpotential $W$ as
\begin{eqnarray}
\nonumber
m_\psi&=&W'(\bar\rho)\\
\label{V_bosonic}
V&=&\bar\rho\, \left[W'(\bar\rho)\right]^2\\
\nonumber
\lambda_Y&=&\s012 W''(\bar\rho)\,.
\end{eqnarray}
All salient features of the classical theory are encoded in the 
functions \eq{V_bosonic}. For a polynomial superpotential the scalar field potential 
always has a minimum at $V(0)=0$ implying that global supersymmetry is unbroken.

\subsection{Renormalization group}
 
Including the effects of quantum and thermal fluctuations implies that the classical 
action \eq{eq:action} is modified and replaced 
by a ``coarse-grained" or ``flowing'' effective action $\Gamma_k$. In
the next-to-leading order in the super-derivative expansion
\begin{align}
\label{Gammak}
\Gamma_k[\Phi] =& \int\! d^3x\, 
\left.
\left(-\frac{1}{2} \Phi \,Z_k\,\bar\cD\cD\, \Phi + 2\,W_k
\right)\right|_{\bar{\theta}\theta}
\end{align}
interpolates between the classical action at the
high-energy cutoff-scale $k=\Lambda$ and the full effective action at $k=0$. 
The fluctuations above $k$ modify both the superpotential, which has turned 
into a scale-dependent superpotential $W_k$, and the kinetic terms, which may 
acquire a non-trivial field- and momentum-dependent wave function 
renormalization factor $Z_k(\frac12\Phi^2,\bar\cD\cD)$.  

The RG momentum scale $k$ is introduced on the level of the path integral by
adding suitable momentum cutoffs $R_k(q^2)$ to the inverse propagators of the
fields. 
The cutoffs regularizes the path integral in the infrared and gives
rise to a finite flow of the scale dependent effective action.
Optimized choices for $R_k$ are available to ensure the stability 
of the resulting RG equations \cite{Litim:2000ci,Litim:2001fd,Litim:2001up}. 
The scale dependence of the effective 
action \eq{Gammak}  is described by a functional differential 
equation \cite{Wetterich:1992yh}
\begin{equation}
{\partial_t}\Gamma_k=\frac{1}{2}{\rm STr} 
\left(\Gamma^{(2)}_k+R_k \right)^{-1} \partial_t
R_k\,,
\label{FRG}
\end{equation}
which emerges as an exact identity from a path integral representation. 
Here, $t=\ln k/\Lambda$ denotes the dimensionless RG ``time" parameter,  $\Gamma^{(2)}_k$ 
the second functional derivative of $\Gamma_k$ with respect to the fields, 
and the supertrace denotes a momentum integration and a sum over all fields, 
including appropriate minus signs for fermions. 

\subsection{Derivative expansion}
Finally we detail our equations to leading order in a super-derivative 
expansion, the so-called local potential approximation (LPA). It amounts 
to setting the wave function factor $Z_k=1$ throughout, which is a good 
approximation  in the large-$N$ limit where RG corrections to the wave 
function renormalization of the relevant degrees of freedom, the Goldstone modes, are suppressed 
as $1/N$. In scalar $O(N)$ theories, 
the LPA gives already very good results for scaling at the Wilson-Fisher 
fixed point \cite{Litim:2002cf}. Here, the LPA does retain the full field- and scale-dependence of 
the superpotential $W_k$. 

In this work, we introduce the momentum cutoff as a supersymmetric 
invariant $F$-term  of the superfield, by adding 
$\Delta S_k=\s012\int d^3x\,\Phi R_k \Phi|_{\bar\theta\theta}$ 
to the action under the path integral, with
\begin{equation}
\Phi\, R_k(\bar\cD\cD)\,\Phi=-\s012
\Phi\,r_k(-\Box)\bar\cD\cD
\,\Phi\,.
\label{eq:regulator}
\end{equation}
The dimensionless function $r_k(p^2)$ describes the shape of the momentum 
cutoff.
The momentum trace is performed analytically for specific optimized choices 
for $r_k$ \cite{Litim:2000ci,Litim:2001fd,Litim:2001up}. Following \cite{Litim:2001up, Synatschke:2010ub}, we adopt 
\begin{equation}
 r_k(p^2) = \left(\frac{k}{|p|}-1\right)\theta (k^2-p^2)\,.
 \label{eq:LPA7}
   \end{equation}
The flow in LPA for the superpotential is obtained by 
projecting \eq{FRG} onto the term linear in the auxiliary field $F$, and
this yields
  \begin{equation}
    \begin{split}
 \frac{N}{k^2} \partial_t W=
 -  \left(N\!-\!1\right)I\left(\frac{W'}{k}\right)
 -I\left(\frac{W'\!+\!2\bar\rho\,W''}{k}\right),
 \label{eq:LPA5}
     \end{split}
   \end{equation}
 where 
$I(x)={x}/(1+x^2)$.
It is understood that $W$ and its derivatives are functions of the 
RG scale $k$ and the fields, and we will omit the index $k$.  The first
term on the RHS is the contribution of the $N-1$ Goldstone modes and
the last term is the contribution of the radial mode. Note that the RHS 
of the flow vanishes for $W'\equiv 0$, and for $1/|W'|\to 0$, corresponding 
to the classical limit where the couplings and the potential \eq{V_bosonic} 
are independent of the RG scale.

To achieve the simple form \eq{eq:LPA5} we have rescaled the fields and the 
superpotential as
\begin{equation}
\label{rescaling}
\bar\rho\to\frac{N}{8\pi^2}\bar\rho\,,\quad W\to \frac{N}{8\pi^2}\, W\,.
\end{equation}
Note that $W'$ is invariant under the rescaling which absorbs the
redundant overall factor $1/(8\pi^2)$, originating from the momentum integration, 
into the field and the superpotential. The additional rescaling with $N$ also 
removes the leading $N$-dependence from the RG equation \eq{eq:LPA5}.
In these conventions, and with given initial condition
$W_{k=\Lambda}(\bar\rho)$ the RG flow
determines the superpotential in the infrared limit $k \rightarrow 0$.

To study the critical behavior we introduce
a dimensionless field variable $\rho$, a dimensionless 
superpotential $w$ and a dimensionless 
scalar potential $v$ as
\begin{eqnarray}
\rho&=&\frac{\bar\rho}{k},\ \ 
w(\rho)=\frac{W(\bar\rho)}{k^2},\ \ 
 v(\rho)= 
   \frac{\bar\rho}{k}\!\left(\frac{W'(\bar\rho)}{k}\right)^2\!\!.
 \label{eq:LPA11}
\label{v}
\end{eqnarray}
In terms of \eq{eq:LPA11} the flow equation \eq{eq:LPA5} reads
\begin{equation}
\partial_t w+2w-\rho w'=
-\big(1-\frac{1}{N}\big)I(w')-\frac{1}{N}I(w' + 2\rho w'')\,.
\label{eq:LPA13}
\end{equation} 
For completeness we add the flow equation for $w'\equiv u$,
\begin{eqnarray}
\partial_t u
+u-\rho u'&=&-\big(1-\frac{1}{N}\big)u'\, I'(u)
\nonumber\\ &&
-\frac{1}{N}\left(3u' + 2\rho u''\right) I'(u + 2\rho u')\,,\quad
\label{flow-w'}
\end{eqnarray} 
and similarly for higher derivatives of the superpotential.

\section{Effective potential}\label{sec:EffPot}
In this section, we discuss the explicit and exact solution for the  effective potential 
in the limit $1/N\to 0$, and derive the main equations which govern the symmetry 
breaking in this model. 
\subsection{RG flow and boundary condition}
In the large-$N$ limit, the flow equation \eq{flow-w'} for
$u\equiv w'$ simplifies considerably and is given by
\begin{equation}\label{eq:LPA15}
\partial_t u+u-\rho\,u'=-\frac{1-u^2}{(1+u^2)^2}\,u'\,.
\end{equation}
The terms on the LHS encode the canonical 
scaling of the superpotential and the fields and the RHS encode the 
effects due to fluctuations. The integration of \eq{eq:LPA15} with respect to 
the logarithmic RG scale $t=\ln k/\Lambda$ gives
\begin{equation}
\frac{\rho-1}{u}-F(u)=G(ue^t)
\label{eq:LPA21}
\end{equation}
with
\begin{equation}\label{F}
   F(u)=\frac{u}{1+u^2}+2\arctan(u)\,.
\end{equation}
The function $G(x)$ is determined by the initial conditions 
for $u(\rho)$ imposed at some reference scale $k=\Lambda$. We use 
throughout the boundary condition
\begin{equation}\label{initial}
k=\Lambda:\quad 
\left\{
\begin{array}{ll}
\displaystyle
u(\rho)&=\tau\,(\rho-\kappa)\\[1ex]
W'(\bar\rho)&=\tau\,(\bar\rho-\kappa\,\Lambda)\,,
\end{array}
\right.
\end{equation}
where $\tau$ denotes the quartic superfield coupling at the cutoff. 
We recall that it is an exactly marginal coupling, i.e. that $\partial_t \tau = 0$. 
If the UV parameter $\kappa$ is positive, $\kappa\,\Lambda$ is interpreted
as VEV for the scalar field at $k=\Lambda$. 

Following \cite{Litim:2011bf}, the fixed point solutions are  parametrized 
in terms of the parameter
\begin{equation}
\label{c}
c=1/\tau\,. 
\end{equation}
 Then the function $G(x)$ is given by
\begin{equation}
\label{G}
G(x)=c-F(x)+\frac{\kappa-1}{x}
\end{equation}
in terms of the initial parameters. For initial conditions different from 
\eq{initial} the function is modified accordingly.

\subsection{Factorization} \label{subsec:Factorization}
Using the initial condition \eq{initial}, the analytical solution 
\eq{eq:LPA21} takes the form
\begin{equation}
\label{urho}
\begin{array}{rl}
\rho-\rho_0(t)&= c\,{u}+H(u)-H(ue^t)\,e^{-t}\\[1ex]
\rho_0(t)&=1+\delta\kappa\,e^{-t},\quad t=\ln k/\Lambda\,,
\end{array}
\end{equation}
where the non-negative function
\begin{equation}\label{H}
H(u)\equiv u\,F(u)=\frac{u^2}{1+u^2}+2u\arctan u
\end{equation}
encodes the RG modifications due to fluctuations \cite{Litim:2011bf}.
The parameter $\delta \kappa=\kappa-1$ measures the 
deviation of the VEV at the initial
scale $\rho_0(t=0)=\kappa$ from its critical value $\kappa_{\rm cr}=1$.
For any positive deviation we have $\rho_0(t)\to\infty$ in the infrared limit
corresponding to a finite VEV of the scalar field.  Since
the potential $V$ in \eq{V_bosonic} shows a second minimum at $\bar\rho=0$,
the global $O(N)$ symmetry is (not) spontaneously  broken if the finite (vanishing)
VEV is taken. Conversely, for a negative $\delta\kappa$ we have $\rho_0(t)<0$ in
the infrared limit such that the global minimum of the effective potential 
is achieved for vanishing $\bar\rho$.  This leaves the global $O(N)$ 
symmetry intact. The case $\delta\kappa=0$ then corresponds to the boundary 
between the symmetric and broken phases.

From \eq{urho} we conclude that the IR repulsive mode associated with
$\rho_0(t)$ is solely controlled by the initial VEV, independently of the
coupling strength $\tau$. This has been seen previously in purely scalar
theories in the large-$N$ limit \cite{Tetradis:1995br}. All the remaining 
couplings included in the 
potential are either exactly marginal or IR attractive. Their flow
is encoded in the term $H(ue^t)e^{-t}$ in the first equation of \eq{urho}.  
This factorization of the solution is a consequence of the large-$N$ limit, 
and allows for a straightforward analysis of the entire phase structure of 
the model. The global form of solutions $u(\rho,t)$ is mainly determined by 
the coupling $\tau=1/c$ and the function $H$, with $\rho_0$ only entering 
through a shift of the $\rho$-axis.

The non-negative function $H$ appearing in the implicit solution
\eq{urho} will be of importance below.  Expanding $H$ in powers
of $1/u$ leads to 
\begin{equation}
\label{Hlarge}
H=\pi\,|u|-1-\frac{1}{3u^2}+{\cal O}\big(\frac1{u^4}\big)\,.
\end{equation}
Conversely for small $u$ we find the expansion
\begin{equation}
\label{Hsmall}
H=3 u^2-\frac53 u^4+\frac75 u^6+{\cal O}(u^8)\,.
\end{equation}
The  solution \eq{urho} is invariant under $(c,u)\leftrightarrow (-c,-u)$ 
since $H(u)$ is an even function. Furthermore, the scalar field 
potential only depends on $u^2$ and we may restrict the discussion to $c\ge 0$. 

\subsection{Fixed points}
We briefly recall the main results from \cite{Litim:2011bf}. The fixed point 
solutions follow from \eq{eq:LPA21} by setting $G(u\,e^t)$ to a constant $c$, 
\begin{equation}
\rho=1+H(u_*)+cu_*\,.\label{fixpt}
\end{equation}
The constant $c$ is related to the marginal quartic superfield coupling 
$\tau=u'(\rho=\rho_0(t))$ as $c=1/\tau$.
Five characteristic values $c_I<c_L<c_P<c_M<c_G$ for $|c|$ have been identified:
\begin{eqnarray}
\label{cI}
\nonumber
c_I&=&0\\
\label{cL}
\nonumber
c_L&=&\s012(\pi+3)\\
\label{cP}
c_P&=&\pi\\
\label{cM}
\nonumber
c_M&=&\s023\pi+\s0{5}{8}\sqrt{3}\\[1ex]
\label{cG}
c_G&=&\infty\,.
\nonumber
\end{eqnarray}
The extreme values $c_I$ and $c_G$ correspond to the
`would-be' Wilson-Fisher and the Gaussian fixed point solution
$\rho(u_*)$, respectively. The solutions exist and extend over all physical field space
$\rho \geq 0$ in the weak coupling regime  $c_P\le |c|< c_G$. 
For $|c|\ge c_M$, fixed point solutions are monotonous functions 
of $u_*$ and extend over the entire real axis.
In the
intermediate coupling regime $c_L< |c|\le c_P$, fixed point solutions exists
both with and without a node at $\rho=1$. Finally, in the strong coupling regime $|c|\le
c_L$, the solutions do not extend over all fields $\rho \geq0$. Numerically,  the ranges
\begin{equation}
\label{range}
\frac{c_M-c_P}{c_P}\simeq 0.011\,,\quad
\frac{c_P-c_L}{c_P}\simeq 0.023
\end{equation}
are very small. The fixed points are non-Gaussian except for $|c|=c_G$, yet 
they displays Gaussian scaling for all physical fixed points except 
for $|c|=c_P$ or $c_I$. 

\subsection{Non-analyticities}
\label{Cusp}
Finally, we discuss  the appearance of non-analytic behavior in the
integrated flows at intermediate and strong coupling. This discussion completes the 
general description of fixed point solutions in  \cite{Litim:2011bf} and will be of help 
to understand the RG flows away from critical points in the next section. 

\begin{figure}[t]
 \begin{center}
\includegraphics[scale=1]{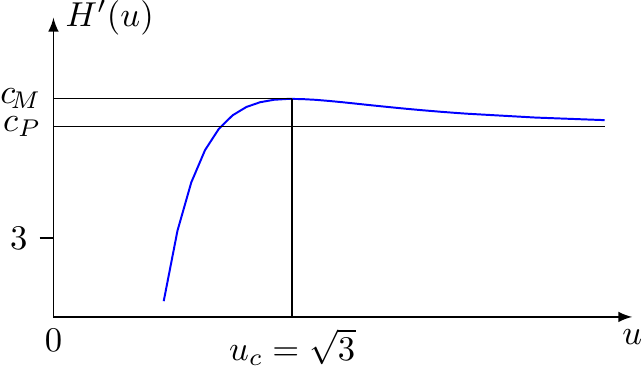}
\end{center}
\caption{The non-monotonic odd function $H'(u)$.}
\label{Hprime} 
\end{figure}

By construction, the basic flow equation \eq{FRG} is well-defined
(finite, no poles). Furthermore, the RHS of the supersymmetric flow  
\eq{eq:LPA15} is bounded, provided that the superpotential remains real.
Incidentally, this is in contrast to the standard purely bosonic flows, 
which potentially may grow large in a phase with spontaneous symmetry breaking. 
Despite their boundedness, the supersymmetric fixed point solutions display 
Landau-type poles at strong coupling due to non-analyticities, such as cusps, 
of the integrated RG flow. This can be appreciated as follows: consider the 
field-dependent dimensionless mass term $u'(\rho)$. 
From the fixed point solution (\ref{fixpt}) we conclude that it diverges provided that
\begin{equation}
\label{drho}
\frac{d\rho}{du}\big\vert_{u_s}=c+H'(u_s)=0.
\end{equation}
This condition determines the singular value $u_s$ and from
(\ref{fixpt}) we obtain the value of the singular field,
\begin{equation}
\label{rhos}
\rho_s= 1+ H(u_s)-u_s\,H'(u_s)\equiv \frac{1-u_s^2}{(1+u_s^2)^2}\,.
\end{equation}
The function $H'(u)$ is odd and bounded 
by $H'(u_c)=\pm c_M$.
Asymptotically, we 
have $|H'(u\to\pm\infty)|=c_P<c_M$, see Fig.~\ref{Hprime}.
Hence, with decreasing $|c|$ a 
divergence for $u'$ is first encountered for $|c|=c_M$. Performing 
an expansion of (\ref{fixpt}) up to the first non-trivial order, 
we find that 
\begin{equation}
\label{rhoc}
\rho-\rho_c =\frac{1}{6} H'''(u_c)(u_*-u_c)^3
\end{equation}
up to subleading corrections. In the expansion we used (\ref{drho}) and 
that $H''$ vanishes at $u_c$.
We note that \eq{rhoc} is continuous across $(u_*,\rho)=(u_c,\rho_c)$. Therefore, 
the non-analyticity in the solution can be written as
\begin{equation}
u_*-u_s=\mp\, {\rm sgn}(\rho-\rho_c)\left|\frac{\rho-\rho_c} {\frac16 H'''(u_c)}\right|^{1/3}\,,
\end{equation}
where the signs refer to $c=\mp c_M$, leading to a non-perturbative Landau pole
in $u'_*$, 
\begin{equation}
\label{cubic}
\frac{1}{u'_*}=
\mp \frac{9}{2} \left\vert H'''(u_c)\right\vert^{1/3}\, \left|{\rho-\rho_c}\right|^{2/3}\,.
\end{equation}
At a fixed point solution, the Landau pole remains invisible, because it 
is achieved at the negative $\rho_c=-1/8$. Increasing the coupling by lowering $|c|$ 
below $c_M$, the expansion in the vicinity of $d\rho/du=0$ becomes
\begin{equation}
\label{rhoq}
\rho-\rho_s=\frac{1}{2}H''(u_s)(u_*-u_s)^2
\end{equation}
up to subleading terms, where $u_s$ is determined through (\ref{drho}). In this
regime, $H''(u_s)$ is non-zero throughout. In the parameter range $c_P\le |c|<c_M$ 
we find two solutions for $u_s$ with $|u_{s1}|<|u_c|<|u_{s2}|$
and $H''(u_{s1})<0<H''(u_{s2})$. 
Effectively, the solution for the superpotential 
becomes multi-valued in a limited region of field space. For $|c|<c_P$ we 
find one solution for $u_s$ with $H''(u_s)>0$.  In contrast to \eq{cubic}, 
the non-analyticity has turned into a square root,
\begin{equation}
\label{quadratic} 
\frac{1}{u'_*}=\pm 2\left\vert H''(u_s)\right\vert^{1/2}\, \left({\rho-\rho_s}\right)^{1/2}\,.
\end{equation}
The non-analyticity \eq{quadratic} is stronger than \eq{cubic} and  the solution
\eq{rhoq} cannot be continued continuously beyond the point
$(u_*,\rho)=(u_s,\rho_s)$. For $|c|< c_L$, we have that $\rho_s(c)>0$ and the
pole appears in the physical regime. In contrast, the solutions extend over
all fields provided that $\rho_s\le 0$ which is the case for $|c|\geq c_L$.

It is interesting to note that non-analyticities, such as cusps, have been 
detected previously in the context of the random field Ising model, where 
disorder is technically introduced with the help of Parisi-Sourlas supersymmetry. Using 
functional renormalization, it has been argued that a cusp behavior
at finite ``Larkin scales" $k=k_L>0$ is at the origin for
the spontaneous breaking of Parisi-Sourlas supersymmetry \cite{Tissier:2011zz,Tissier:2011mu,Tissier:2011mv}.
 
At this point it should be mentioned that the superpotential $W^{\prime}$ shows
another non-analytic behavior: It is not differentiable at its node $\bar\rho_0$ in the exact IR limit for
arbitrary couplings $c>0$. This issue is discussed in detail in Sec.~\ref{subsec:SSB} and \ref{subsec:effectivep} below.

\section{Renormalized field theory}\label{sec:RenormalizedTheory}  
In this section, we discuss the spontaneous breaking of symmetry and the 
phase structure of the model in the limit where the UV scale $\Lambda$ is removed. 

\subsection{Renormalization}
The solution \eq{urho}  is valid for all $k$ and  $\Lambda$, and we may take 
the `continuum  limit' $1/\Lambda\to 0$.  The term containing the explicit $t$-dependence
drops out in the continuum limit,  in consequence  of the 
limit $k/\Lambda\to 0$ for fixed and finite $k$ and (\ref{Hsmall}). The remaining scale-dependence 
solely reduces to the implicit scale-dependence of $\rho_0(k)$ in
\begin{equation}
\label{urhoinf}
\begin{array}{rl}
\rho-\rho_0(k)&= c\,{u}+H(u)\\[1ex]
\rho_0(k)&=1+\bar\rho_0/k\,.
\end{array}
\end{equation}
The dimensional parameter $\bar\rho_0$ has taken over the 
role of $\delta\kappa\,\Lambda$ in \eq{urho}. In the above, the VEV (or the mass 
term, respectively)  is the only quantity which is non-trivially renormalized 
in the continuum limit by requiring that
\begin{equation}
\label{continuum}
\bar\rho_0\equiv\lim_{\Lambda\to\infty}(\delta\kappa(\Lambda)\, \Lambda )<\infty\,.
\end{equation}
Consequently, the canonical dimension of fields remain unchanged (no anomalous 
dimension). The continuum limit maps the original set of free 
parameters $(\tau,\kappa,\Lambda)$ to the parameters $(\tau,\bar\rho_0)$.    
Note that all couplings of the superfield derivative -- the 
marginal coupling $c$ and the IR attractive higher-order couplings $u^{(n)}(\rho_0)$
 -- have settled on their fixed point values.  The only `coupling' which 
has not settled on a fixed point is the UV attractive dimensionless VEV $\rho_0$.
With this perspective, $\bar\rho_0$ and the non-renormalized parameter $c$ 
should be viewed as a free parameters of the model, fixed by the microscopic 
parameters of the theory. In terms of the dimensional fields $\bar\rho = \rho\,k$ 
and superfield  derivative $W'(\bar\rho)=u(\rho)\,k$, the integrated RG flow becomes  
\begin{equation}
\label{UVdim}
\begin{array}{rl}
\bar\rho-\bar\rho_0(k)&= \displaystyle
c\,W'+k H\left(W'\!/k\right)\\[1ex]
\bar\rho_0(k)&=k+\bar\rho_0\,.
\end{array}
\end{equation}
We note that $\bar\rho_0$ also has the interpretation of the physical VEV in 
the infrared limit of the theory, provided it is positive.
Below, we find it is useful to switch between the representations \eq{urhoinf} 
and \eq{UVdim}.

\subsection{Characteristic energy}
The RG flow \eq{urhoinf}, \eq{UVdim} carries a characteristic energy scale $E$, 
meaning that  the theory changes its qualitative behavior depending 
on whether  fluctuations have an energy larger or smaller than $E$. The energy
scale is set by the UV renormalization of the model \eq{continuum} and given by 
\begin{equation}
E=|\bar\rho_0|\,.
\end{equation}
For $k\gg E$, the dimensionful VEV scales proportional to $k$,  and the 
dimensionless parameter $\rho_0$ becomes a constant. This corresponds to a fixed point. 
All other dimensionless couplings equally have stopped to evolve with RG scale
and thus the entire solution approaches a high-energy (UV) fixed point. This 
fixed point would persist for all $k$ provided that $E=0$. It then has also 
the interpretation of an IR fixed point. This regime is most conveniently 
described using \eq{urhoinf}. For $E>0$, and with decreasing $k$, deviations 
from the fixed point become visible once $k$ reaches $E$. Here the VEV 
displays a cross-over from linear scaling $\bar\rho(k)\propto k$ for $k\gg E$ 
to the constant value $\bar\rho_0$ for $k\ll E$. In full 
analogy, the dimensionless VEV displays a cross-over from a constant value to 
scaling inversely proportional to the RG scale. In addition, the running of 
all dimensionful couplings in the potential is switched on once $k\approx E$ 
and below. This regime is conveniently described using \eq{UVdim} which governs 
the remaining RG running through its RHS.

\subsection{Gap equations}

\begin{figure}[t]
\begin{center}
\unitlength0.001\hsize
\begin{picture}(900,1050)
\put(0,130){\includegraphics[width=.415\textwidth]{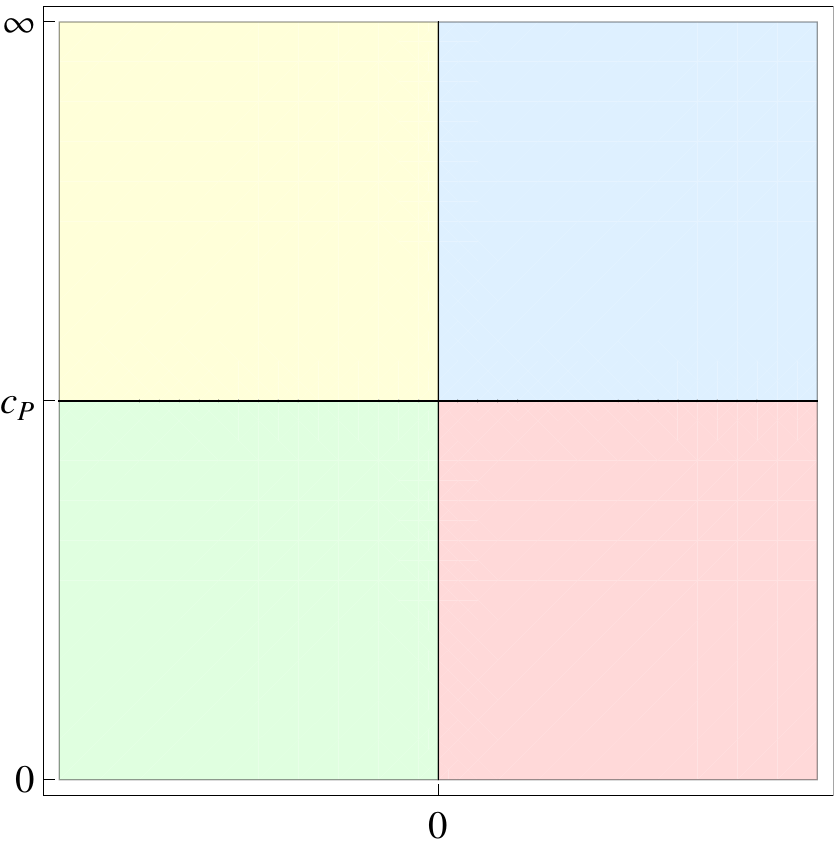}}
\put(-30,920){\Large $\frac{1}{\tau}$}
\put(250,750){$m$}
\put(600,750){$M\ \ \ M_\rho$}
\put(210,400){$m\ \ \ M$}
\put(650,400){$M_\rho$}
\put(220,920){SYM}
\put(620,920){SSB}
\put(850,130){\large $\bar\rho_0$}
\end{picture}
\vskip-1cm
\caption{Schematic phase diagram of the supersymmetric model based on the gap
equation \eq{gap} in the infinite cutoff limit. Results agree with earlier
findings  in \cite{Bardeen:1984dx,Moshe:2003xn}.}
\label{pGap} 
\end{center}
\end{figure}
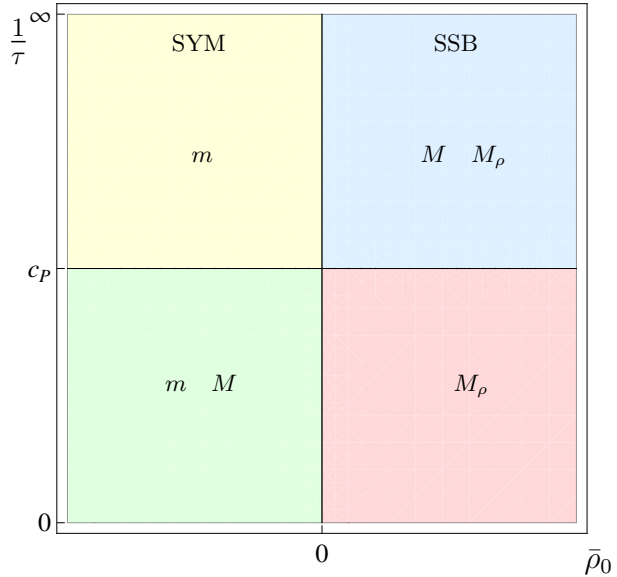

We first discuss the phase structure based on the integrated RG equations in the IR limit $k=0$, see Fig.~\ref{pGap}. 
This allows for a direct comparison with earlier results based on gap equations and 
Schwinger-Dyson equations \cite{Bardeen:1984dx,Moshe:2003xn}. 
In the infrared limit we may use (\ref{Hlarge}) in \eq{UVdim} and
obtain
\begin{equation} 
\label{Inf}
\bar\rho-\bar\rho_0= c\,W'+c_P\left|W'\right|\,.
\end{equation}
Since the potential shows a local minimum at vanishing field,
the squared particle masses are given by
\begin{equation}
\bar{\mu}^2=\left.V''(\phi)\right|_{\phi=0}=\left.W^{\prime\,
2}(\bar\rho)\right|_{\bar{\rho}=0}.
\label{defmass}
\end{equation}
Thus, \eq{Inf} becomes a gap equation for
the mass parameter $\bar\mu\equiv W'(\bar\rho=0)$,
\begin{equation}
\label{gap}
\bar\rho_0= -c\,\bar\mu -c_P\left|\bar\mu\right|\,.
\end{equation}
The significance of \eq{gap} is as follows. For fixed $\bar\rho_0$ and
$c$ it yields the possible infrared solutions for the masses at
vanishing field. Without loss of generality we restrict the discussion to $c\geq
0$. For non-vanishing $\bar\rho_0$ we find two solutions
\begin{equation}
\label{gapBrancha}
\begin{array}{rl}
m\,=\bar\mu=&
\displaystyle
-\frac{\bar\rho_0}{c_P+c}\geq 0\\[2ex]
M\!=-\bar\mu=&
\displaystyle
-\frac{\bar\rho_0}{c_P-c}\geq 0\,.
\end{array}
\end{equation}
In the symmetric regime with negative $\bar\rho_0$ the mass $m$ is always present
and the second mass $M$ is available as long as $c<c_P$.
In the SSB regime with positive $\bar\rho_0$ there are two degenerate ground
states: As expected, we find a non-symmetric ground state with a radial
mass $M_\rho$, see sections ~\ref{subsec:effectivep} and \ref{subsec:critical}.
However, for $c>c_P$, the gap equations show an additional symmetric ground state, 
characterized by  the mass $M$. Note that changing the sign of $c$ leads to equivalent results 
under the following replacements 
\begin{equation}\label{symmetry}
(c, m, M,M_\rho)\leftrightarrow (-c, M, m, -M_\rho)\,.
\end{equation}
At the phase transition, i.e. for $\bar\rho_0= 0$, the gap equation \eq{gap}
states that either $c=\pi$ with the mass $M>0$ undetermined, or $c=-\pi$ and 
the mass $m$ undetermined. These findings agree with the earlier ones 
from \cite{Bardeen:1984dx,Moshe:2003xn}. The sole difference is that the 
value for the critical coupling, $c_P$, depends on the regularization.
The precise link to the conventions used in \cite{Bardeen:1984dx,Moshe:2003xn} 
is given in Tab.~\ref{tab:translation}.

\label{sec:translationGuide}
\begin{table}[t]
\begin{ruledtabular}
\begin{tabular}{lcc}
this paper&$\bar\rho_0$&$\tau=1/c$\\
Bardeen et.\,al. \cite{Bardeen:1984dx} & $-4\pi^2{\mu}{\lambda}^{-1}$&$({4\pi^2})^{-1}
\lambda$\\
Moshe and Zinn-Justin \cite{Moshe:2003xn}&
$-4\pi^2({\mu-\mu_c})u^{-1}$&$({4\pi^2})^{-1}u$
\end{tabular}
\end{ruledtabular}
\caption{``Translation guide" between the conventions used in \cite{Bardeen:1984dx},  \cite{Moshe:2003xn}, and this paper.\label{tab:translation}}
\end{table}

\subsection{RG phase diagram}

Next we discuss the phase diagram implied by the integrated RG equations for all scales $k$, and compare with the results based on the $k=0$ limit.

\subsubsection{Graphical representation}

We begin with a useful graphical representation of the renormalized  RG trajectories \eq{urhoinf}. 
For vanishing $\bar\rho_0$, we note that the trajectories \eq{urhoinf} 
reduce to the fixed point solutions  $u_*(\rho)$ analyzed in \cite{Litim:2011bf}.
The only difference with the fixed point solutions is related to a shift of the 
argument, 
\begin{equation}
\label{link}
u(\rho)=u_*(X)\,,\quad X\equiv \rho+1-\rho_0(k)=\rho-\frac{\bar\rho_0}{k}
\end{equation}
in terms of the fixed point solutions.

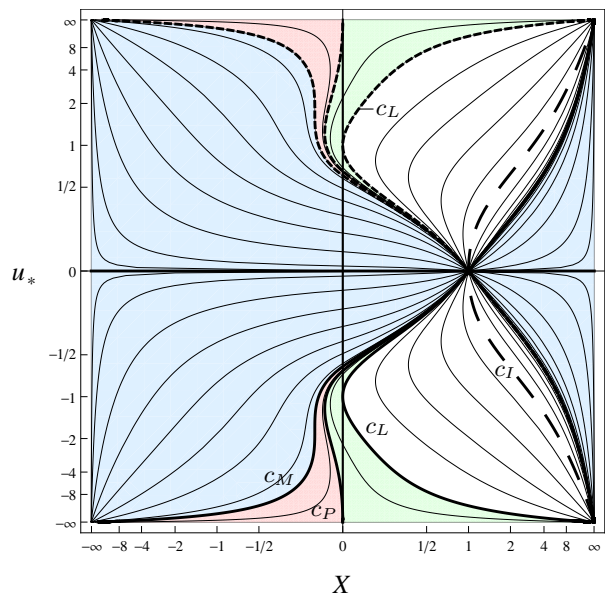
\begin{figure}[t]
\begin{center}
\unitlength0.001\hsize
\begin{picture}(900,1050)
\put(0,130){\includegraphics[width=.44\textwidth]{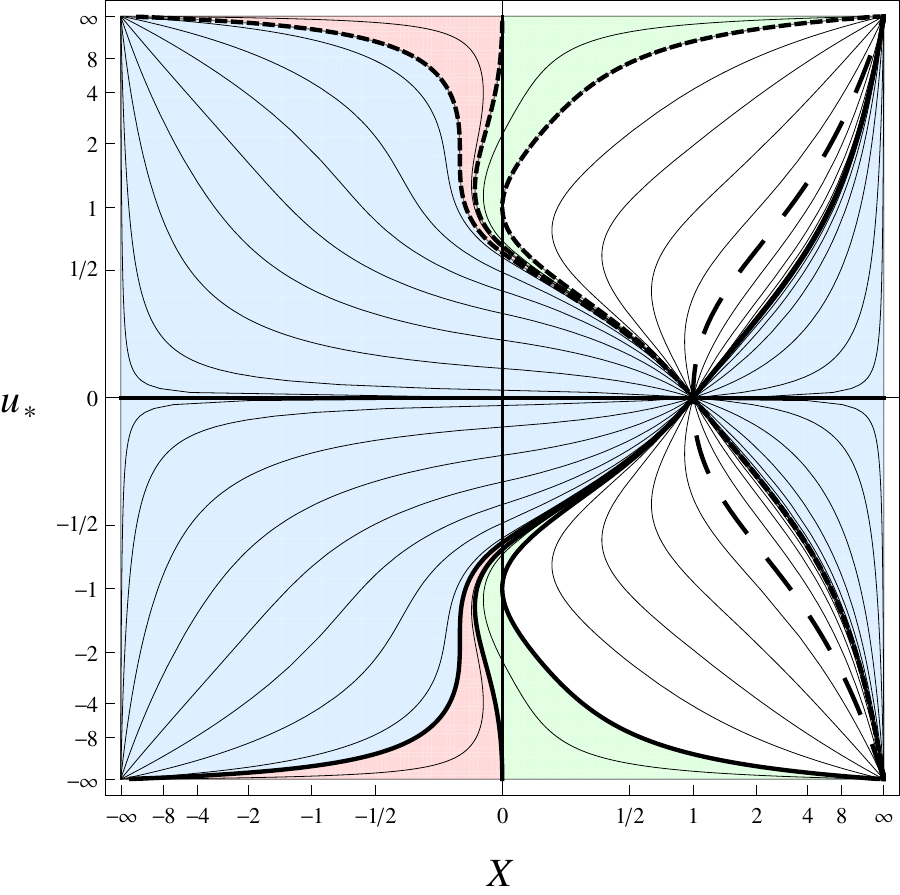}}
\put(527,869){$-c_L$}
\put(540,383){$c_L$}
\put(385,307){$c_M$}
\put(457,255){$c_P$}
\put(737,470){$c_I$} 
\end{picture}
\vskip-1cm  
\caption{Graphical representation of the solutions $u_*(X)$ of \eq{urhoinf},
		where  $X=\rho-\bar\rho_0/k$. The
		shaded areas are separated by thick lines at $|c|=c_I,c_L,c_P,c_M$ and $c_G$.} \label{pXY}
\end{center}
\end{figure}

The structure of the solutions and their dependence on the constant $c$ is 
shown in Fig.~\ref{pXY}.  
Once the free parameters are fixed, the RG evolution of a particular solution 
stays on a curve with constant $c$, indicated by the curves given in  the Figure.
Rotating counter-clockwise around $(X,u_*)=(1,0)$ from the horizontal $c_G$-line 
to the $c_I$-curve (from the $c_I$-curve to the $c_G$-line) covers all curves with 
positive (negative) $c$. Both sets connect through the point $(1,0)$. We recall that
$(c,u_*)\leftrightarrow (-c,-u_*)$ describe  equivalent physics.

Using \eq{link} and \eq{urhoinf}, we conclude that for $u(\rho)$ to cover all 
physical fields $\rho\in[0,\infty]$, we need that 
\begin{equation}
\label{X}
X\in \left[-\bar\rho_0/k,\infty\right]\,.
\end{equation} 
The curves $u_*(X)$ in Fig.~\ref{pXY} define monotonous (and invertible)
functions provided that $X>1$. A unique classification of curves is then achieved 
by choosing a value for $u_*$ on a line of constant $X>1$, together with 
fixing $\bar\rho_0$. Interestingly, two different values for $u_*$ may correspond 
to one and the same parameter $c$. Below, we mostly stick to the classification 
in terms of $c$, and we will highlight situations where this is no longer sufficient.

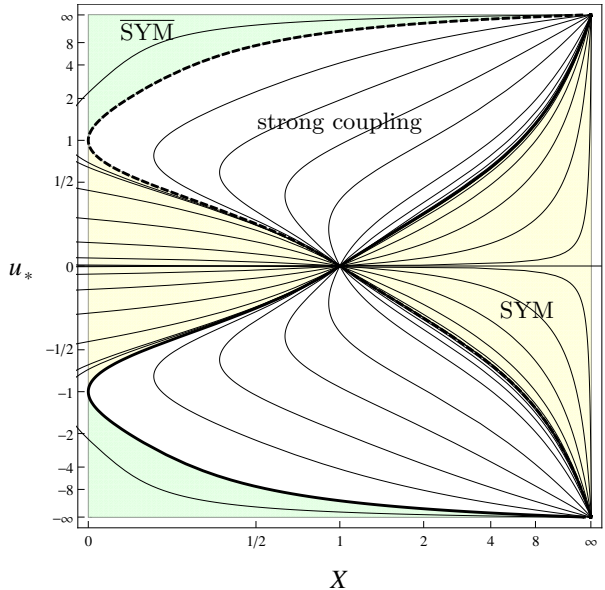
\begin{figure}[t]
\begin{center}
\unitlength0.001\hsize
\begin{picture}(900,1050)
\put(0,130){\includegraphics[width=.44\textwidth]{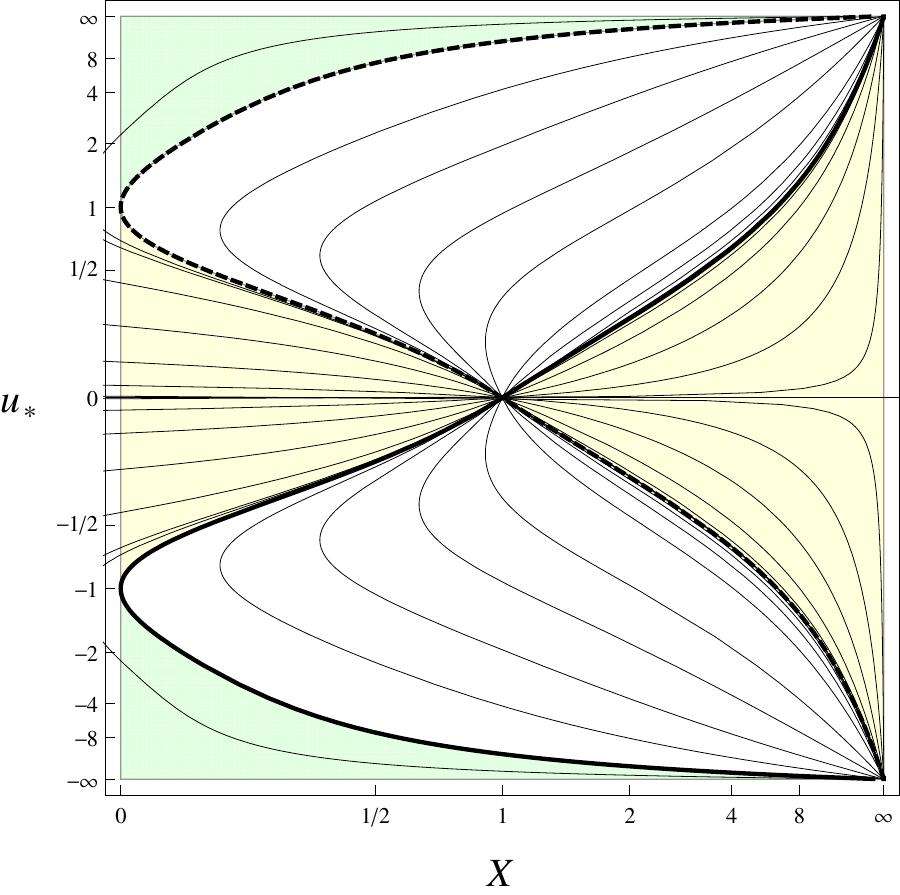}}
\put(380,840){strong coupling}
\put(170,975){$\mathrm{\overline{SYM}}$}
\put(750,550){SYM}
\end{picture}
\vskip-1cm
\caption{RG trajectories in the $O(N)$ symmetric phase: at weak coupling, 
trajectories either  show a non-vanishing VEV for large scales (SYM, yellow shading), 
or a vanishing VEV  for all scales ($\overline{\rm SYM}$, green shading). At strong 
coupling trajectories  terminate at Landau poles.}
\label{pSYM} 
\end{center}
\end{figure}

\begin{figure}[t]
\begin{center}
\unitlength0.001\hsize
\begin{picture}(900,1050)
\put(0,130){\includegraphics[width=.44\textwidth]{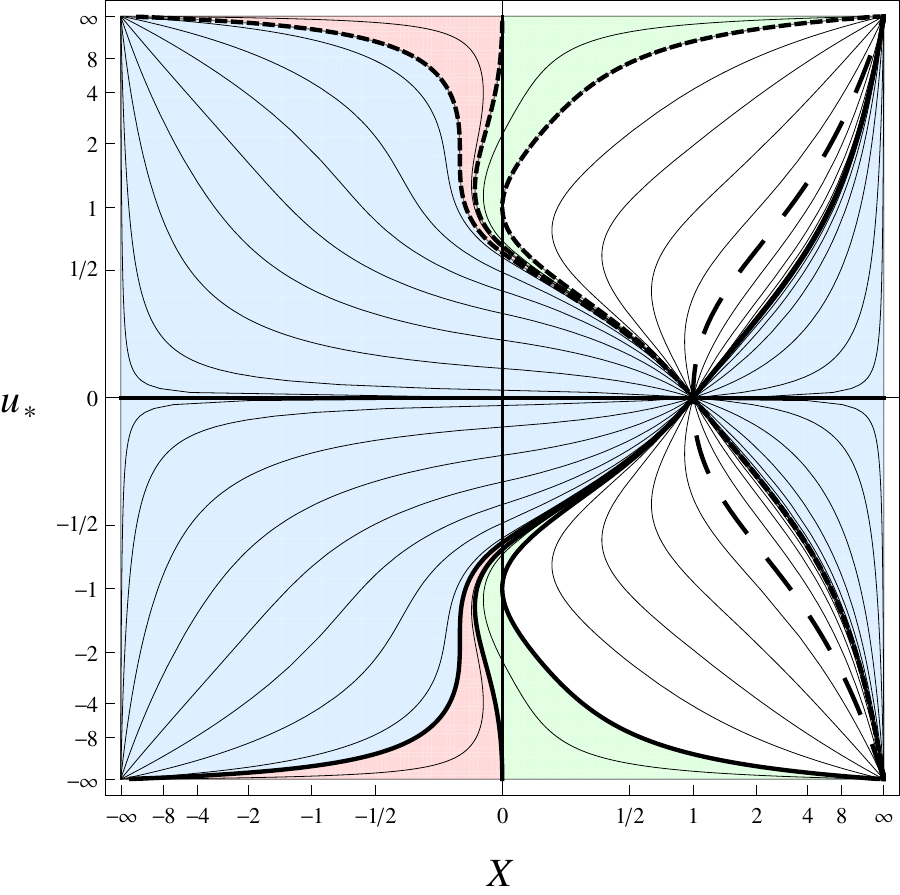}}
\put(250,900){SSB}
\put(377,992){\small{Landau}}
\put(545,850){strong coupling}
\end{picture}
\vskip-1cm
\caption{RG trajectories $u_*(X)$ according to \eq{urhoinf}  in the regions with
spontaneous breaking of the $O(N)$ symmetry in the parametrization \eq{link}. Couplings are either 
finite for all $k$ (SSB, blue shading), or run into a singularity (Landau, 
red shading). Some trajectories cannot be continued beyond the Landau pole
(magenta shading). The SSB phase cannot be defined for strong coupling (white area).}
\label{pSSB} 
\end{center}
\end{figure}

\subsubsection{Symmetric regime} 
The symmetric phase is characterized by a finite and 
negative $\bar\rho_0$ and for large scales
$X$ reduces to $\rho$. A restriction on the coupling parameter $c$ is imposed if 
we require that the solution $u$ should exist for all $\rho$. For weak coupling,
\begin{equation}
\label{cSYMweak}
\bar\rho_0<0\,,\quad  |c|\ge c_P
\end{equation}
all $u_*$ are single-valued for non-negative arguments such that the 
$u_*(X)$ stay well-defined for all scales, see Fig.~\ref{pSYM}. For intermediate coupling
\begin{equation}
\label{cSYMinter}
\bar\rho_0<0\,,\quad  c_P\ge |c|\ge c_L
\end{equation}
the theory admits two distinct effective potentials, and two scalar mass parameters. 
They are related to  trajectories which either run through a node, or not, depending 
on whether $u(0)$ for $k\gg E$ is larger or smaller than 1, see Fig.~\ref{pSYM}. 
The theory is then characterized by the coupling and the scalar mass at vanishing 
field. This peculiar structure has been found previously and we discuss it in 
more detail below. 

\subsubsection{Symmetry broken regime}
Spontaneous symmetry breaking is possible for positive $\bar\rho_0$. 
This requires that $u_*(X)$ has to be 
well-defined for all real $X$. In view of the analytical solution 
in Fig.~\ref{pXY}, this limits the achievable couplings to
\begin{equation}
\label{cSSBP}
\bar\rho_0>0,\quad |c|>c_P.
\end{equation}
Smaller $|c|$ do not lead to a well-defined physical theory in the IR. For 
\begin{equation}
\label{cSSBweak}
\bar\rho_0>0,\quad |c|\ge c_M
\end{equation}
the function $u_*$ is one-to-one and the theory described by $u(\rho)$ 
in \eq{urhoinf} remains  well-defined  even in the IR limit. 
The theory is then 
characterized by two mass scales. The first one is given by the scalar mass at vanishing
field corresponding to an $O(N)$ symmetric phase, whereas the second mass 
scale is given by  the radial mass at $\bar\rho=\bar\rho_0$ allowing for SSB.

\subsubsection{Strong coupling and Landau regime} \label{sec:Strong}
\begin{figure}[t]
\begin{center}
\unitlength0.001\hsize
\begin{picture}(900,700)
\put(0,130){\includegraphics[width=.42\textwidth]{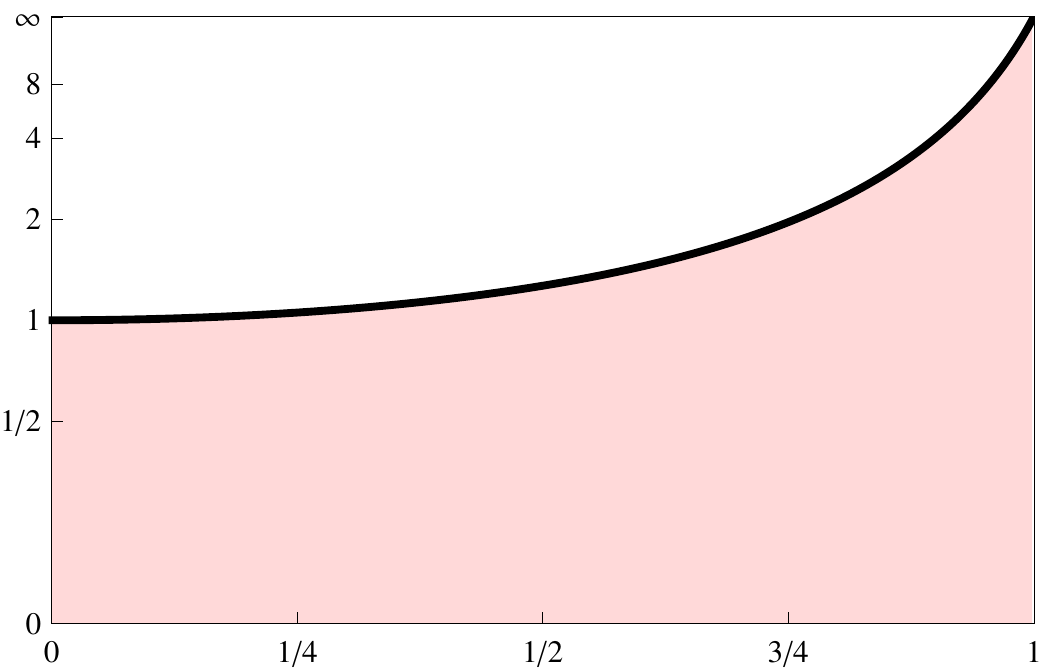}}
\put(-60,610){ $\displaystyle \frac{k}{E}$}
\put(420,70){\large $c/c_L$}
\put(320,460){$k_L$}
\put(420,300){supersymmetry}
\end{picture}
\vskip-.5cm
\caption{Location of the UV Landau pole for trajectories in the SYM phase at 
strong coupling with \eq{cSYMstrong} (see text).} 
\label{pLandau} 
\end{center}
\end{figure}
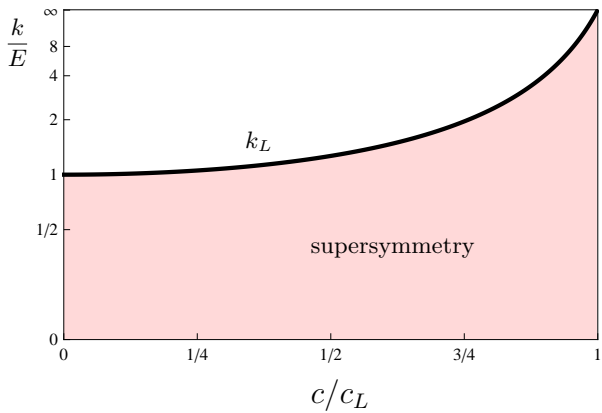

It remains to discuss the strong coupling and Landau regimes in Figs.~\ref{pSYM} 
and~\ref{pSSB}.  
We begin with trajectories in the SYM regime, with
\begin{equation} 
\label{cSYMstrong}
\bar\rho_0<0\,,\quad  |c|<c_L\,.
\end{equation}
We take a `bottom-up'  view according to which the couplings evolve from the
infrared towards higher scales, parametrizing the effective potential in terms of local
couplings in an expansion about vanishing field. Trajectories with
\eq{cSYMstrong} emanate from the upper/lower-right corner in Fig. \ref{pSYM}
for $k\approx 0$ and increasing $k$ corresponds to decreasing $X$. With 
increasing $k$, the running mass term and the fermion-boson coupling at 
vanishing field $u'(\rho=0)\equiv u_*'(-\bar\rho_0/k)$  diverge at  $k=k_L$, 
and the renormalized RG flow comes to a halt: the solutions \eq{urhoinf} 
cannot be continued beyond these points, because $X$ cannot decrease any 
further along the integral curve $u_*(X)$. 
Interestingly, the potential is double-valued for $k<k_L$ with two different 
trajectories terminating at the same Landau pole.  Using \eq{rhoq} together 
with \eq{link}, the non-analyticity in $u$ reads
\begin{equation}
\label{rhost}
\begin{array}{rl}
\rho-\rho_s(k)&=\frac{1}{2}H''(u_s)(u(\rho)-u_s)^2
\end{array}
\end{equation}
and the Landau poles are located at
\begin{equation}
\label{rhosk}
\rho_s(k)=\rho_s-1+\rho_0(k)=\rho_s+\frac{\bar\rho_0}{k}\,.
\end{equation}
From the fixed point solution we know that $\rho_s\le 1$ and therefore 
$\rho_s(k)\le \rho_0(k)$ for all $k$. In the IR limit, this implies 
that $\bar\rho_s(k)\to \bar\rho_0(k)$ from below.  Here, the values 
for $\rho_s$ are fixed by the coupling strength $c$ via \eq{rhos}
and is positive in the regime \eq{cSYMstrong}.  
From \eq{rhosk} it follows that $k_L=-\bar\rho_0/\rho_s$ is positive, see Fig.~\ref{pLandau}. 
We conclude that the parameters \eq{cSYMstrong} allow for a supersymmetric model 
with linearly realized $O(N)$ symmetry up to scales $k=k_L$.  

Next we discuss the SSB regime starting with intermediate couplings 
\begin{equation}
\label{cLandau}
\bar\rho_0>0\,,\quad  c_P\le |c|\le c_M\,.
\end{equation}
Here all curves of constant $c$ contain two Landau points 
with $|u_{s1}|<|u_c|<|u_{s2}|$ and $H''(u_{s1})<0<H''(u_{s2})$ (see Sec.~\ref{Cusp}). 
Each of them is described by \eq{rhost} with \eq{rhosk} and parameters 
$0>\rho_{s2}(c)>\rho_{s1}(c)$. The singularity at $(\rho_{s1},u_{s1})$ corresponds 
to  an IR Landau pole (`top-down'), whereas the one at $(\rho_{s2},u_{s2})$ 
corresponds to an UV Landau pole (`bottom-up').  In the infrared limit, the 
domain where $u$ is multi-valued, collapses to a point with 
$\Delta \bar\rho= k\,(\rho_{s1}-\rho_{s2})\to 0$. The location of both 
discontinuities approach the VEV $\bar\rho_s(k)\to \bar\rho_0$ from below, 
and the discontinuity in the superpotential derivative
\begin{equation}
\label{disc}
\Delta W'\equiv W'(\bar\rho_{s1})-W'(\bar\rho_{s2})=k(u_{s1}-u_{s2})
\end{equation}
 then also becomes arbitrarily small.  Interestingly, the UV and IR Landau poles become degenerate on the integral curve for $|c|=c_M$ where $\rho_{s1}=\rho_{s2}=-1/8$. The non-analyticity evolves with
\begin{equation}
\label{rhoct}
\begin{array}{rl}
u(\rho)-u_s&\displaystyle
=\mp\, {\rm sgn}(\rho-\rho_s(k))\left|\frac{\rho-\rho_s(k)} 
{\frac16 H'''(u_s)}\right|^{1/3}
\end{array}
\end{equation}
together with \eq{rhosk}. In this case, the quartic scalar self-coupling 
$u'(\rho=0)$ still diverges at the Landau pole, but the renormalized RG 
flow continues non-perturbatively rendering $u'(0)$ again finite. The 
non-analyticity \eq{rhoct} first appears for vanishing field at the 
scale $k_L=-\bar\rho_0/\rho_s$ and evolves up to the VEV $\bar\rho_0$ in the IR limit.

Next we consider  the SSB regime at strong coupling,
\begin{equation}
\label{cStrong}
\bar\rho_0>0\,,\quad  |c|<c_L\,.
\end{equation}
The model has a radial mass proportional to the VEV. Curves of constant $c$ 
in Fig.~\ref{pSSB} have a Landau pole with (\ref{rhost},\ref{rhosk}) and
parameter $\rho_s>0$. The integral curves have no continuation beyond the pole, 
which occurs within the physical regime for all $k$. In particular, the 
effective potential is not defined for the entire inner part $\bar\rho<\bar\rho_0$ 
in the IR limit and a scalar mass $W'(0)$ cannot be defined. 

Finally we consider  trajectories in the SSB regime, with
\begin{equation}
\label{cStrong2}
\bar\rho_0>0\,,\quad  c_L<|c|<c_P\,.
\end{equation}
Here, in contrast to \eq{cStrong},  solutions \eq{link} cover all positive 
values for $X$ even for large $k$; see  \eq{X}. In a `top-down' perspective 
(with decreasing $k$) trajectories in the regime \eq{cStrong2} emanate 
at $X\approx 0$ and continue towards smaller $X$. Again, all trajectories 
reach a Landau pole for the quartic (and higher) superfield coupling at 
vanishing field, given by \eq{rhost} and \eq{rhosk} with the parameter 
$\rho_s(c)$ taking negative values. The Landau scale reads $k_L=-\bar\rho_0/\rho_s>0$, 
and the effective potential does not exist for fields below 
$\bar\rho_s(k)=k(\rho_s-1)+\bar\rho_0(k)\le\bar\rho_0(k)$. As in \eq{cStrong}, 
the theory still has a radial scalar mass set by the VEV and the quartic coupling, 
because the one-sided derivative  $\frac{dW'}{d\bar\rho}|_{\bar\rho_0}$ 
with $\bar\rho\ge\bar\rho_0$ can be taken for fields larger than the VEV. 
In turn, a scalar mass at vanishing field cannot be defined. 
Therefore we conclude that the renormalized RG flow cannot be continued 
towards the infrared for scales below the Landau scale $k<k_L$ for 
parameters \eq{cStrong2}.

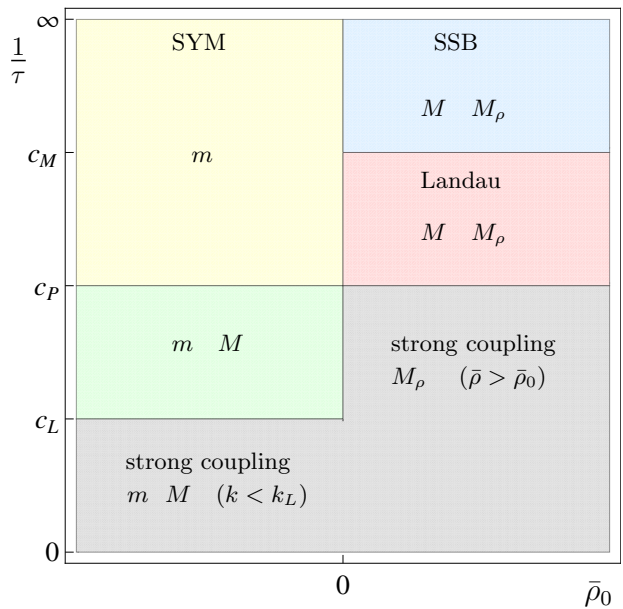
\begin{figure} 
\begin{center} 
\unitlength0.001\hsize
\begin{picture}(900,1050)
\put(0,130){\includegraphics[width=.44\textwidth]{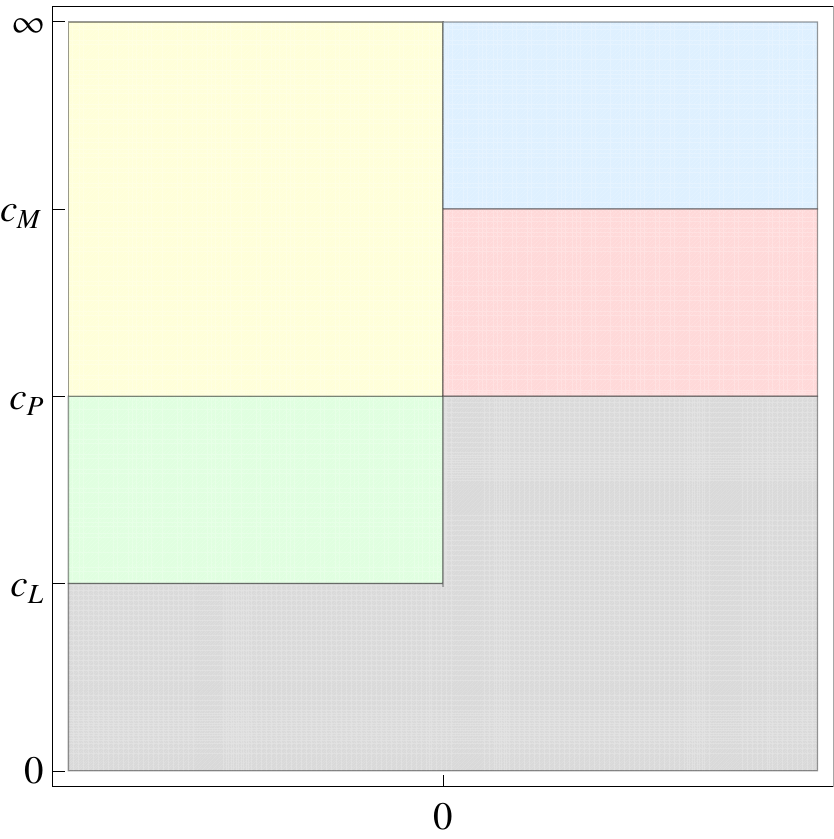}}
\put(-30,940){\Large $\frac{1}{\tau}$}
\put(250,800){$m$}
\put(220,510){$m\ \ \ M$}
\put(150,330){strong coupling}
\put(152,280){$m\ \  M\ \ \ (k<k_L)$}
\put(600,870){$M\ \ \ M_\rho$} 
\put(600,760){Landau}
\put(600,680){$M\ \ \ M_\rho$}
\put(555,510){strong coupling}
\put(555,460){$M_\rho\ \ \ \ (\bar\rho>\bar\rho_0)$}
\put(220,970){SYM}
\put(620,970){SSB}
\put(850,130){\large $\bar\rho_0$}
\end{picture}  
\vskip-1cm 
\caption{Schematic phase diagram based on the RG in the infinite cutoff limit.
The scale $k_L$ is given in Fig.~\ref{pLandau}. The tighter constraints as
opposed to Fig.~\ref{pGap} arise from the inspection of the full effective
potential at all scales $k$. The parameter range between $c_M$, $c_P$ and $c_L$ 
is very narrow \eq{range}.}
\label{pRG} 
\end{center}
\end{figure}

\subsection{Discussion}
Our results are summarized in Fig.~\ref{pRG} and should be compared with 
Fig.~\ref{pGap}. The phase diagram is given in dependence on the coupling 
parameter $c$ and the scale parameter $\bar\rho_0$. 

In the SYM regime, the theory has a weakly coupled phase  with a scalar mass 
$m$ where both the $O(N)$ symmetry and supersymmetry are preserved \eq{cSYMweak}.
With increasing coupling parameter $\tau$, the theory admits two $O(N)$
symmetric phases with two mass scales $m$ and $M$ \eq{cSYMinter}. This regime 
has a very narrow width in parameter space, see \eq{range}, which is sensitive 
to the underlying regularization. For strong coupling \eq{cSYMstrong}, the 
theory displays two mass scales $m$ and $M$. However, it  is also plagued by 
Landau-type singularities which admit no solution for the superpotential at 
scales above the Landau scale $k_L$. This is not visible from an evaluation 
of the IR gap equations alone, see Fig.~\ref{pGap} for comparison.

In the SSB regime, the theory has a weakly coupled phase \eq{cSSBweak} where the
$O(N)$ symmetry could be spontaneously broken and the effective potential for
the scalar has two degenerate minima corresponding to two mass scales $M$ and
$M_{\rho}$.  The first mass scale is associated with an $O(N)$ symmetric phase,
whereas the second mass scale emerges from a finite VEV allowing for SSB. 
Furthermore, global supersymmetry remains intact. With increasing coupling 
parameter $\tau$, the theory enters a narrow parameter range where 
RG trajectories would run through a series of Landau poles at intermediate 
energies \eq{cLandau}. Here, the discontinuity in field space and in the 
superpotential derivative shrinks to zero in the IR limit, the details of 
which are sensitive to the underlying regularization. For even larger 
couplings $|c|<c_P$ \eq{cStrong} and \eq{cStrong2}, the theory is so strongly 
coupled that RG trajectories terminate at Landau poles in the physical regime. 
The effective potential does not exist for fields below the non-trivial 
VEV $\bar\rho<\bar\rho_0$ in the IR limit. Still, the potential does admit 
a radial mass $M_\rho$.

Unbroken global supersymmetry requires a ground state with vanishing energy, and an elsewise positive dimensionful effective potential for all fields and all RG scales. Strictly speaking, the non-existence of an effective potential for small fields means that we cannot decide, based on the potential alone, whether supersymmetry is spontaneously broken at strong coupling, or not. However, the occurrence of a Landau scale $k_L$ makes it conceivable that supersymmetry may be spontaneously broken in the strongly coupled regime.
This interpretation  would be consistent with the picture  for the spontaneous breaking of Parisi-Sourlas supersymmetry  in disordered Ising models   \cite{Tissier:2011zz}, which is triggered  by cusp-like non-analyticities of the RG flow at a finite "Larkin scale" $k_L$. 
At strong coupling, these limitations of the 
full effective potential and the occurrence of Landau poles are not directly 
visible from the infrared limit only, see  Figs.~\ref{pGap} and~\ref{pRG}. It is a virtue 
of the fully integrated RG flow for all scales $k$ that the structure of the effective 
potential at strong coupling has become transparent.

\section{Effective field theory} \label{sec:EffectiveTheory}
In this section we discuss the integrated RG flow from an effective theory perspective. 
We assume that the UV scale $\Lambda$ is finite, and that the boundary 
condition at $k=\Lambda$ has been achieved by integrating-out the fluctuations 
with momenta above $\Lambda$.  The RG equations then detail the remaining 
low-energy flow of couplings for all scales $k<\Lambda$.  In terms of dimensional 
quantities, the solution \eq{urho} reads 
\begin{equation}
\label{W'}
\begin{array}{rl}
\bar\rho-\bar\rho_0(k)&= 
\displaystyle
c\,W'+H\left(\frac{W'}{k}\right)\,k-H\left(\frac{W'}{\Lambda}\right)\,\Lambda\\[1ex]
\bar\rho_0(k)&=k+\bar\rho_0\,.
\end{array}
\end{equation} 
The parameter $\bar\rho_0$ is given by $\bar\rho_0=\Lambda(\kappa-1)$ in terms 
of the microscopic (UV) parameters. Our motivation for studying \eq{W'} is twofold. 
Firstly, we want to further clarify the origin of the ``peculiar" phases discussed in the
previous section. Second, we want to evaluate the effect of changes in the 
boundary condition and higher-order couplings on the phase structure and critical phenomena
 
\begin{figure}[t]
\begin{center}
\includegraphics{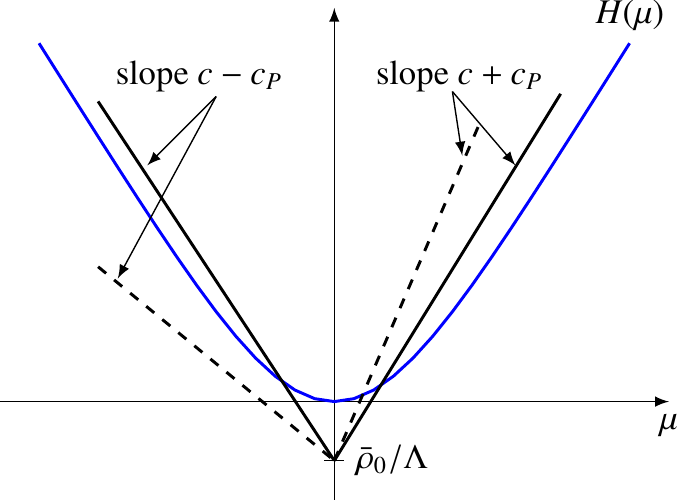}
\caption{The graphical solution of the gap equation (\ref{gapBranch}) in
the symmetric phases with $\bar\rho_0<0$. For positive $\mu$ there exist
one, two or three solutions.}
\label{fig:gapeq} 
\end{center}
\end{figure}

\subsection{Gap equations}
We begin with the IR limit of the integrated RG flow. The corresponding 
gap equations for the scalar masses at vanishing field
$W'(0)\equiv \mu\,\Lambda$, i.e. in the $O(N)$ symmetric phases,  are given 
in terms of the dimensionless parameter $\mu$ by
\begin{equation}
\label{gapL}
\begin{array}{rl}
\displaystyle
-\frac{\bar\rho_0}{\Lambda}&= 
c\,\mu+c_P|\mu|-H(\mu).
\end{array}
\end{equation}
where we used expansion \eq{Hlarge} for $H$.
For $\bar\rho_0\neq 0$ we find two possible branches of solutions with
\begin{equation}
\label{gapBranch}
\begin{array}{rl}
H\left(\mu\right)&=\displaystyle (c+c_p)\mu +\frac{\bar\rho_0}{\Lambda},\quad (\mu>0)\\[2ex]
H\left(\mu\right)&=\displaystyle (c-c_p)\mu +\frac{\bar\rho_0}{\Lambda},\quad (\mu<0).
\end{array}
\end{equation}
We consider $c\ge 0$ since changing the sign amounts to interchanging 
$\mu\leftrightarrow-\mu$ in \eq{gapBranch}. The main difference with \eq{gap} 
in the infinite cutoff limit is the appearance of the term $H(\mu)$.

In the SYM regime with negative $\bar\rho_0$ we find one, two, or three solutions to 
\eq{gapBranch} with $m=\Lambda\mu>0$, and none, one, or two solutions
$M=-\Lambda\mu>0$, see Fig.~\ref{fig:gapeq}. Three solutions for positive $\Lambda\mu$ 
can only exist if the slope $c+c_P$ is inbetween $c_P$ and $c_M$, cf.
Fig. \ref{Hprime}.

For most parts of the parameter
space we only have a single scalar mass $m$, similar to the weak coupling phase
of the renormalized theory. For small $\bar\rho_0/\Lambda$ and strong coupling, a
``triangle" opens up allowing for two additional mass scales of the type $M$.
The borderline $c(\bar\rho_0/\Lambda)$ is found analytically, starting at the 
point $(c,\bar\rho_0/\Lambda)=(c_P,0)$ and ending at 
$(c,\bar\rho_0/\Lambda)\approx(0,-1.077)$, see Fig. ~\ref{pPhaseGap}. 
Furthermore, we find two more masses 
of the type $m$ in a tiny ``spike"-like region at very strong coupling, 
bordered by the curves connecting $(c,\bar\rho_0/\Lambda)=
( c_M-c_P,-9/8)\approx (0.035,-1.125)$ with $(c,\bar\rho_0/\Lambda)=(0,-1.077)$ 
and $(0,-1)$
as indicated in the same Figure. 
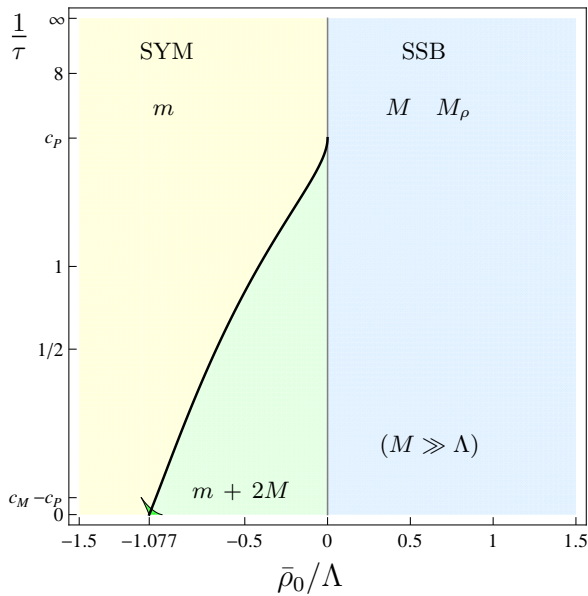
\begin{figure}
\begin{center}
\unitlength0.001\hsize
\begin{picture}(900,1060)
\put(0,200){\includegraphics[width=.43\textwidth]{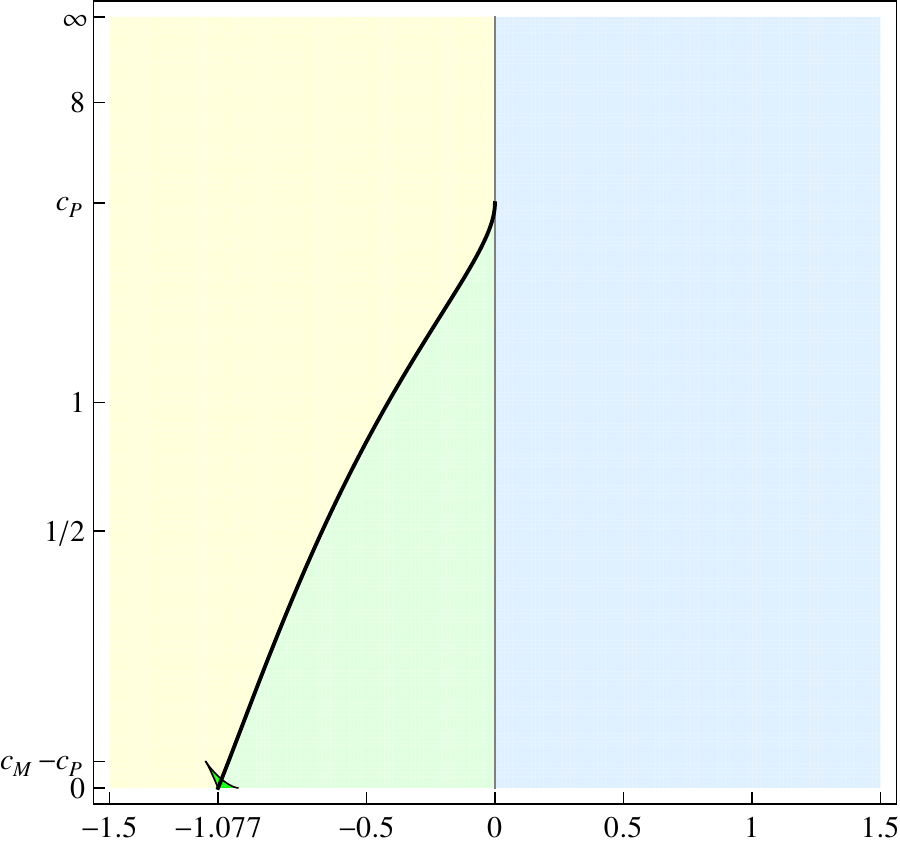}}
\put(0,970){\Large $\frac{1}{\tau}$}
\put(220,865){$m$}
\put(280,280){$m\, +\, 2 M$}
\put(575,865){$M\ \ \ M_\rho$}
\put(566,350){$(M\gg\Lambda)$} 
\put(410,145){\large $\bar\rho_0/\Lambda$}
\put(600,950){SSB} 
\put(200,950){SYM} 
\end{picture}
\vskip-1cm  
\caption{Phases of the supersymmetric model  according to the gap equations at
finite UV scale. The SYM phase displays either a single mass scale $m$, or
several ones.  The SSB regime displays two scalar mass scales  $M$ and
$M_{\rho}$ for all couplings. Note that the mass scale $M$ represents an $O(N)$
symmetric state within the regime where we would normally only expect SSB to
occur. The phase transition between the SYM phase and the SSB phase is
continuous with Gaussian exponents.}
\label{pPhaseGap} 
\end{center}
\end{figure}   
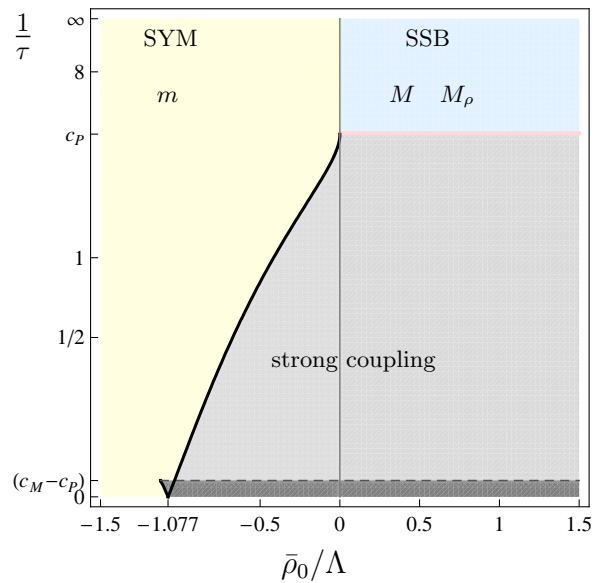
\begin{figure}[t]
\begin{center}
\unitlength0.001\hsize
\begin{picture}(900,1040)
\put(0,200){\includegraphics[width=.43\textwidth]{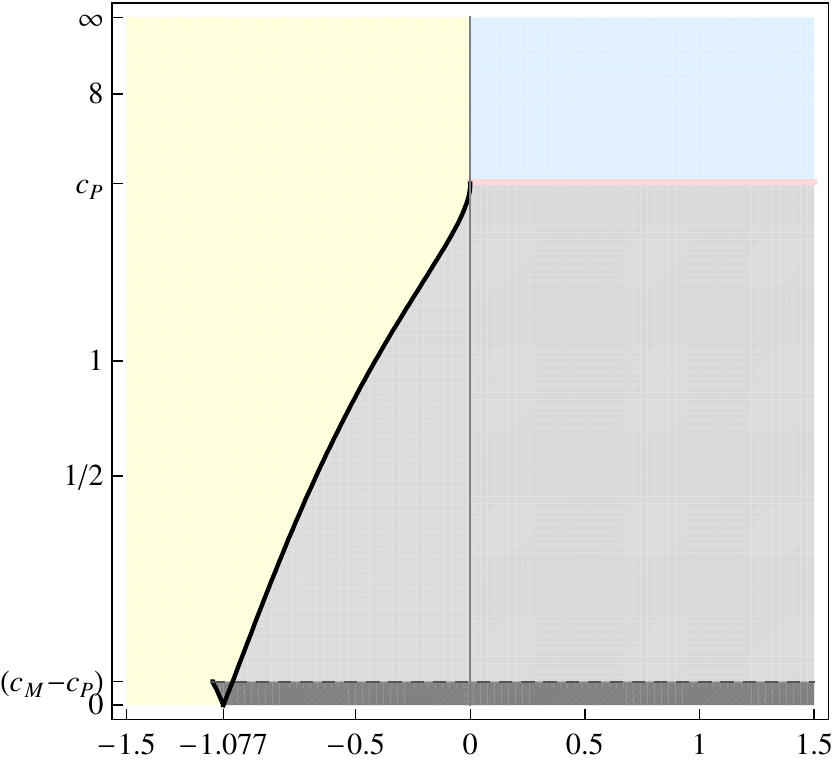}}
\put(0,950){\Large $\frac{1}{\tau}$}
\put(410,145){\large $\bar\rho_0/\Lambda$}
\put(220,865){$m$}
\put(575,865){$M\ \ \ M_\rho$}
\put(395,460){strong coupling}
\put(600,950){SSB}
\put(200,950){SYM}
\end{picture}
\vskip-1cm
\caption{Phases of the supersymmetric model according to the RG equations at 
finite UV scale. The SSB regime is quite similar to the result at $1/\Lambda=0$, 
see Fig~\ref{pRG}. The SYM phase is substantially larger (see text). The 
phase transition between the SYM phase and the SSB phase is continuous with 
Gaussian exponents. Note that there exists a very tiny  Landau phase for couplings
$c_P<|c|<c_M$ in the SSB regime (red line).}
\label{pPhaseRG} 
\end{center}  
\end{figure}
The bordering lines $c(\bar\rho_0/\Lambda)$ are known analytically. In total, we
either have a single mass $m$, or three masses $m+2M$ or $3m$, or five
different mass scales of the type $3m+2M$ in the region where the triangle and
the spike overlap. Some of the masses are parametrically large in the strong 
coupling domain. We believe that these masses in the very strongly coupled domain 
are an artifact of the regularisation and should not be trusted.  

In the  regime $\bar\rho_0>0$ allowing for SSB, a unique scalar mass 
solution $M$  to \eq{gapBranch} is achieved from the branch with 
negative $\mu$, for all couplings. In addition, the theory shows the expected 
radial mass $M_\rho$. 
However, we emphasize that  some of the solutions found here, in particular 
those at strong coupling, have parametrically large masses suggesting that 
these may be spurious.

\subsection{RG phase diagram}

Next we turn to the phase diagram of the integrated RG flow at finite $\Lambda$ for all scales $k$.

\subsubsection{Symmetric regime}  

\begin{figure} 
\begin{center} 
\unitlength0.001\hsize
\begin{picture}(900,1000)
\put(0,130){\includegraphics[width=.41\textwidth]{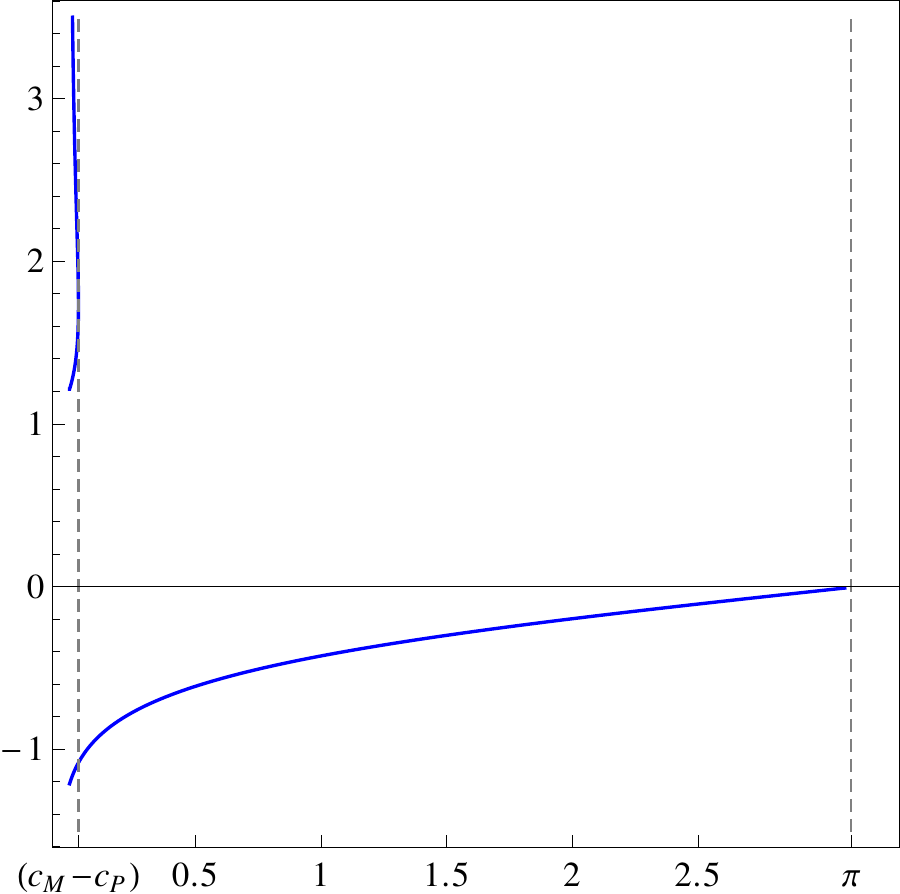}}
\end{picture} 
\put(-980,600){{\large $\frac{W_L^{\prime}}{\Lambda}$}}
\put(-480,80){{\large $c$}}
\vskip-1cm
\caption{Possible values  $W_L^{\prime}/\Lambda$ as a function of
the inverse superfield coupling coupling $c=1/\tau$ associated with  a Landau
pole  in the IR limit, i.e. with $d\bar\rho/dW^{\prime}=0$.}
\label{WUni} 
\end{center} 
\end{figure}

The phase
diagram corresponding to \eq{W'} is given in Fig.~\ref{pPhaseRG}, where the axes
denote the (inverse) quartic superfield coupling $1/\tau$ and the parameter
$\bar\rho_0$ in units of the initial scale $\Lambda$. 
For $\bar\rho_0<0$ the theory is in the symmetric phase, provided that the
coupling is small enough. There is also a strong coupling regime where the RG
flow develops a Landau pole and the effective potential becomes multi-valued in
the physical regime $\bar\rho>0$. The boundary between the two regimes is
marked  by a curve $c_{\rm cr}(\bar\rho_0/\Lambda)$. The latter  is
determined as follows: In the IR limit, the solution \eq{W'} reads 
\begin{equation}
\bar\rho-\bar\rho_0=cW^{\prime}+\pi
|W^{\prime}|-H\left(\frac{W^{\prime}}{\Lambda}\right)\Lambda
\label{infasc}
\end{equation}
and shows a Landau pole, if $d\bar\rho/dW'$ vanishes. Using (\ref{infasc}) in
the condition $d\bar\rho/dW'=0$ at $\bar\rho=0$ yields
\begin{eqnarray}
\frac{\bar\rho_0}{\Lambda}=
H\left(\frac{W'_L}{\Lambda}\right)
-\frac{W_L'}{\Lambda}\, H'\left(\frac{W_L'}{\Lambda}\right),\label{landpolWp}
\end{eqnarray}
where $W'_L$ is equal to $W'(0)$ when the Landau pole enters
the physical region at $\bar{\rho}=0$. The real roots of this polynomial
equation are
\begin{equation}
\frac{W_L^{\prime}}{\Lambda}=\pm
\left(\frac{\sqrt{9+8 \bar\rho_0/\Lambda}-\left(3+2\bar\rho_0/\Lambda\right)}
{2\left(1+\bar\rho_0/\Lambda\right)}\right)^{1/2},
\label{criticalC}
\end{equation}
where the plus (minus) sign belongs to the critical coupling characterizing
a Landau pole at $\bar\rho=0$ in the  positive (negative)
half-plane of $W^{\prime}$.
Inserting this into \eq{infasc}, evaluated at $\bar\rho=0$, yields
the critical couplings 
\begin{equation}
c_{\rm cr}
=\left.\frac{1}{W'}\left(
-\bar\rho_0
+H\left(\frac{W'}{\Lambda}\right)\Lambda
-c_P|W'|\right)\right|_{W'_L}
\end{equation}
as a function of the VEV $\bar\rho_0$.
 In general, we find that the
occurrence of Landau poles is only possible in the parameter range\footnote{Note
that we allow for negative $\kappa=\bar{\rho}_0/\Lambda+1$, i.e.~classical potentials with a single minimum at $\bar\rho=0$ (symmetric 
phase).} $\bar\rho_0/\Lambda \in(-1.125,0)$ and $c \in (0,c_P)$, i.e. the 
strong coupling regime.  Besides ambiguities with $W_L^\prime<0$ for couplings
$c<c_P$, we also find Landau poles with $W_L^{\prime}>0$ in the very narrow 
strong coupling regime with  $c<(c_M-c_P)\approx 0.035$, see Fig.~\ref{WUni}.

Hence, we interpret the different regimes  of the symmetric phase as
follows (see Fig.~\ref{pPhaseRG}): We observe Landau poles in the physical
regime with $W_L^{\prime}<0$, if the superfield coupling $\tau$ is larger
than $c^{-1}_{\rm cr-}$, i.e. $c<c_{\rm cr-}$. The corresponding  
borderline starts at the point $(c,\bar\rho_0/\Lambda)=(c_P,0)$ and ends at
$(0,-1.077)$, similar to borderline resulting from the gap-equation analysis. 
Furthermore, for very strong couplings $c<c_M-c_P\ll1$ we
observe ambiguities with $W_L^{\prime}>0$ in the physical regime (dark shaded
area in Fig.~\ref{pPhaseRG}). However, this area is bounded by $c_{\rm cr+}$ 
from below, where $c_{\rm cr+}$ starts at $(c_M-c_P,-1.125)$ and ends at 
$(0,-1.077)$.  
 
Interestingly, the available domain of couplings is substantially larger than 
in Fig.~\ref{pRG}. The reason for that is quite intuitive, since decreasing the VEV
$\bar\rho_0$ comes along with a shift of the solution $W^{\prime}$ to the left
and thus the Landau pole may enter the unphysical regime $\bar\rho<0$. 
In addition, the equations do not admit a second mass $M$, unlike the case 
for $1/\Lambda=0$. We emphasize that the RG study of the phase diagram also  
allows for a simple descriptive explanation of the occurrence of the various 
masses as shown in Fig.~\ref{pPhaseGap}. The two additional masses $M$, observed
in the strong coupling domain (see big triangle, Fig.~\ref{pPhaseGap}) result
from an ambiguity of the solution $W^{\prime}$ in the negative half-plane. The
borderline connecting $(c_P,0)$ and $(0,-1.077)$ in Fig.~\ref{pPhaseGap}
represents the special solution  with $c=c_{cr-}$ showing a Landau pole in the 
IR exactly at $\bar\rho=0$ and  this corresponds to an additional infinitely large 
mass $M$. Similarly, the two additional masses of type $m$ in the spike-like strong
coupling region  result from ambiguities of the solution
for positive $W^{\prime}$.

\subsubsection{Symmetry broken regime} \label{subsec:SSB}

For $\bar\rho_0>0$, the theory is in a phase featuring spontaneous
$O(N)$ symmetry breaking. For sufficiently weak coupling with $|c|\geq c_M$, the
theory displays a well-defined low-energy regime with two mass scales $M$ and
$M_\rho$. The first one is associated to the curvature at vanishing field and
thus represents an $O(N)$ symmetric phase, whereas the second mass 
is given by the curvature at the non-vanishing  VEV $\rho_0$ and implies SSB.

\begin{figure*}[t]
\unitlength0.001\hsize 
\begin{picture}(900,420)
\put(-20,0){\includegraphics[width=.31\textwidth]{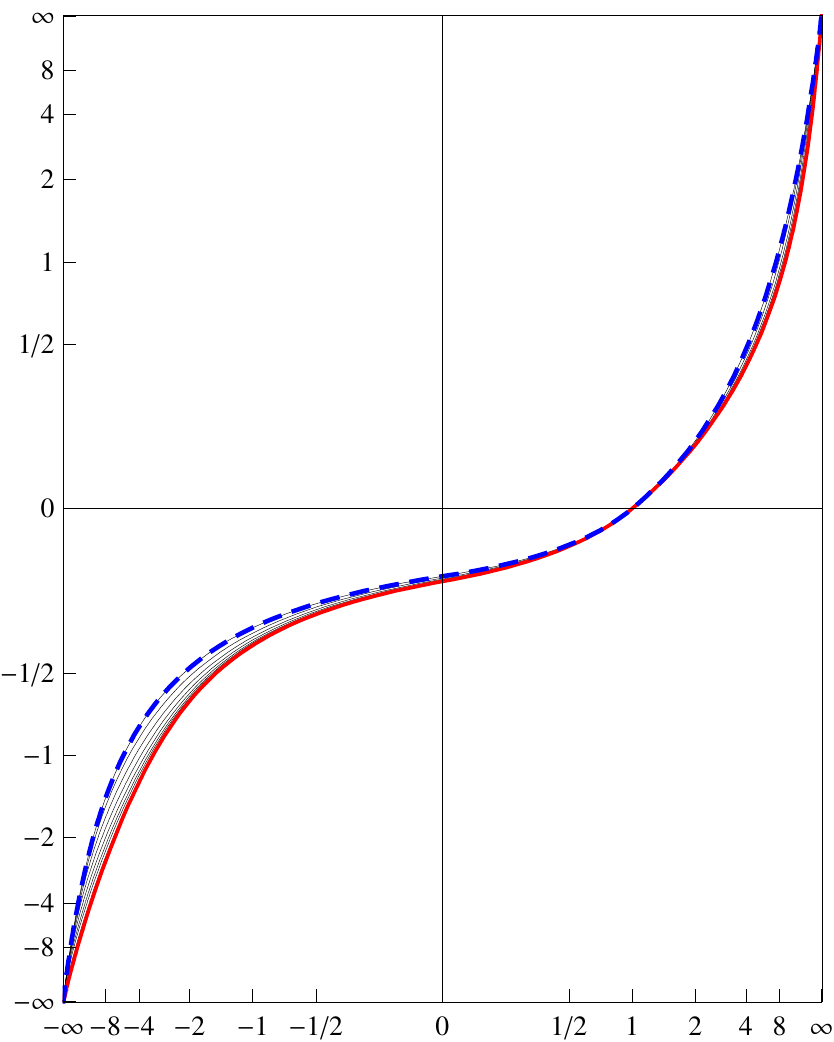}}
\put(300,0){\includegraphics[width=.31\textwidth]{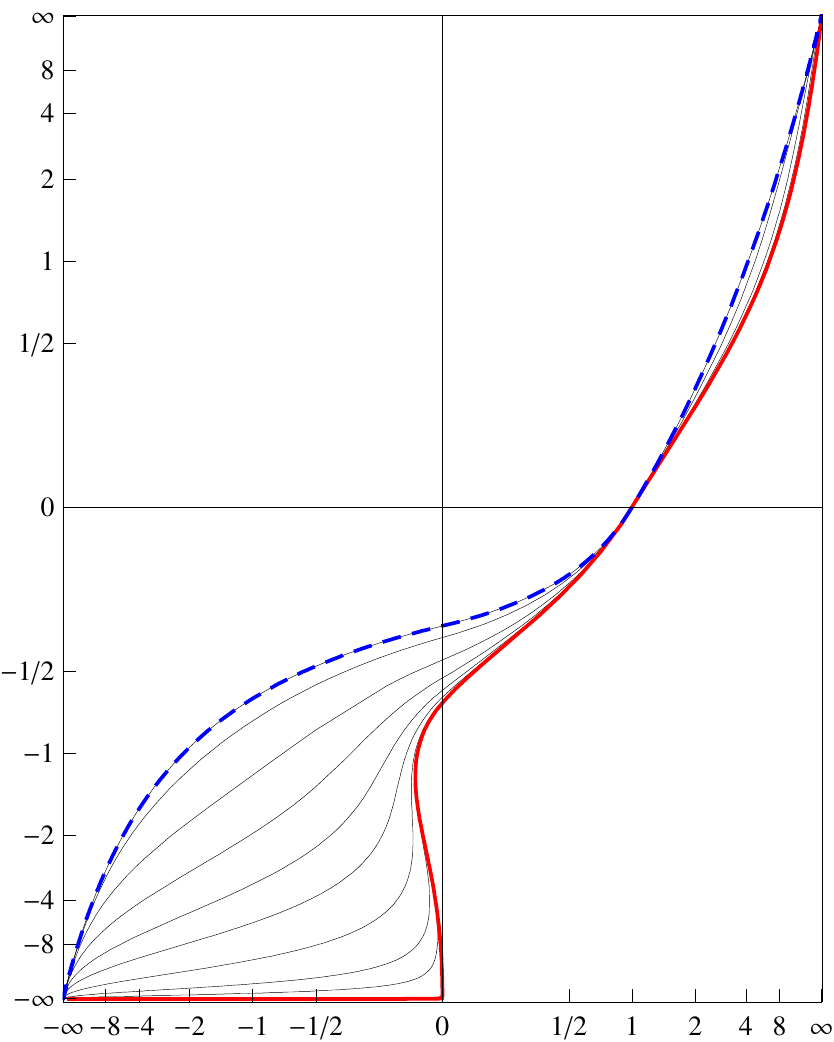}}
\put(625,0){\includegraphics[width=.31\textwidth]{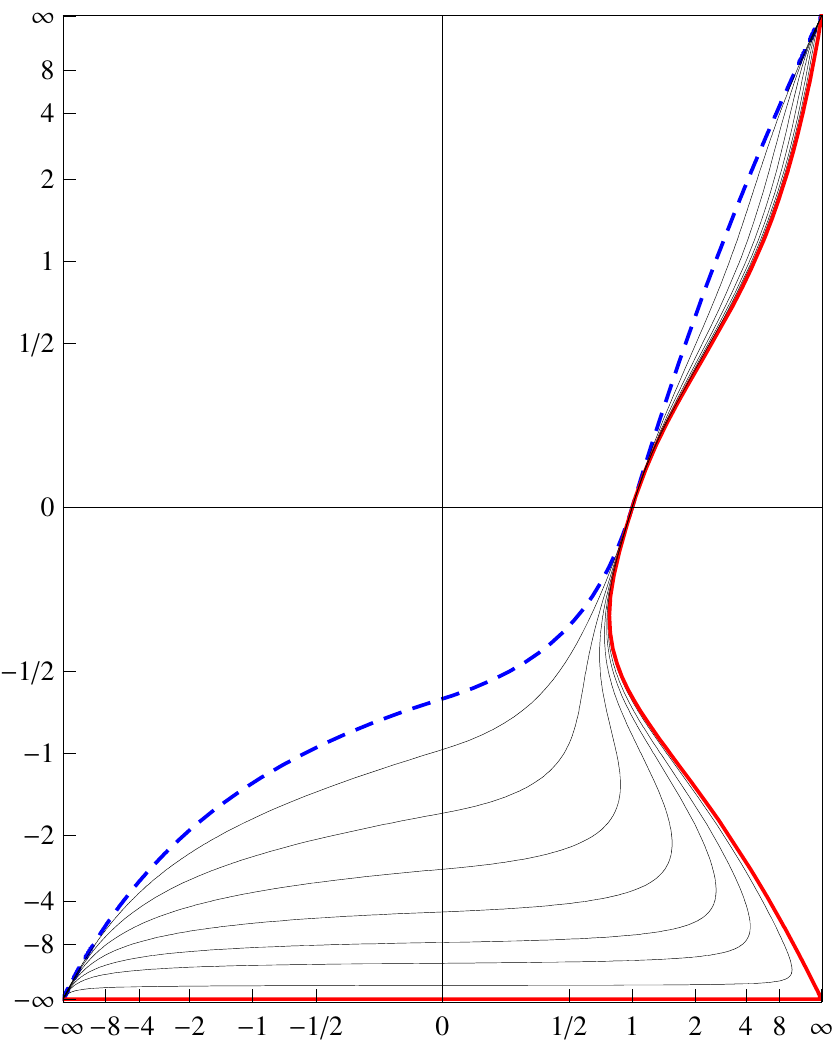}}
\put(-40,190){{\large $u$}}
\put(460,-30){{\large $X$}}
\put(15,350){{weak coupling}}
\put(15,325){{($c=2c_P$)}}
\put(335,350){{int.~coupling}} 
\put(335,325){{($c=c_P$)}}
\put(660,350){{strong coupling}}
\put(660,325){{($c=c_P/2$)}}
\put(25,150){{UV}} 
\put(60,100){{IR}}
\put(410,160){{UV}} 
\put(470,100){{IR}}
\put(690,120){{UV}} 
\put(900,100){{IR}}
\end{picture} 
\vskip.5cm 
\caption{Effective field theory with finite $\Lambda$: Graphical representation
of the  dimensionless superpotential derivative $u(X)$ as a function of
$X=\rho-\bar{\rho}_0/k$ in  \eq{link} for weak, intermediate and strong
superfield coupling (from left to right). Each panel shows the RG flow, starting with $u_{\Lambda}(\rho)=\tau\left(\rho-\rho_0(\Lambda)\right)$
according to \eq{initial} in the UV up to the IR limit.}
\label{LaufenDL} 
\end{figure*}
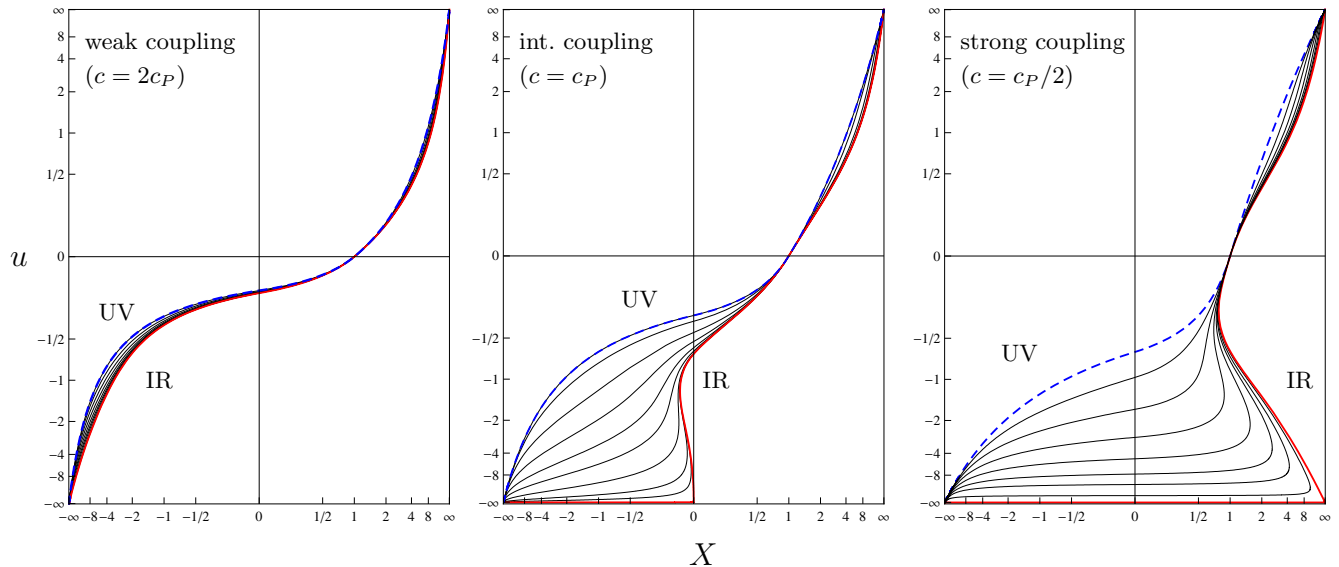

In  the very narrow coupling-regime $c_P<|c|<c_M$ there occur IR Landau poles
at  $(\rho_s(k), u_s(k))$ with
\begin{align}
\rho_s(k)&=1+c\, u_s +H(u_s)-H\left(u_s
e^t\right)e^{-t}+\frac{\bar{\rho}_0}{k},\notag\\
0&=c+H^{\prime}\left(u_s\right)-H^{\prime}\left(u_s e^t\right)
\label{rhoseff}
\end{align}
within the physical regime for scales $k<k_L$. 
However,
similar to the renormalized theory, the poles approach the VEV
$\lim_{k\rightarrow 0}\bar{\rho}_s(k)\rightarrow \bar{\rho}_0$ in the IR limit
from below and the domain, where $W^{\prime}$ is multi-valued collapses to a
point. 
Hence,  the effective Potential is well defined and
unique. 

For stronger couplings $c<c_P$,  the effective potential is plagued by Landau
poles and  becomes multi-valued even in the IR. This becomes apparent by
considering the second derivative $W^{\prime\prime}$ of the superpotential. 
The latter  shows a non-analyticity at $\bar\rho_0$ exactly in the IR limit with
\begin{equation}
\lim_{\;\bar\rho\rightarrow
\bar\rho_{0 \pm}}W^{\prime\prime}(\bar\rho)=\frac{1}{c\pm \pi},
\label{nonana}
\end{equation}
where  $W^{\prime}(\bar\rho_{0\pm})\rightarrow \pm 0$.
Apparently, the solution $W^{\prime}$ shows a cusp 
with positive $W^{\prime\prime}$ for $W^{\prime}\rightarrow  +0$ and negative
$W^{\prime\prime}$ for $W^{\prime}\rightarrow -0$  in the vicinity of the node
if $|c|<c_P$.  Since there exists at most one Landau pole with  $W^{\prime}_L<0$
in the IR limit (Fig.~\ref{WUni})  and  since $W^{\prime}(\bar\rho\rightarrow
-\infty)=-\infty$, it becomes apparent that there  has to be a Landau pole
located in the physical regime for $k\rightarrow 0$ if and only if $|c|<c_P$.

\subsection{Discussion}
 
Now we  compare and discuss the phase diagrams obtained by (a)
considering the renormalized theory with  $\Lambda \rightarrow
\infty$ and (b) looking at the effective theory with $\Lambda$ finite. 

Firstly, let us compare the phase diagrams Fig.~\ref{pGap} and  \ref{pPhaseGap}
as derived from the gap equations \eq{gap} and \eq{gapBranch}. 
Apparently, the gap equations \eq{gapBranch} of the effective theory contain an
additional, cutoff (and regulator)-dependent  contribution $H(\mu)$ compared to
\eq{gap}. The term $H(\mu)$ thus leads to the following
modifications of the phase diagram of the renormalized theory: In the symmetric
phase, it diminishes the parameter-range where we observe further masses in addition  to
$m$. Besides, we find up to five different $O(N)$ symmetric phases in the very
strong-coupling regime $|c| \ll 1$ and for certain VEV $\bar{\rho}_0$. In the
spontaneously broken regime, the function $H(\mu)$ enlarges the parameter range
to infinitely large couplings $\tau=1/c$, where we observe a second mass $M$ in 
addition to $M_{\rho}$. However, since the masses in the  very strong coupling regime
are quite large, i.e. of the order of the cutoff $\Lambda$, we believe them 
to be  regulator-dependent and unphysical. 
 
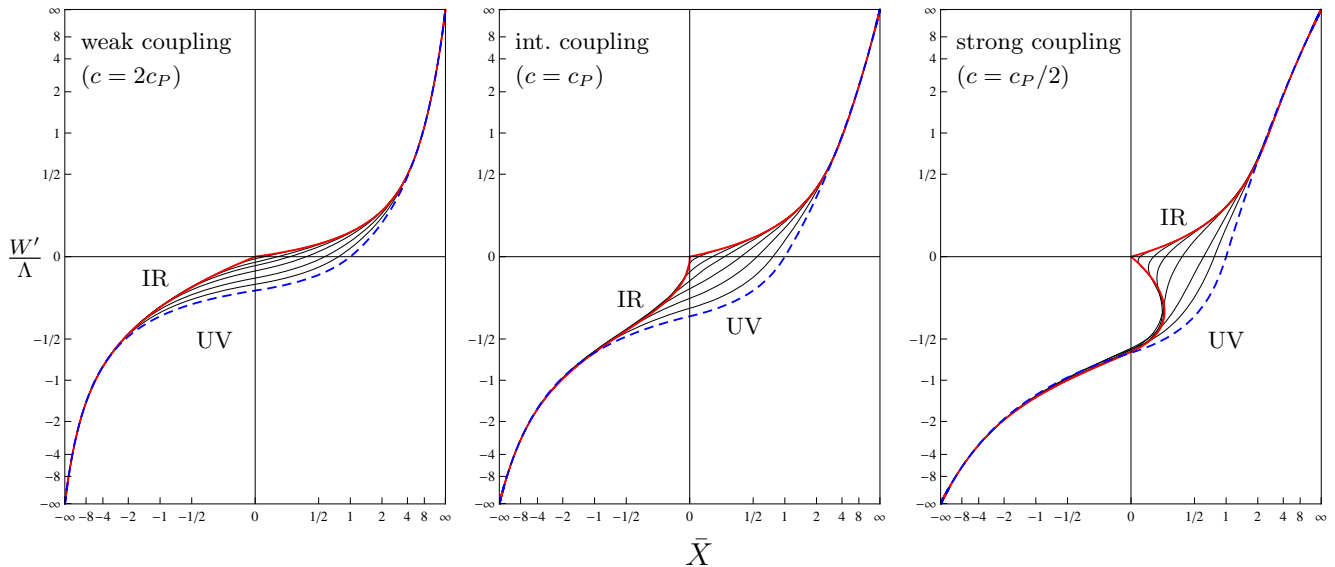
\begin{figure*}[t]
\unitlength0.001\hsize 
\begin{picture}(900,420)
\put(-20,0){\includegraphics[width=.31\textwidth]{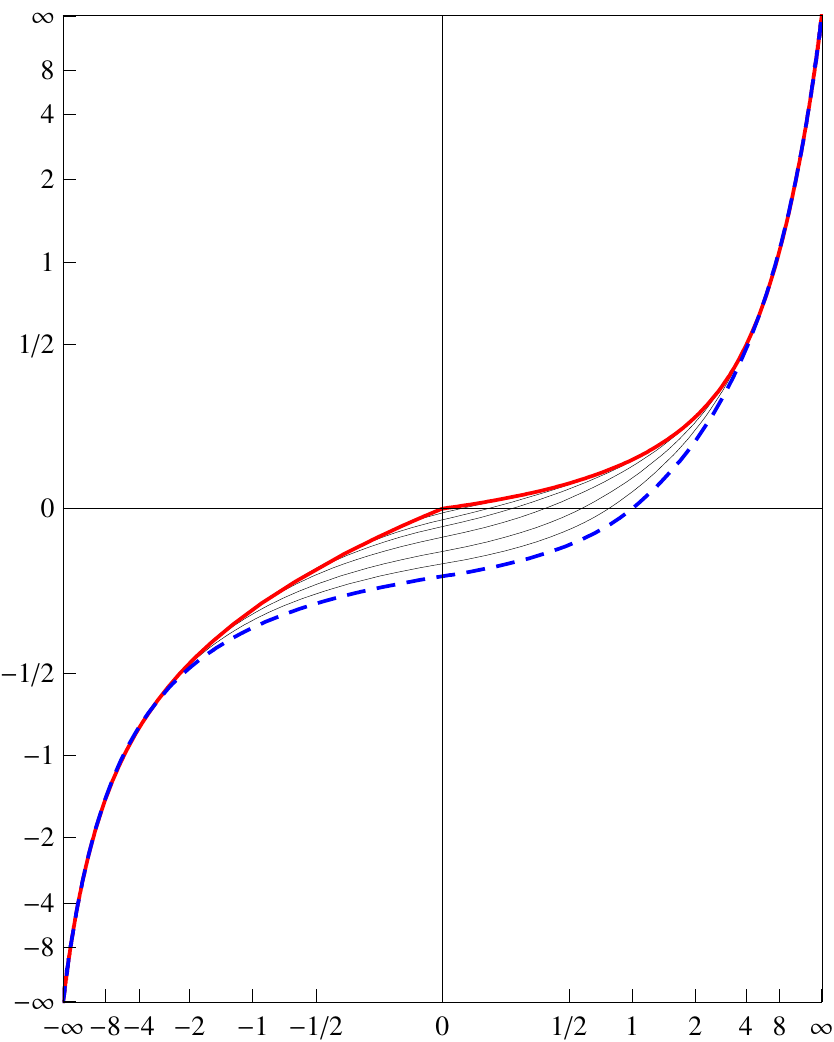}}
\put(300,0){\includegraphics[width=.31\textwidth]{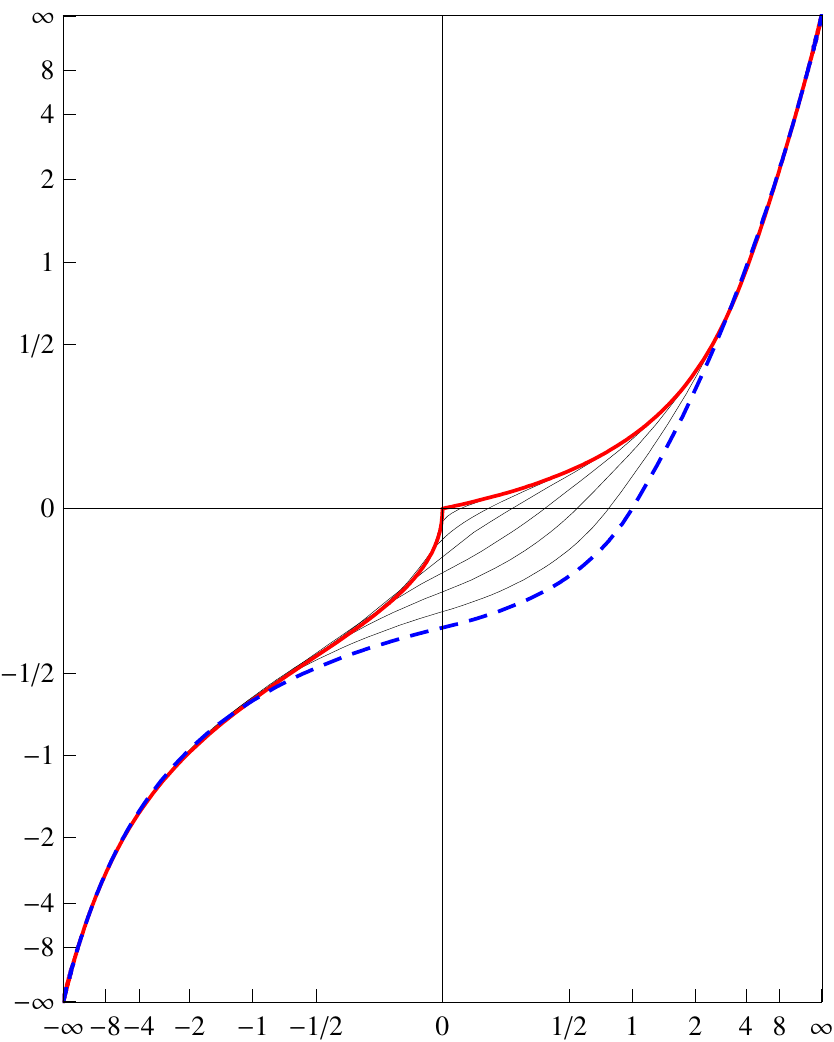}}
\put(625,0){\includegraphics[width=.31\textwidth]{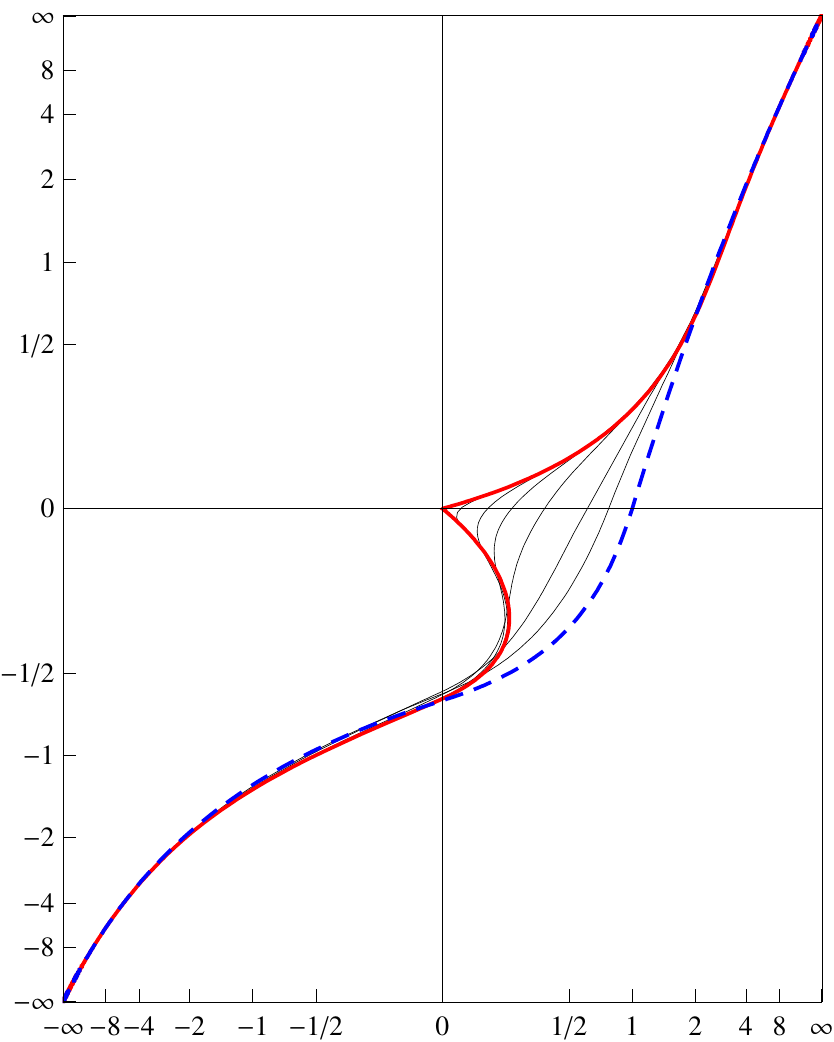}}
\put(-40,190){{\large $\frac{W^{\prime}}{\Lambda}$}}
\put(460,-30){{\large $\bar{X}$}}
\put(15,350){{weak coupling}}
\put(15,325){{($c=2c_P$)}}
\put(335,350){{int.~coupling}}
\put(335,325){{($c=c_P$)}}
\put(660,350){{strong coupling}}
\put(660,325){{($c=c_P/2$)}}
\put(100,130){{UV}}
\put(60,175){{IR}}
\put(410,160){{IR}}
\put(490,140){{UV}} 
\put(810,220){{IR}}
\put(845,130){{UV}}
\end{picture} 
\vskip.5cm  
\caption{Effective field theory with finite $\Lambda$: Graphical representation
of the  dimensionful superpotential derivative
$W^{\prime}/\Lambda$ as a function of
$\bar{X}=\left(\bar\rho-\bar{\rho}_0\right)/\Lambda$ for weak,
intermediate and strong superfield coupling (from left to right). Each panel
shows the RG flow of the superpotential, starting with
$W^{\prime}_{\Lambda}(\bar\rho)=\tau\left(\bar\rho-\bar{\rho}_0(\Lambda)\right)$
according to \eq{initial} in the UV up to the IR limit. Note the non-analyticity
of the effective superpotential at $\bar{X}=0$ in the IR limit
$k \rightarrow  0$ (see Secs.~\ref{subsec:SSB} and
\ref{subsec:effectivep}). Note further that the running potential with $c=c_P$
(middle panel) shows ambiguities for very small $|\bar{X}|\ll1$ for small scales
$k$ which are not visible in the figure.}
\label{LaufenDB} 
\end{figure*}

Next, let us compare the phase diagrams Fig.~\ref{pRG} and \ref{pPhaseRG} as
deduced from our RG studies. Here, we claimed solutions $W^{\prime}(\bar{\rho})$ 
to be physically relevant, if there exists no Landau pole characterized by an  
infinitely large fermion-boson coupling $W^{\prime \prime}$ in the physical domain.

Let us first consider the SYM regime. Here, the narrow window between the
couplings $c_L$ and $c_P$, where there exist two masses $m$ and $M$ vanishes
for finite $\Lambda$ and the effective theory  shows  only a single mass
$m$.  Furthermore, the strong coupling domain is  reduced and becomes
$\bar{\rho}_0$-dependent for $\Lambda$  finite.  The different structure of the
SYM regimes become apparent by comparing Fig.~\ref{pXY} with
Fig.~\ref{LaufenDL}, \ref{LaufenDB}.  In the renormalized theory
(Fig.~\ref{pXY}), there exists an UV Landau pole in the physical domain for 
superfield couplings stronger than $\tau=c_L^{-1}$ and the
potential is not even defined for all fields $\rho>0$. In contrast, the
effective theory always features a well-defined UV limit, given by the
superpotential
$W^{\prime}_{\Lambda}=\tau\left(\bar{\rho}-\bar{\rho}_0(\Lambda)\right)$ at
the UV scale $k=\Lambda$. The potential is defined for all fields but may show
ambiguities for sufficiently strong couplings. This is illustrated in 
Fig.~\ref{LaufenDB}, where  the superpotential
$W^{\prime}/\Lambda$ is plotted as a function of
$\bar{X}=\left(\bar{\rho}-\bar{\rho}_0\right)/\Lambda$. In the SYM phase, the
origin $\bar{\rho}=0$ corresponds to $\bar{X}=|\bar{\rho}_0|/\Lambda>0$. Now,
let us consider the strongly coupled domain with $|c|<c_P$
fixed  (Fig.~\ref{LaufenDB}, right panel) and $|\bar{\rho}_0|\ll1$. Here, a IR
Landau pole occurs at $k_L>0$ in the physical regime and
additional masses at the origin appear by approaching the IR. However, if we
choose  $|\bar{\rho}_0|$ large enough, the IR Landau pole drifts out of the
physical domain and the effective potential is unique and well-defined for all
$\bar{\rho}\geq0$ with a single mass $m$. This upper limit of
$|\bar{\rho}_0|(c)$ simply corresponds to the borderline connecting $(c_P,0)$
and $(0,-1.077)$ in Fig.~\ref{pPhaseRG}.

We find  identical weak, Landau and strong coupling SSB regimes  for the
renormalized and the effective theory, see Fig.~\ref{pRG}, \ref{pPhaseRG}. 
The  renormalized as well
as the effective  theory exhibit an IR Landau pole for all $|c|<c_P$
(Fig.~\ref{pXY} and Fig.~\ref{LaufenDB}).  The existence of an IR  Landau pole
within the effective theory is shown as follows, see Fig.~\ref{LaufenDB}, right
panel: The origin $\bar{\rho}=0$ corresponds to
$\bar{X}=-\bar{\rho}_0/\Lambda<0$ and thus there always emerges an IR Landau
pole in the physical domain at $k_L>0$ for $|c|<c_P$. Independent of the
superfield coupling and the VEV $\bar{\rho}_0>0$, there always exists only a
single mass $M$ at the origin representing an $O(N)$ symmetric phase. 
Again, the effective potential is always defined
for all fields, but may show ambiguities, whereas  the potential is not defined for 
all fields $\bar{\rho}>0$ in the strong coupling regime $|c|<c_P$ in the infinite
cutoff-limit $\Lambda \rightarrow \infty$.

\begin{figure*}[t]
\unitlength0.001\hsize 
\begin{picture}(900,420)
\put(-20,0){\includegraphics[width=.298\textwidth]{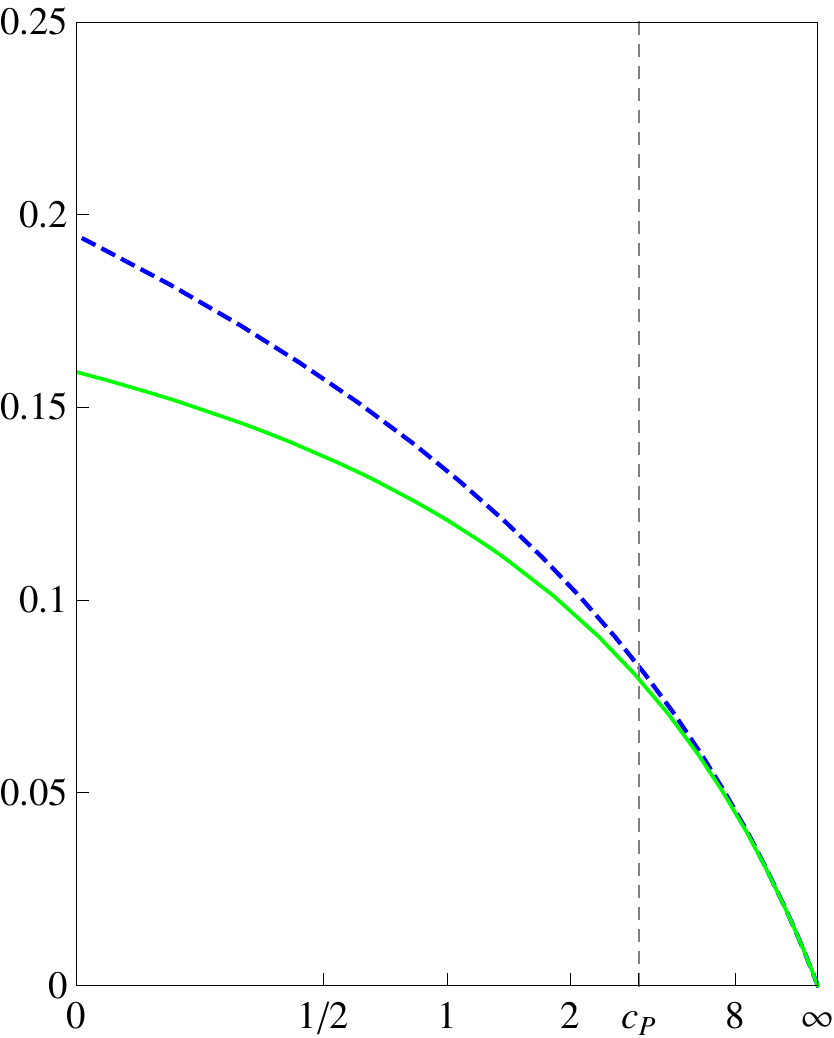}}
\put(300,0){\includegraphics[width=.29\textwidth]{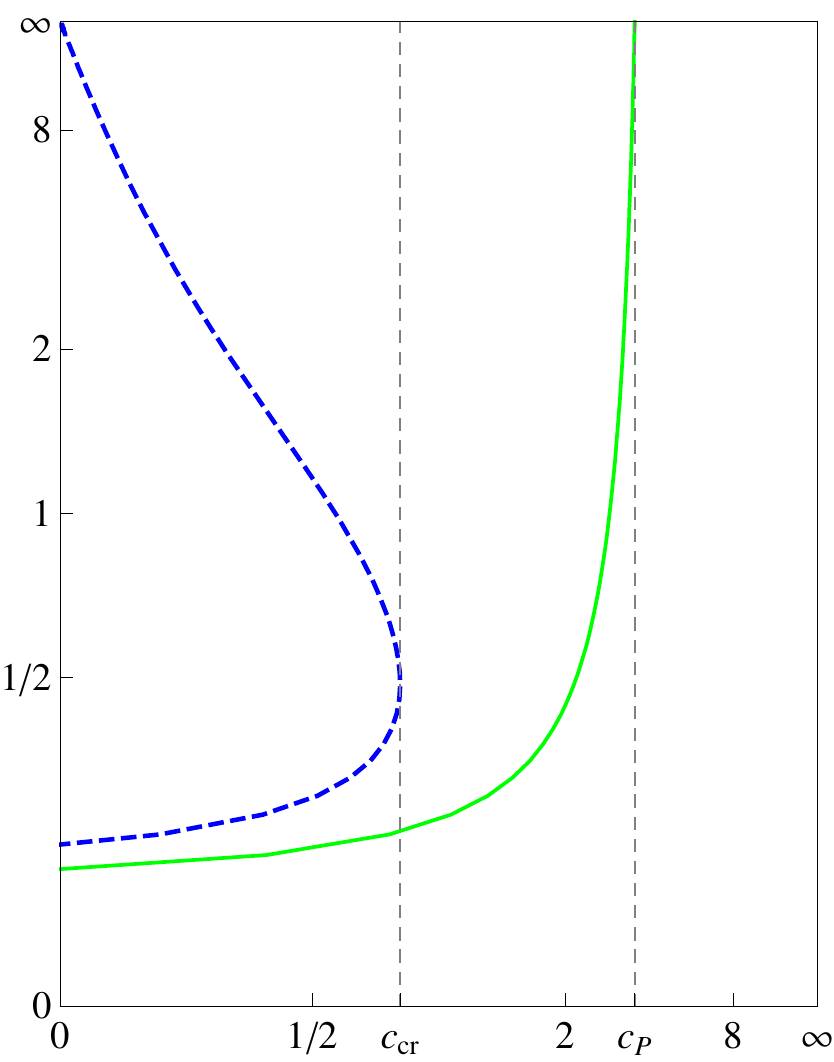}}
\put(625,0){\includegraphics[width=.29\textwidth]{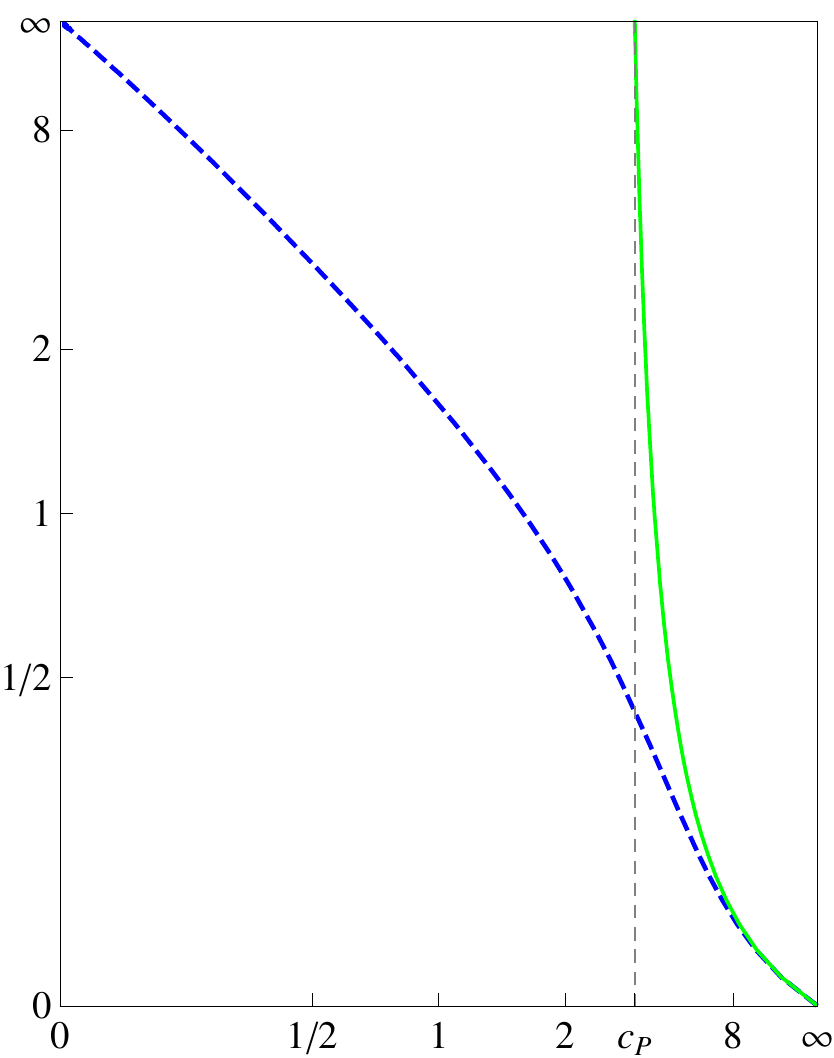}}
\put(20,313){$\frac{m}{\Lambda},\,(\Lambda < \infty)$
\;\,$\textcolor{blue}{\textbf{-\! -\! -\! -\! -}}$} 
\put(20,340){$m,\,(\Lambda \rightarrow
\infty)$}
\put(130,340){\thicklines \textcolor{green}{\line(1,0){50}}}
\put(350,313){$\frac{M}{\Lambda}$
\,\,$\textcolor{blue}{\textbf{-\! -\! -\! -\! -}}$} 
\put(350,340){$M$\,\, \thicklines \textcolor{green}{\line(1,0){50}}} 
\put(660,40){$\frac{M}{\Lambda}$
\,\,$\textcolor{blue}{\textbf{-\! -\! -\! -\! -}}$} 
\put(660,67){$M$\,\, \thicklines \textcolor{green}{\line(1,0){50}}} 
\put(130,-30){{\large $c$}}
\put(445,-30){{\large $c$}}  
\put(770,-30){{\large $c$}}
\put(100,369){{\large $\bar\rho_0<0$}}
\put(425,369){{\large $\bar\rho_0<0$}}
\put(755,367.5){{\large $\bar\rho_0>0$}} 
\end{picture}
\vskip.5cm   
\caption{\textit{Green, solid lines}: mass scales $m$, $M$  of the renormalized
theory ($\Lambda \rightarrow \infty$) as functions of the coupling $c$ according to the gap equations 
\eq{gapBrancha} for fixed $\bar{\rho}_0=\{-0.5,-0.5,0.5\}$ (left, middle, right panel).
\textit{Blue, dashed lines}: Masses $m/\Lambda$, $M/\Lambda$ of the effective
theory ($\Lambda < \infty$) as functions of the coupling $c$ according to the gap equations \eq{gapL} for fixed $\bar{\rho}_0/\Lambda=\{-0.5,-0.5,0.5\}$. }
\label{masses} 
\end{figure*}
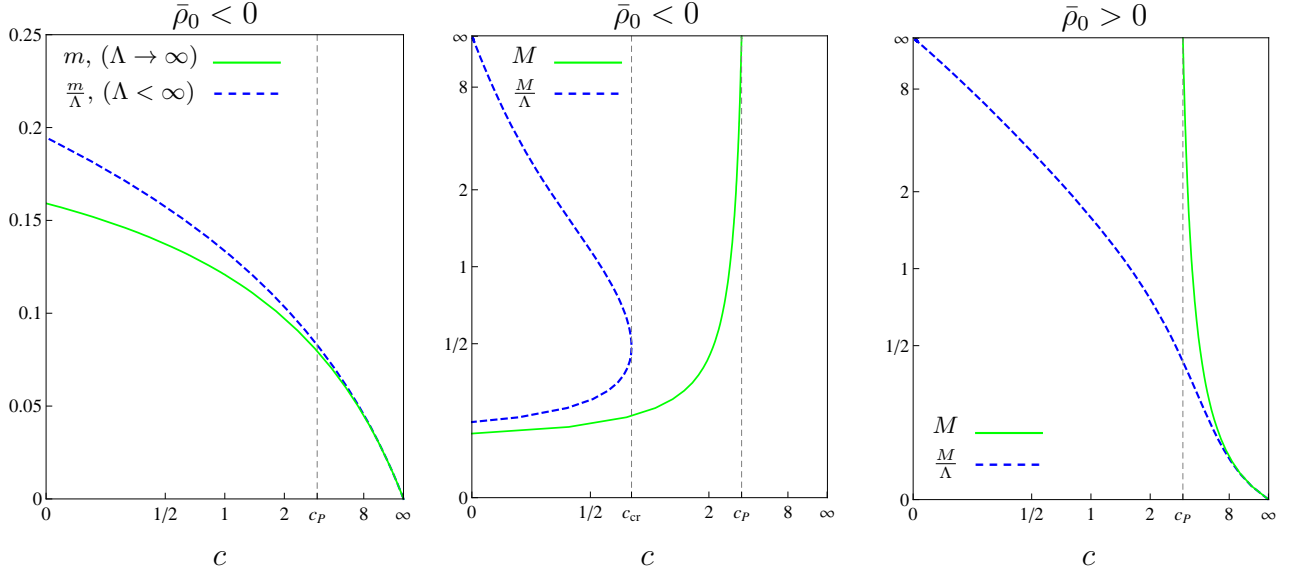    

Finally, Fig.~\ref{masses} compares the different mass scales  of the renormalized  
and the effective model. Notice that these
masses represent $O(N)$ symmetric phases of the model, since they emerge from
the curvature of the potential at vanishing field $\bar{\rho}=0$. The
parametrically large masses $m$ observed in the spike-like region (see
Fig.~\ref{pPhaseGap}) are not included in Fig.~\ref{masses}, since we believe 
them to be an artifact of the chosen regularization. 

In summary, in the SSB phase, and in the symmetric phase at weak coupling, the difference between the $(\Phi^2)^2_{d=3}$ theory at finite and infinite UV cutoff is minute, resulting in equivalent phase diagrams. In the symmetric phase for $c<c_P$, the difference is more pronounced: At finite UV cutoff the fluctuations of the Goldstone modes  have less  ``RG time" available to built-up non-analyticities in the effective potential. This leads to a shift in the effective boundary between weak and strong coupling, allowing for a substantially larger domain of a regular $O(N)$ symmetric phase. At strong coupling, we also conclude that the absence of an $O(N)$ symmetric phase at infinite cutoff arises from the theory at finite UV scale through an $O(N)$ symmetric phase with anomalously large mass of the order of the UV scale itself.

\subsection{Effective potential}\label{subsec:effectivep}

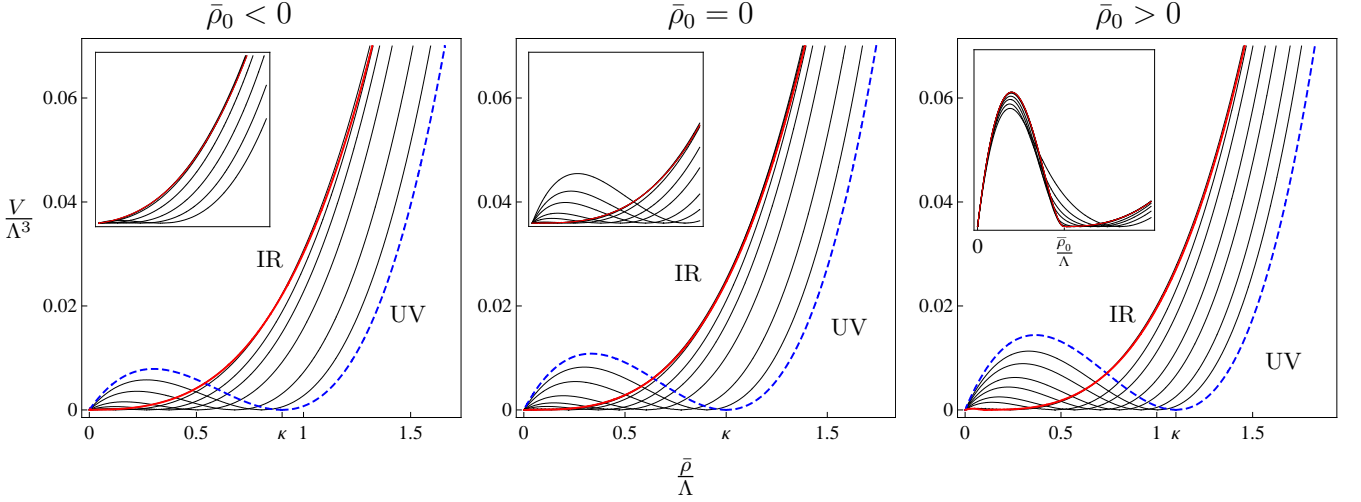
\begin{figure*}[t]
\unitlength0.001\hsize 
\begin{picture}(900,400)
\put(-20,0){\includegraphics[width=.31\textwidth]{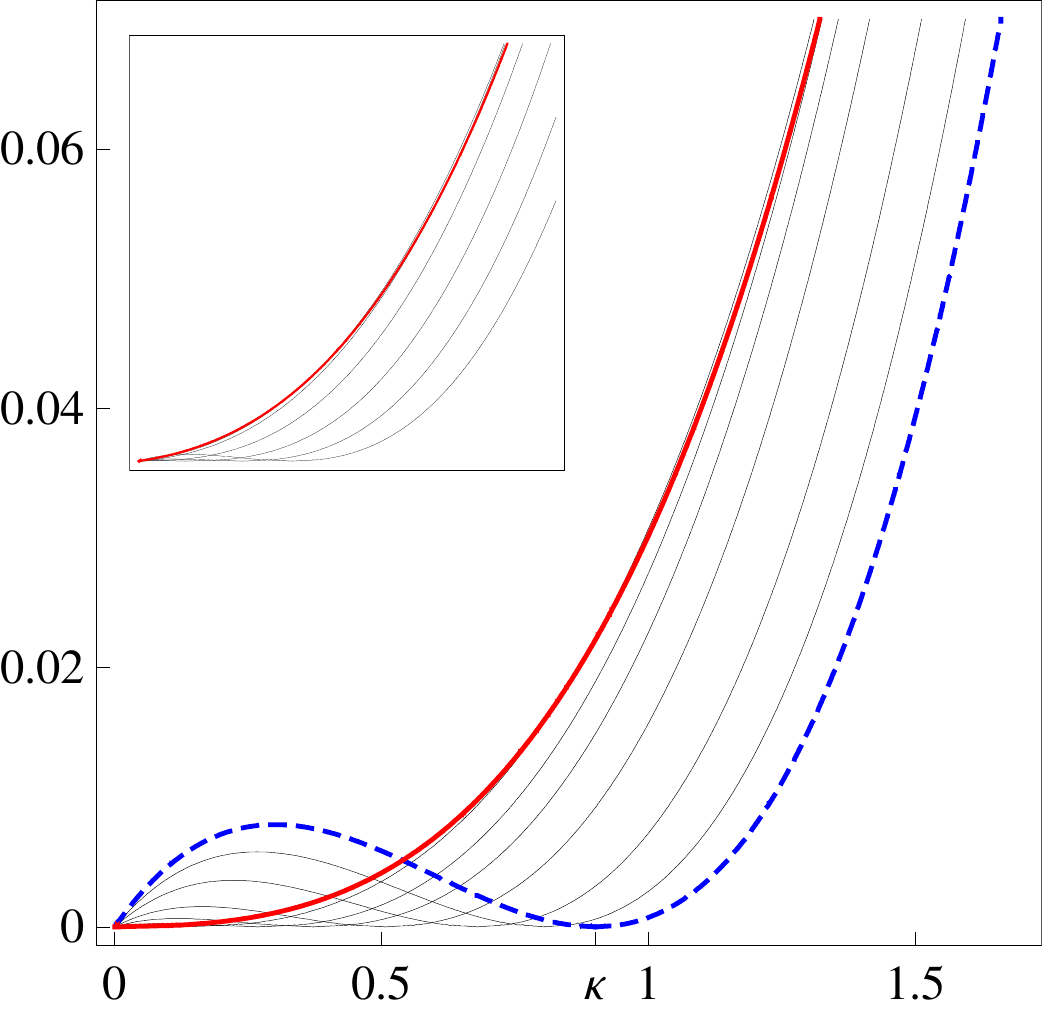}}
\put(300,0){\includegraphics[width=.31\textwidth]{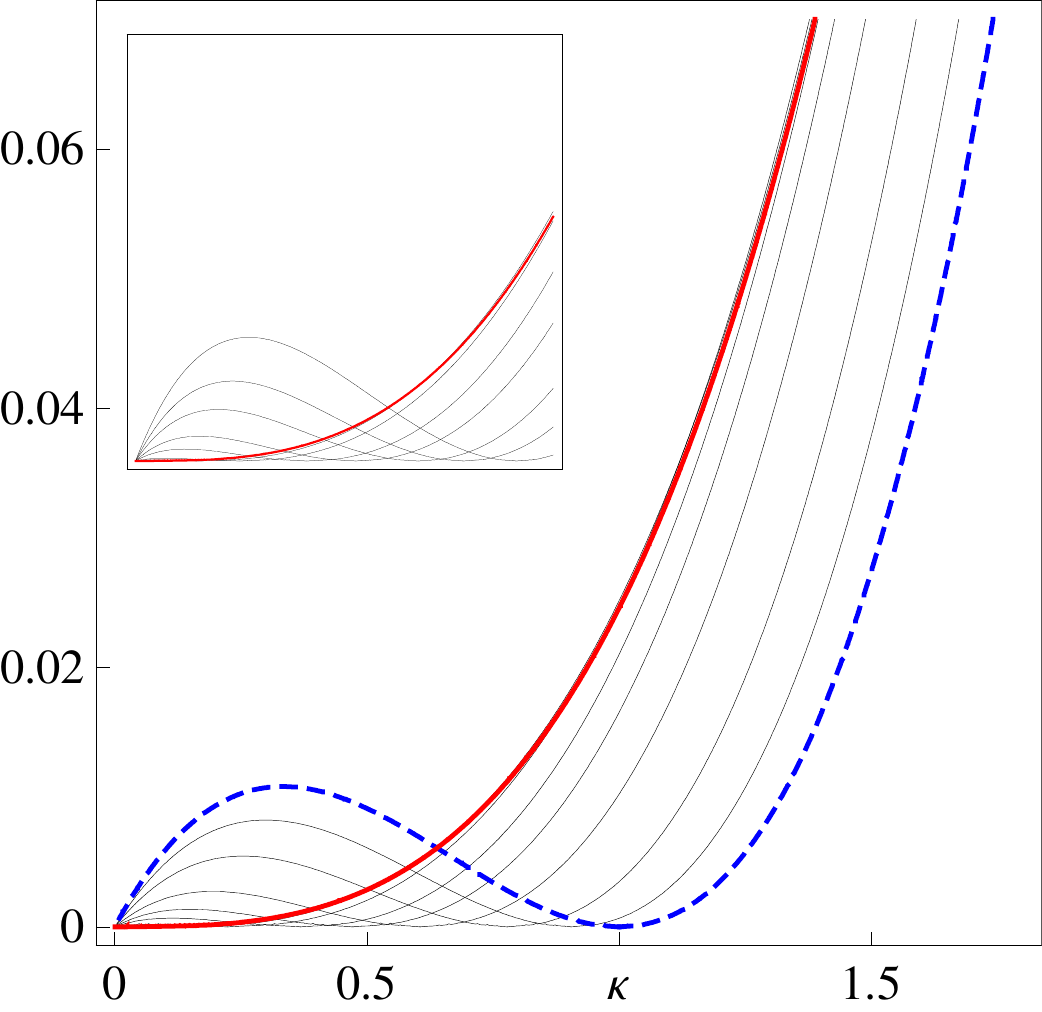}} 
\put(625,0){\includegraphics[width=.31\textwidth]{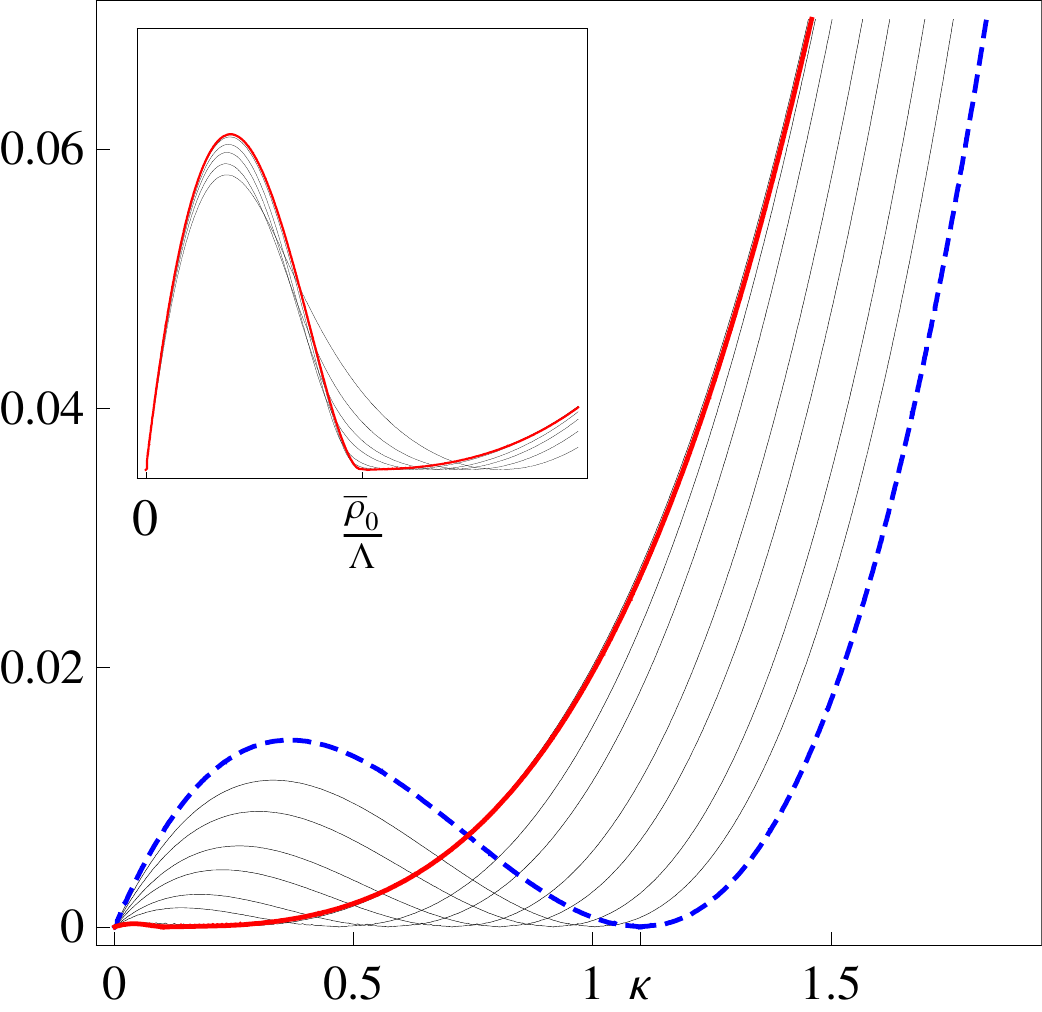}}
\put(-50,160){{\large $\frac{\!V}{\Lambda^3}$}}
\put(445,-30){{\large $\frac{\bar{\rho}}{\Lambda}$}}  
\put(137,130){{IR}}
\put(235,90){{UV}}
\put(100,310){{\large $\bar\rho_0<0$}}
\put(440,310){{\large $\bar\rho_0=0$}}
\put(755,310){{\large $\bar\rho_0>0$}}
\put(445,120){{IR}}
\put(560,80){{UV}}
\put(880,55){{UV}}
\put(765,90){{IR}}
\end{picture}
\vskip.5cm     
\caption{RG flow of the effective average potential $V_k/\Lambda^3$ as a
function of $\bar{\rho}/\Lambda$ according to \eq{W'} for different values of 
$\delta \kappa=\bar{\rho}_0/\Lambda=\{-0.1, 0, +0.1\}$ at weak coupling 
$c=3.7$. If $\bar{\rho}_0<0$, the system evolves into an $O(N)$ symmetric 
phase (left panel). Vanishing $\bar{\rho}_0$ corresponds to the phase
transition between the $O(N)$ symmetric and the SSB phase and the scale
invariant solution is approached in the IR limit (middle panel). If
$\bar{\rho}_0>0$, the macroscopic theory is characterized by a
non-vanishing VEV $\bar{\rho}_0(k\rightarrow0)=\bar{\rho}_0>0$ (right
panel). The insets show the potential at small fields approaching the IR limit.
} 
\label{potential}    
\end{figure*}

As already mentioned in Sec.~\ref{subsec:Factorization}, the relevant
microscopic coupling $\kappa=\kappa_{\rm cr}+\bar\rho_0/\Lambda$ determines the macroscopic 
physics of the model: if $\kappa<\kappa_{\rm cr}$ ($\bar\rho_0<0$), 
the effective potential preserves global $O(N)$ symmetry. Contrary, 
if $\kappa>\kappa_{\rm cr}$ ($\bar\rho_0>0$), the symmetry may be spontaneously 
broken, if the VEV $\bar\rho_0>0$ is taken.  The specific UV coupling 
$\kappa_{\rm cr}=1$ marks the phase transition between the two regimes.
 Fig.~\ref{potential} shows the flow of the effective average potential
$V_k(\bar\rho)$  for different values of $\kappa$, starting in the UV at
$k=\Lambda$ with 
\begin{equation}
V_{\Lambda}=\bar\rho\left(W^{\prime}_{\Lambda}\right)^2=\tau^2\bar\rho
\left(\bar\rho-\kappa\Lambda\right)^2
\label{UVP}
\end{equation}  
according  to  \eq{initial}, up to the IR limit $k \rightarrow
0$. Three  aspects of the potential need to be discussed further:

Firstly, there exists a  strong coupling domain, where the effective potential
shows ambiguities within the physical domain, both in the infinite cutoff limit (Fig.~\ref{pXY}) and in the effective theory limit (Fig.~\ref{LaufenDB},
right panel). At strong coupling, the effective potential admits no physical solution for small fields, except for an unphysical one with $1/|u'|\ll 1$ in the effective theory description. This result indicates that a description of the theory in terms of an effective superpotential  is no longer viable, possibly hinting at the formation of bound states with or without the breaking of supersymmetry. Incidentally, for the same parameter values the effective potential admits two solutions for large fields, except for an unphysical third solution one in the effective theory description. The theory admits two different effective potentials associated to the same microscopic parameters, which has been discussed in \cite{Litim:2011bf} in the context of fixed point solutions.

Secondly, the effective potential at $k=0$ is non-analytic at its nontrivial minimum
$\bar\rho_0$. Consider therefore the second derivative of the superpotential
\begin{equation}
W^{\prime\prime}\left(\bar\rho\right)=\frac{1}{c+H^{\prime}
\left(W^{\prime}/k\right)-H^{\prime}\left(W^{\prime}/\Lambda\right)}
\label{secD}
\end{equation}
in the vicinity of $\bar{\rho}_0(k)$,  where $W^{\prime}(\bar{\rho}_0(k))=0$
according to \eq{W'}. By approaching the IR, \eq{secD} simplifies to \eq{nonana}.
Apparently, this non-analyticity does not  appear until the exact IR limit $k=0$
is approached. Contrary, for all finite scales $k>0$ we find
$W^{\prime\prime}(\bar\rho_0(k))=1/c$,  which simply
represents the exactly marginal superfield coupling $\tau$. Since the radial
mass is given by
\begin{equation}
M_{\rho}^2=\left.V^{\prime\prime}(\phi)\right|_{\phi=\phi_0}
=\left(2\bar{\rho}W^{\prime\prime}(\bar{\rho})\right)^2\big\vert_{\rho_0(k)},
\label{mradial}
\end{equation}
a uniquely defined radial mass only
exists for finite scales $k>0$ and reads
\begin{equation}
M_{\rho}(k)=2\tau\bar\rho_0(k)=2\tau(k+\bar\rho_0), \quad\quad \bar{\rho}_0>0.
\label{mrad}
\end{equation}
First studies at finite $N$ indicate that the 
non-analyticity of $W^{\prime}$ for  $k=0$ is
solely due to the large-$N$ limit.

Thirdly, the effective scalar field potential in the SSB phase with non-vanishing 
VEV is not convex, even in the IR limit $k \rightarrow 0$. As it has 
already been mentioned in \cite{Synatschke:2008pv}, the supersymmetric analogon
of the potential term in the classical action is the superpotential $W$, \eq{eq:action}. 
Consequently, a flow of the superpotential is derived 
which drives the approach to convexity of the superpotential $W$, but not
necessarily of the potential $V=\bar\rho\, W^{\prime 2}$.
The superpotential $W$ is a convex function if and only if the first derivative
$W^{\prime}(\bar\rho)$ represents a  monotonically increasing function of
$\bar\rho$. According to \eq{infasc}, this condition is satisfied as long as
$c>c_P$, i.e. in the weakly coupled domain. This fact supports the conjecture
that supersymmetry may be broken spontaneously 
in the strongly coupled domain exhibiting
Landau poles.

\subsection{Phase transition \& critical exponents}\label{subsec:critical}

The supersymmetric $O(N)$ model in $d=3$ is an effective field theory that
features the large-distance properties of statistical models near a
second order phase transition.  According to \cite{Litim:2011bf}, the fixed-point solution characterizing the phase transition shows Gaussian scaling for all 
finite couplings $c$, except for $|c|=c_P, c_I$. 
Following \cite{Berges:2000ew} we can also extract the thermodynamical critical
exponents. The expectation value of the field $\langle \phi \rangle $ serves as order parameter,
and in the SSB regime it is
related to the  VEV $\bar{\rho}_0$ via (choose $\phi_i=\delta_{i1}
\phi$)
\begin{equation}
\langle \phi \rangle = \lim_{k \rightarrow 0}\sqrt{2\bar{\rho}_0(k)} \equiv
\sqrt{2\bar{\rho}_0} =\sqrt{2\delta\kappa\Lambda}.
\label{eq:Ordnung}
\end{equation}
We may associate the deviation of $\kappa$ from its critical value
$\kappa_{cr}=1$ with the deviation of the temperature $T$  from the critical
temperature $T_c$ according to $\delta\kappa\Lambda\sim(T_c-T)$.
Thus we have 
\begin{equation}
\langle \phi \rangle \sim \left(\bar{\rho}_0\right)^{\beta} \quad \mbox{with}
\quad \beta =\frac{1}{2}\,.
\end{equation}
Next, consider the critical exponent $\nu$ describing the manner in which the
correlation length $\xi$ diverges (the mass vanishes) by approaching the phase
transition. We thereby distinguish between
\begin{align}
\xi^{-1}&=m\sim \left(-\bar{\rho}_0\right)^\nu \quad (\mbox{SYM regime},\,
\bar{\rho}_0<0)\notag \\
\xi^{-1}&=m\sim \left(\bar{\rho}_0\right)^{\nu^{\prime}} \quad \;\; (\mbox{SSB
regime},\;\, \bar{\rho}_0>0).
\label{defnu}
\end{align}
Let's consider first the  squared masses corresponding to $O(N)$ symmetric ground
states as given by (\ref{defmass}). We are interested in how the
superpotential $W^{\prime}$ vanishes at the origin when
$\bar{\rho}_0\rightarrow 0$. We begin with the parameter range $c>0$ and $c\neq c_P$.
Using (\ref{infasc}) and (\ref{Hsmall}) we have
\begin{equation}
\bar{\rho}-\bar{\rho}_0 =c\, W^{\prime} + \pi
|W^{\prime}|-\frac{3}{\Lambda}W^{\prime\,2}+\mathcal{O}\left(\frac{W^{\prime\,4}}{\Lambda^3}\right)
\label{smallmass}
\end{equation}
for small masses. In the SYM regime, this gives 
(\ref{gapBrancha}) for $W^{\prime}/\Lambda\ll 1$, where the second mass in
(\ref{gapBrancha}) only exists in the strong coupling region $c<\pi$. Hence,
according to (\ref{defnu}) we have  
\begin{equation}\label{nu}
\nu=1\,.
\end{equation}
In the SSB regime, there exists a unique $O(N)$ symmetric ground state with mass $M$
given by (\ref{gapBrancha}) for all $c>\pi$, implying 
\begin{equation}\label{nu'}
\nu^{\prime}=1\,.
\end{equation}
We also observe a spontaneously $O(N)$ broken ground state, characterized by
its radial mass according to (\ref{mrad}). Since $M_{\rho}\sim \bar{\rho}_0$, 
this also leads to \eq{nu'}.

Now consider the exponent $\delta$, given by
$\left.J\right|_{\bar{\rho}_0=0}\sim \phi^{\delta}$, where
$J=\partial V/\partial\phi$. Close to the phase transition, where we may assume
the cutoff to be much larger than the mass scale, i.e. $W^{\prime}/\Lambda\ll1$,
the effective potential reads
\begin{equation}
V(\bar{\rho})=
\frac{1}{A^2}\,\bar{\rho}\left(\bar{\rho}-\bar{\rho}_0\right)^2
	\label{effpotential}
\end{equation}
with
$A=c+\pi\operatorname{sgn}(\bar\rho-\bar\rho_0)$
and $\operatorname{sgn}(0)=0$.
This leads to 
\begin{equation}
\left.J\right|_{\bar{\rho}_0=0}=\frac{3}{4A^2} \phi^{\delta}\quad{\rm with}\quad
\delta=5\,.
\label{delta}
\end{equation}
Finally, we discuss the critical exponent $\gamma$ associated with 
the susceptibility $\chi=\partial \phi/\partial
J=(\partial^2V/\partial\phi^2)^{-1}$ near the phase transition,
\begin{align}
\left.\chi(J)\right|_{J=0}&\sim (-\bar{\rho}_0)^{\gamma}\quad (\mbox{SYM phase},\,
\bar{\rho}_0<0)\notag\\
\left.\chi(J)\right|_{J=0}&\sim (\bar{\rho}_0)^{\gamma^{\prime}}\quad\,\;
(\mbox{SSB phase},\;\; \bar{\rho}_0>0).
\label{defgamma}
\end{align}
Using (\ref{effpotential}) and (\ref{defgamma}) we get
\begin{equation}\label{gamma}
\gamma=\gamma^{\prime}=2\,.
\end{equation}
Note that the results \eq{nu}, \eq{nu'}, \eq{delta} and \eq{gamma} are invariant under changing $c\leftrightarrow -c$, see \eq{symmetry}. The thermodynamical scaling exponents derived here can equally be obtained from the leading RG  exponent  together with scaling relations by using $\nu=1/\theta$, where $\theta=1$ is the IR relevant eigenvalue due to the VEV. The scaling exponents in the special case where $c=\pm c_P$ are discussed in the following section.

\section{Spontaneous breaking of scale invariance}  

\label{sec:ScaleInvariance}

In this section we discuss the supersymmetric analogon of the Bardeen-Moshe-Bander (BMB) phenomenon, the spontaneous breaking of scale invariance and the associated non-classical scaling.

\subsection{Bardeen-Moshe-Bander phenomenon}
\begin{figure}[t]
 \centering
\includegraphics[scale=0.55]{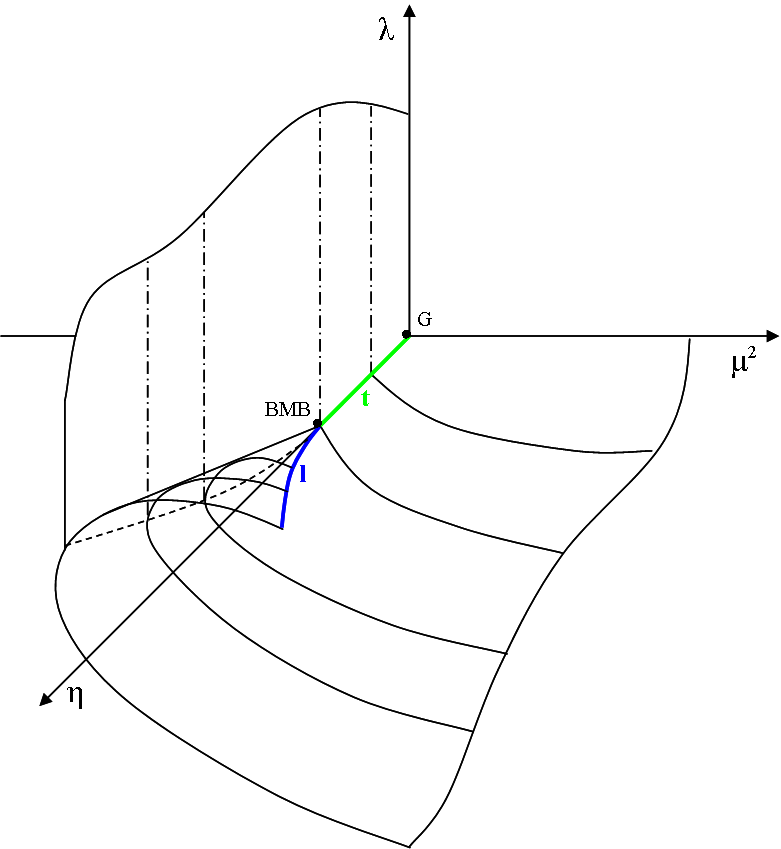}
 \caption{Phase structure of the scalar $O(N)$ model at infinite $N$ including the BMB fixed point, according to
 \cite{David:1985zz} (see text).}
 \label{PhasenD}
\end{figure}

We first recall the BMB phenomenon for scalar $O(N)$ symmetric theories.
Linear $O(N)$ models serve as perfect testing ground for studying  
critical phenomena. For large $N$ the solvable spherical model gives 
a qualitatively accurate picture of the phase structure of the
theory. The  $(\phi^2)^2_{d=3}$ theory  exhibits an IR-attractive Wilson-Fisher fixed point corresponding to a second
order phase transition between the $O(N)$ symmetric and the spontaneously
broken phase \cite{Tetradis:1995br}.
In contrast, the scalar $(\phi^2)^3_{d=3}$
model shows a more complex phase structure \cite{David:1984we, David:1985zz,
Bardeen:1983rv, Tetradis:1995br}. Depending on the renormalized couplings
$\mu^2, \lambda$ and  $\eta$ of the operators $\phi^2, \phi^4$ and $\phi^6$, 
one observes a first-order phase transition without universal behavior 
or a second-order phase transition with universal behavior. 
Both regimes are separated by a tricritical line $t$, 
characterized by vanishing couplings $\mu^2$ and $\lambda$
as depicted in Fig.~\ref{PhasenD}. A surface of first-order
transitions continues into the $O(N)$ symmetric phase for couplings
with $\eta>\eta_c$ and ends at a
gas-liquid transition line $l$. Scale invariance is an exact symmetry of the
tricritical theory, but at the end point $(0, 0, \eta_c)$, scale 
invariance is spontaneously broken. The free coupling $\eta$ is 
dimensionally transmuted to an undetermined mass scale $m$
and a massless Goldstone-boson (dilaton) shows up. In the large-$N$ limit
this non-trivial and UV-stable BMB fixed-point marks the point
where the tricritical line $t$ and the gas-liquid line $l$ meet. 
Hence, the tricritical line connects the Gaussian fixed point and the 
BMB fixed point. One expects that at finite $N$ the tricritical line
extends all the way to infinite $\eta$ and the BMB point disappears \cite{Karsch:1987uj}. 
We note that the BMB fixed point is also of interest as a fundamental  UV fixed point, allowing for a non-Gaussian continuum limit for the $(\phi^2)^3_{d=3}$ theory  with non-classical scaling.

\subsection{Supersymmetric BMB phenomenon}

In the supersymmetric theory, the BMB phenomenon has first been discussed in \cite{Bardeen:1984dx} with variational methods. Here, the critical $(\Phi^2)^2_{d=3}$ theory with a quartic superfield potential corresponds, in the scalar sector,  to a critical  $(\phi^2)^3_{d=3}$ with a sextic potential. The main new addition due to supersymmetry is that the scalar quartic and sextic couplings are no longer independent of each other. 

Using the fully integrated RG flow, the following picture for the BMB phenomenon
emerges: If we fine-tune the classical coupling $\kappa=\kappa_{\rm cr}$, the solution \eq{urho} at the origin $\rho=0$ reads
\begin{equation}
  -1=c\,u_0+H\left(u_0\right)\,
  \label{eq:algebraic}
\end{equation}  
in the IR limit, where $u_0\equiv u(\rho=0)$. 
This equation simply represents the fixed point solution at vanishing field. The 
$O(N)$ symmetric ground state is characterized by the mass 
\begin{equation}
M^2=\left(W^{\prime}(0)\right)^2= \bar{\mu}^2 =\left(u_0\,k\right)^2,
\label{massdef}
\end{equation}
where $M=-\bar{\mu}>0$. Evidently, $u_0$ has to diverge as $1/k$ in order to
allow for  spontaneous breaking of scale invariance with  a finite mass scale 
$M$ in the IR limit $k \rightarrow 0$. Now we find the transcendental equation
\eq{eq:algebraic} to have always a single zero mass solution $M=0$, except for 
$c=c_P$, where it shows an
additional, infinitely large solution $u_0\to-\infty$. Note that this limit
emerges from $u(\rho)$ through negative field squared values $\rho\to 0^-$, which is a consequence of our regularisation. Hence, the specific microscopic parameters 
\begin{equation}\label{sBMB}
(\kappa, \tau)=(1,1/\pi)
\end{equation}
lead to a macroscopic theory, where the mass of the $O(N)$
bosonic and fermionic quanta is left undetermined. Thus, scale
invariance is spontaneously broken in accordance with
\cite{Bardeen:1984dx, Moshe:2003xn, Gudmundsdottir:1984yk} and a mass  is
generated  by dimensional transmutation. The  coupling parameter $\tau$ takes the value
 \eq{sBMB} in our conventions, and the associated degree of freedom is `transmuted' to an
arbitrary  mass scale $M$. Spontaneously broken
scale invariance leads to the appearance of a Goldstone boson (dilaton) which is
accompanied by a Goldstone fermion (dilatino), since supersymmetry is left
unbroken. Note that these particles are exactly massless, since $\tau$ is not
renormalized.

\subsection{BMB scaling exponents}
Next we turn to the scaling exponents of the supersymmetric BMB fixed point. 
The critical exponents \eq{defnu} and \eq{defgamma}
become double-valued due to a different scaling
behavior of the different mass scales $m, M$ near the fixed point.
These, in turn, originate from the finite and the infinite $u_0$ solutions detected at $|c|=c_P$, see Fig.~\ref{pXY}. The latter is responsible for the special nonanalytic behavior of the
solution at the BMB fixed point. We first consider $c=c_P$, and the critical exponents $\nu$ and $\nu^{\prime}$ defined in \eq{defnu}. By approaching the
phase transition from the SYM regime, we find $m=-\bar{\rho}_0/2\pi$ 
and hence 
\begin{equation}
\label{BMBnu}
\nu=1\,.
\end{equation} 
In turn,  approaching the fixed point
from the SSB regime, the expression for $M$ in (\ref{gapBrancha}) is not applicable
since the contribution linear in $W^{\prime}$ in (\ref{smallmass}) vanishes. The subleading quadratic terms take over and we are lead to
$M^2=
\frac{\Lambda}{3}\bar{\rho}_0$, 
implying that the supersymmetric BMB exponent $\nu'$ is given by
\begin{equation}
\label{BMBnu'}
\nu_{{}{\rm BMB}}=\frac{1}{2}\,.
\end{equation}
We now consider $c=-c_P$. By virtue of the symmetry \eq{symmetry} we note that the mass scales $m\leftrightarrow M$ interchange their roles under $c_P\leftrightarrow -c_P$. Consequently, the scaling exponents \eq{BMBnu} and \eq{BMBnu'} also interchange their values. Therefore we conclude that the theory at $|c|=c_P$ displays conventional scaling with \eq{BMBnu} as well as un-conventional scaling with \eq{BMBnu'}. The former is a consequence of the smooth `non-BMB-type' scaling  related to finite $u_0$, whereas the latter is the BMB scaling associated to infinite $u_0$. In either case, and under the above identification, we conclude that the scaling indices from the symmetric and symmetry broken regimes agree. We also stress that the BMB scaling exponent \eq{BMBnu'} is non-classical. Furthermore, it cannot be derived from the RG scaling alone, as they are due to non-analyticities in the field dependences.  As a final comment we note that an infinite $u_0$, the fingerprint for spontaneous breaking of scale invariance, is stable under alterations of the RG scheme.

\section{Radial mode fluctuations}
\label{sec:FiniteN}
In this section,  we give a first account of the phase transition in a theory with finitely rather than infinitely many
supermultiplets $N$, focussing on the existence of a fixed point, the phase
transition, and the fate of the supersymmetric BMB phenomenon to leading order in a gradient expansion.

\subsection{Exact fixed point}
The main new addition to the supersymmetric RG flow at finite $N$ are the fluctuations of the radial mode. They imply that the quartic coupling $\tau$ is
no longer an exactly marginal coupling with an identically vanishing
$\beta$-function. Instead,  the flow of this coupling is
governed by terms of order $1/N$. The absence of an 
exactly marginal coupling implies that the line of fixed points found at 
infinite $N$ will collapse into a finite, possibly empty set of fixed points.
Furthermore, the running of the VEV no longer factorizes from the other 
couplings of the theory resulting in a more complex structure of the RG flow.

In order to study the supersymmetric $O(N)$ model at finite $N$ we return to
the full RG flow  \eq{flow-w'}, which in terms of  $u\equiv w'$ takes the form
\begin{align}
\partial_t u=  &-u +
\rho\,u'-(1-\frac{1}{N})u'\,\frac{1-u^2}{(1+u^2)^2}\notag\\
&-\frac{1}{N}(3u^{\prime}+2u^{\prime\prime}\rho)\frac{1-(u+2\rho
u^{\prime})^2}{(1+(u+2\rho u^{\prime})^2)^2}\,.
\label{fullflow}
\end{align}
A global, analytical, solution of the RG flow \eq{fullflow} is presently not at hand, and 
we have to resort to approximate solutions instead \cite{Bervillier:2007rc}. We start with a polynomial 
approximation to order $n$ for the `potential' $u$, writing
\begin{equation}
u(\rho,t)=\sum_{i=1}^{n}a_i(t)(\rho-\rho_0(t))^{i}\,.
\label{polynomial}
\end{equation}
It expresses the potential in terms of $(n+1)$ couplings $(\rho_0, a_1,\cdots,a_n)$ to determine its fixed points. Inserting the ansatz \eq{polynomial} into the PDE \eq{fullflow} we find a
tower of ordinary, coupled differential equations for the couplings,
\begin{align}
\partial_t \rho_0(t)&=
-\rho_0(t)+ \left(1-\frac{1}{N}\right)\notag\\
&\,+\frac{1}{N}\left(3+4\rho_0(t)\frac{a_2(t)}{a_1(t)}\right)
\frac{(1-(2\rho_0(t)a_1(t))^2)}{(1+(2\rho_0(t)a_1(t))^2)^2}\notag\\
&\vdots\notag\\
\partial_t a_n(t)
&=f_n\left(\rho_0(t),a_1(t), a_2(t), \cdots, a_{n+2}(t) \right).
\label{system}
\end{align}
Note that the functions $f_n$ depend on the couplings $a_{n+1}$ and $a_{n+2}$, because the RHS of \eq{fullflow} involves up to second derivatives of $u$. 
The fixed point solution requires the flow of all couplings to vanish and
hence we set the LHS of \eq{system} equal to zero, leading to an algebraic
system of  $(n+1)$ equations for $(n+3)$ unknowns. These may be solved, tentatively,
by setting the last two couplings $a_{n+1}$ and $a_{n+2}$ to zero. 
We find 
\begin{align}
\rho_{0*}(N)&=1-\frac{1}{N}\notag\\
a_{1\,*}(N)&= 
\frac{1}{2}\frac{N}{N-1}\notag\\
a_{2\,*}(N)&= - \frac{3}{8}\frac{N^2}{(N-1)^2}
\label{FPCoupl}
\end{align}  
for the first three couplings.  The solution bifurcates into two independent fixed points starting with $a_3$. Intriguingly, the recursive relation leads to an exact analytical 
solution of the full system for all $N$ to arbitrarily high expansion order $n$. The reason for this unlikely outcome is that the fixed point \eq{FPCoupl} is independent of the boundary condition which we have imposed initially on the higher order couplings. This follows from noticing that all fixed point equations \eq{system} with $n\geq2$ are of the form 
\begin{eqnarray*}
0&=&f_n(\rho_0,a_1,\dots,a_{n+2})\\
&=&\tilde f_n(\rho_0,a_1,\dots,a_n)\\
&&\;+\,(n+1)\left(\rho_0-1+1/N + \partial_t\rho_0\right)a_{n+1}\\
&&\;-\,\frac{n+1}{N}\frac{1-\xi^2}{(1+\xi^2)^2}\left[(3+2n)a_{n+1}+2(n+2)\rho_0a_{n+2}\right]\\
&&\;-\,\frac{4\rho_0\xi(n+1)^2}{N}\frac{(3a_1+4a_2\rho_0)(\xi^2-3)}{(1+\xi^2)^3}a_{n+1}\,.
\end{eqnarray*}
Here $\xi=2a_1\rho_0$, and $\partial_t \rho_0$ is given according to
\eq{system}. At the fixed point  \eq{FPCoupl} we have $\xi_*=1$, and all terms
proportional to $a_{n+1}$ and $a_{n+2}$ vanish.
Thus, the fixed point equation for every $ a_n$ $(n>2)$ is independent of $a_{n+1}$ and $a_{n+2}$ provided the first three couplings have the values \eq{FPCoupl}, and we are lead to a closed system of $(n+1)$ equations for $(n+1)$ couplings allowing for an exact solution order by order.

\subsection{Exact scaling exponents}
The new fixed point \eq{FPCoupl} has two branches one of which is IR attractive in all couplings except for the running VEV which remains an IR relevant operator. The second fixed point is UV relevant in all couplings and is not pursued any further. The universal scaling exponents of the Wilson-Fisher type fixed point can be determined analytically. 
From the eigenvalues of the stability matrix $B_i^j =
\partial(\partial_ta_i)/\partial a_j|_*$ we read off that the lowest coupling $(a_0\equiv\rho_0)$ defines 
an IR unstable direction with a critical index
\begin{equation}
\theta_0=1\,.
\label{Theta0}
\end{equation}  
Note that the leading critical exponent $\nu=1/\theta_0$ in \eq{Theta0}  is super-universal and identical to the result at infinite $N$.   The exponent does not receive corrections due to the fluctuations of the radial mode and therefore cannot be used to distinguish universality classes of different $N$. 
All other couplings $a_i$, $i=1,2,3,\dots$ define IR attractive directions with subleading critical exponents
\begin{equation}
\theta_{i}= 1-i-\frac{i(i+1)}{6}\left(\sqrt{\frac{N+17}{N-1}}-1\right)\,.
\label{Thetai}
\end{equation}
The universal eigenvalues $\theta_i$ are strictly negative for all $N>1$.
 Furthermore, the Gaussian critical exponents
$\theta_{{\rm G},i}=1-i$ for integer $i\ge 0$ of the theory in the large-$N$ limit are recovered
from \eq{Theta0}, \eq{Thetai} in the limit $1/N \rightarrow 0$. In particular the formerly exactly marginal $\phi^6$ coupling has 
now become irrelevant.

Similarly, the fixed-point values of the couplings (\ref{FPCoupl}) converge to
the large-$N$ fixed-point values  for $N \rightarrow \infty$.
In the presence of the radial fluctuations, the $N$-dependent quartic superfield couplings $\tau_*(N)$ is given by the coefficient $a_{1\,*}(N)$, see \eq{FPCoupl}. 
Taking the limit of infinite $N$ singles out a unique value for the quartic superfield coupling, 
\begin{equation}\label{tauN}
\lim_{N\rightarrow \infty}
\tau_{*}(N)=\frac12\,,
\end{equation}
meaning that  the line of non-trivial fixed points  parametrized by the exactly
marginal  superfield coupling $\tau$ has shrunk
to a single point.  Notice also that the fixed point value \eq{tauN} is different from the supersymmetric BMB value $\tau=1/c_P$  in the infinite $N$ limit, see \eq{cM}. This serves as a strong indication for the non-existence of a  supersymmetric BMB fixed point in the presence of the radial fluctuations and $N>1$.

\subsection{Global scaling solution}

The infinite $N$ limit \eq{tauN} belongs to the strong coupling regime where the fixed point solution for the superpotential derivative $u_*$ displays two branches, neither of which extends towards arbitrarily small fields  \cite{Litim:2011bf}. The latter, signalled through the divergence of $du_*/d\rho$ at some finite field value $\rho\ge 0$, is responsible for the occurrence of a Landau scale. It remains to be seen whether the fixed point at finite $N$ continues to belong to the strongly coupled regime or not. 

To answer this question, and to compare the fixed points at finite and infinite $N$,  we need to study the finite $N$ potentials at small fields numerically. The Taylor series \eq{polynomial} of the scaling solution has a finite radius of convergence.
Alternatively, one may expand the inverse fixed point solution $\rho(u)$ in powers of $u$. At infinite $N$,
the analytical scaling solution 
$\rho=1+c u_* +H(u_*)=\sum_{i=0}^{\infty} b_i u_*^{i}$ 
has a finite radius of convergence $r$ set by the gap of the inverse propagator (here: $r=1$) \cite{Litim:2000ci}. Either expansion is limited to a finite range in field space.
In order to cover the full field space, and to make potential
non-analyticities of the form $u_*^{\prime}(\rho) \rightarrow \infty$ visible,
we numerically integrate  the differential equation of the inverse function $\rho(u_*)$ 
instead of $u_*(\rho)$.  It
reads
\begin{align}
0=&\rho-u_*
\rho^{\prime}-\left(1-\frac{1}{N}\right)\frac{(1-u_*^2)}{(1+u_*^2)^2}\notag\\
&-\frac{1}{N}(3\rho^{\prime\,2}-2\rho\rho^{\prime\prime})
\frac{\rho^{\prime\,2}-(u_*\rho^{\prime}+2\rho)^2}{(\rho^{\prime\,2}+(u_*\rho^{\prime}+2\rho)^2)^2}
\label{inverseDGL}
\end{align} 
subject to suitable boundary conditions. The  boundary conditions $\rho(0)=\rho_{0*}$ and $\rho^{\prime}(0)=2\rho_{0*}$
correspond to a singular point of \eq{inverseDGL} and cannot be used. 
Instead, we extract boundary conditions for
$\rho(u_{*}), \rho^{\prime}(u_{*})$ for  $|u_{*}|=0.01\ll 1$ from the polynomial
approximation to $u_*(\rho)$ of the order $n=9$. The combined use of polynomial expansions and subsequent numerical integration  is a well-tested technique in critical scalar theories \cite{Bervillier:2007rc}.

\begin{figure}[t]
\begin{center}
\unitlength0.001\hsize
\begin{picture}(900,1000)
\put(0,130){\includegraphics[width=.40\textwidth]{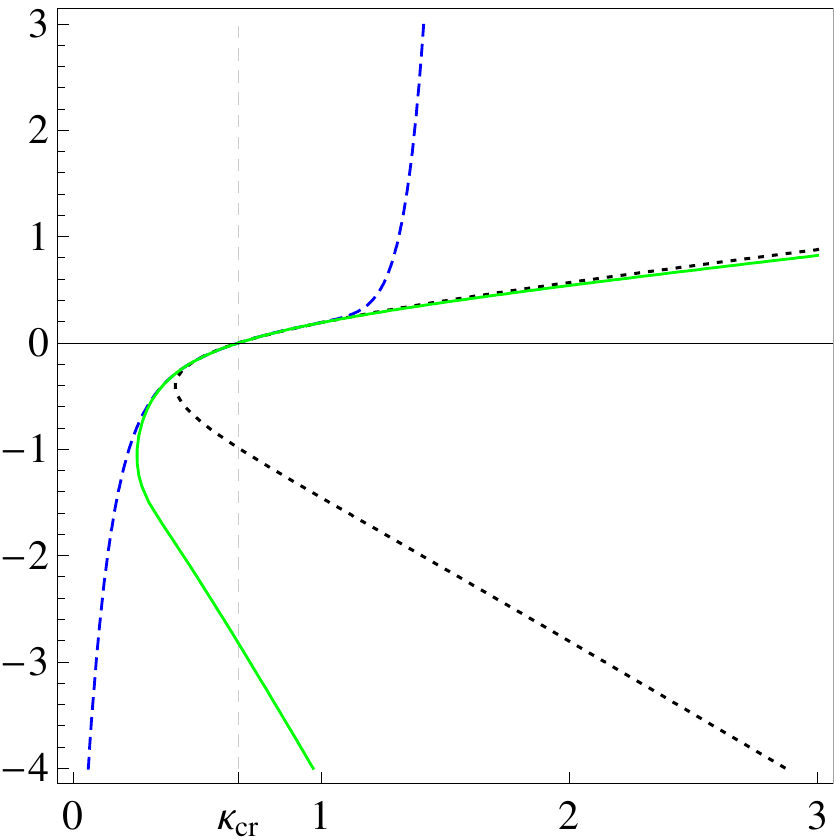}}
\put(-45,600){\large $u_*$}
\put(470,105){\large $\rho$} 
\end{picture}
\vskip-1cm
\caption{Fixed point solution $u_*(\rho)$ for $N=3$.
The figure compares the polynomial approximation (blue, dashed line) 
with the non-perturbative integration (green, solid line) and a large-$N$ 
like solution (black, dotted line).}
\label{NumPoly} 
\end{center}
\end{figure}

Fig.~\ref{NumPoly} compares the polynomial approximation of  the  scaling
solution with the numerical one for $N=3$. 
The graph also contains the analytical solution of the theory at infinite $N$.
 The latter
is given by the fixed-point equation of (\ref{fullflow}), where we neglect the contribution of the radial mode (the term in
the second line) and fix the free parameter of the solution to \eq{tauN}.
\begin{figure}[t]
\begin{center}
\unitlength0.001\hsize
\begin{picture}(900,1000)
\put(0,130){\includegraphics[width=.40\textwidth]{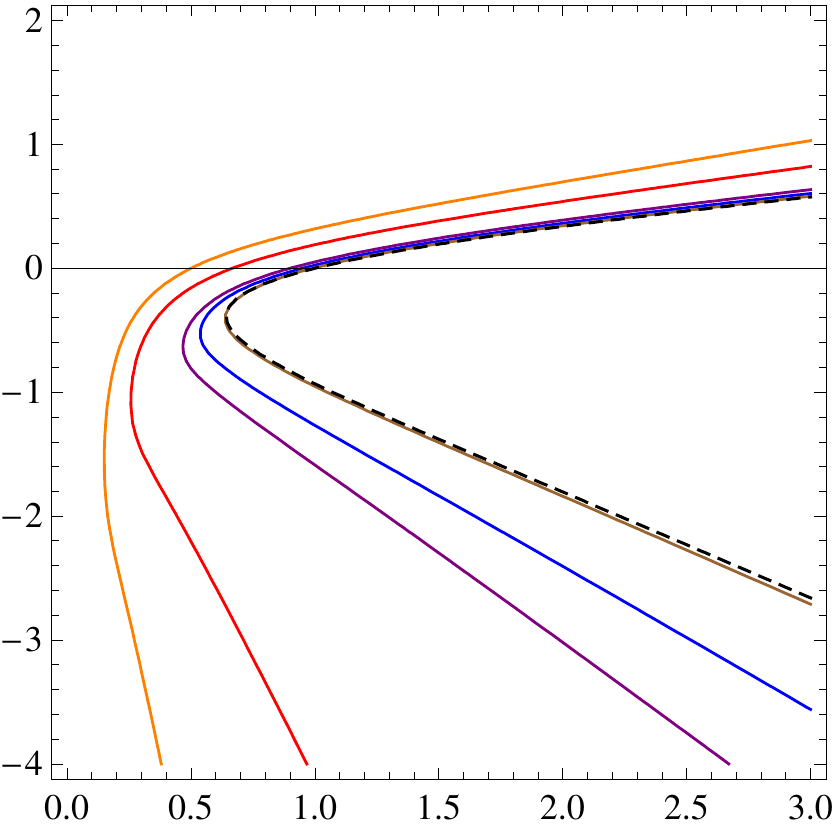}}
\put(-45,600){\large $u_*$}
\put(470,105){\large $\rho$} 
\end{picture}
\vskip-1cm
\caption{Fixed point solution 
$u_*(\rho)$ 
for  various $N>1$, showing $N=2, 3, 10, 20$ and $ 100$ from left to right   (full lines) in comparison with the infinite $N$ result (dashed line). With increasing $N$ the 
solutions converges to the exact infinite $N$ result with
$\tau(N)$ approaching \eq{tauN}.}
\label{NumN} 
\end{center}
\end{figure}
We find that the large-$N$ solution approximates the finite-$N$ solution very well in the
vicinity of the node $\rho_{0*}$ and above, largely independently of  
the chosen value for $N>1$. This is entirely due to the structure of the fixed point
\eq{FPCoupl}, where $2a_{1\,*}\rho_{0*}\equiv 1$.
The numerical solutions illustrate further that the fixed point solution at
finite $N$ shows a similar non-analytic behavior characterized by a  diverging mass term
$u_*'\rightarrow \infty$, as it appears in the large-$N$ limit for strong
quartic superfield coupling (cf. Sec.~\ref{Cusp}).

We now discuss the $N$-dependence of the scaling solution \eq{FPCoupl}.
Fig.~\ref{NumN} shows that the fixed point solution, displayed for various
integer $N\ge 2$, always generates a diverging $du/d\rho$ for some positive
field values $\rho=\rho_c(N)$, with $0<\rho_c(N)<\rho_{0*}(N)$. The solution
$u_*(\rho)$ does not exist for small $0\le \rho<\rho_c$, for all $N$ considered.
Also, we find that $u_*(\rho_c)$ becomes increasingly large in magnitude with
decreasing $N$.  Hence, the main effect of the competition between the radial
mode and the Goldstone mode  fluctuations, with decreasing $N$,  is a shift of
the end point $\rho_c(N)$ and the VEV $\rho_{0*}(N)$ towards smaller  values.
Continuity in $N$ suggests that this pattern persists for all $N>1$ where
$\rho_{0*}>0$.

For the supersymmetric  Ising model where $N=1$, the Goldstone modes are absent and the RG dynamics is controlled by the fluctuations of the radial mode. In the limit $N\to 1$, \eq{FPCoupl}  predicts a vanishing VEV, $\rho_0=0$ and implies the existence of a supersymmetric Ising fixed point valid for all fields, though  at the expense of a non-analytic behavior of $u_*(\rho)$ at vanishing field. Note that a direct study of the $N=1$ case using the same RG equations \cite{Synatschke:2010ub} has also detected a regular Ising fixed point analytic in the fields, whose critical eigenvalue $\theta_0=3/2$ is different from \eq{Theta0}.
Furthermore,  the diverging of all higher order couplings  \eq{FPCoupl} in the limit $N\to 1$ together with the continuity of the fixed point in $N$ suggests  that $\rho_c\to0$ and $|u_*(\rho_c)|\to\infty$ in this limit. This behavior is intriguing inasmuch as the diverging of $u_*(\rho\to 0)$ is the fingerprint for the spontaneous breaking of scale invariance. It may thus qualify for a novel supersymmetric BMB phenomenon which originates from the radial mode rather than the Goldstone fluctuations.
It would seem worth to test this picture directly in the supersymmetric Ising model  without relying on the limit $N\to 1$ adopted here.

To conclude,  the fixed point \eq{FPCoupl} is of the  strongly-coupled type for all $N>1$ as signalled by the same qualitative behavior seen previously at infinite $N$ \cite{Litim:2011bf}. Furthermore,   the fluctuations of the Goldstone modes are central for the existence of the endpoint in field space $\rho_c>0$ of strongly-coupled fixed point solutions. At infinite $N$, and as a consequence of $\rho_c>0$, 
the phase diagram at strong coupling is governed by non-analyticities at finite
RG scales. Due to $\rho_c(N)>0$ for $N>1$, the same type of non-analyticities
with an associated Landau scale $k_L$ control the phase transition associated
with the fixed point  \eq{FPCoupl}  at finite $N$.  The above behavior at strong
coupling  is  thus generic for supersymmetric $(\Phi^2)^2$ theories with $N>1$,
and to distinguish from the non-analyticities at  infinite $N$ responsible for,
e.g.~the conventional BMB phenomenon.

\section{Summary and conclusions} \label{Conclusions}

Analytical solutions of interacting local quantum field theories are benchmarks for a deeper understanding of concepts and mechanisms in theoretical physics.
In this work, we have provided a global renormalization group study of interacting supersymmetric theories in three euclidean dimensions, the $O(N)$ symmetric $(\Phi^2)^2$ Wess-Zumino theories, continuing a line of research initiated in \cite{Litim:2011bf}.  
These theories are the supersymmetric versions of $O(N)$ symmetric scalar $(\phi^2)^3$ theories, which display 
 first- and second-order phase transitions, and the seminal Bardeen-Moshe-Bander (BMB) mechanism.

The main new features due to  supersymmetry arise through the fluctuations of the
Goldstone modes, in particular at strong coupling, and their competition with the fluctuations of the radial mode. 
In the limit of infinitely many superfields, the radial mode is absent and the theory is solved exactly.
The phase diagram is then controlled by two free parameters,  the exactly marginal quartic superfield coupling and the vacuum expectation value, which takes the role of an infrared relevant coupling. 
Locally, the theory has an interacting fixed point 
for all quartic couplings, yet globally the line of fixed points terminates at a critical value. 
At weak coupling, the theory displays a second order phase transition between an $O(N)$ symmetric and a symmetry broken phase with Gaussian scaling,
and global supersymmetry remains intact. 
At strong coupling,  the global effective potential becomes multi-valued in certain regions of field space, signalled by 
divergences in the local fermion-boson interactions
at a finite Landau scale $k_L$.
The appearance of the characteristic energy scale $k_L$ resolves the long-standing puzzle about peculiar degenerate $O(N)$ symmetric ground states detected previously \cite{Bardeen:1984dx,Moshe:2003xn}, 
 showing that these arise, gradually, from the integrating-out of strongly-coupled  long wave-length  fluctuations. In this regime, supersymmetry may be spontaneously broken. Furthermore, this pattern is largely insensitive to whether 
an infinite or a finite short-distance cutoff is chosen, solely inducing a  shift in the boundary between the weakly and strongly coupled regimes.
At finite $N$, and to leading order in a gradient expansion,
the additional fluctuations of the radial mode lift the degeneracy of the quartic superfield coupling and the line of fixed points collapses to a finite set. Locally, a new Wilson-Fisher type fixed point  appears with non-Gaussian exponents and super-universal scaling 
in its infrared relevant coupling. Globally the fixed point belongs to the strongly coupled regime, in complete analogy to the strong coupling behavior observed at infinite $N$. In its vicinity, and with decreasing $N$, the admixture of radial fluctuations shrinks the domain in field space where a Landau scale occurs. 
The scaling solution extends over all fields  as soon as the Goldstone fluctuations are absent, though at the expense of a square-root type non-analyticity in the effective potential at vanishing field.

The availability of a supersymmetric BMB  phenomenon equally depends  on the competition between Goldstone modes and the radial mode. At infinite $N$,  the Goldstone fluctuations 
 lead to the well-known BMB fixed point 
whose  scaling exponent $\nu=1/2$ 
arises due to non-analyticities of the infinite $N$ limit. Supersymmetry remains intact, and the spontaneous breaking of scale invariance leads to the appearance of an arbitrary mass scale together with an exactly  massless Goldstone boson and fermion.
The fixed point disappears in the presence of both, radial and Goldstone mode fluctuations.  The BMB mechansim may re-appear provided the Goldstone modes are absent altogether, in which case the spontaneous breaking of scale invariance  is driven solely by the radial mode. 
A definite conclusion on this point requires more study.

From a structural point of view, the most distinctive new feature due to supersymmetry 
at strong coupling 
is the build-up of a multi-valued effective potential, accompanied by non-analyticities in the polynomial interactions at a Landau scale $k_L$. Here, we have established that this  phenomenon arises primarily through the fluctuations of the Goldstone modes, irrespective of whether there are finitely or infinitely many of them.
It is worth noting that  similar non-analyticities have recently been observed in the random-field Ising model, where the disorder is implemented with the help of Parisi-Sourlas supersymmetry \cite{Tissier:2011zz}. In these models, the spontaneous breaking of supersymmetry is directly associated to the appearance of cusp-like non-analyticities at a finite Larkin scale $k_L$, analogous to the Landau scale found here.  Provided this similarity persists on a fundamental level, it suggests
that supersymmetry may be spontaneously broken in the $(\Phi^2)^2$ theory at strong coupling. Conversely, our findings make it conceivable that the occurrence of a Larkin scale is the signature of a multi-valued effective potential in disordered Ising models.

Finally, we stress that the availability of  an analytic functional RG for supersymmetry was decisive to achieve our results, allowing for a controlled and global interpolation between the short- and long-distance regimes of the theory even at strong coupling.
It is a virtue of the fully integrated RG flow at all scales  that the structure of the quantum effective theory has become transparent. We expect that the combination of analytical and numerical tools adopted from  \cite{Bervillier:2007rc}  
will prove equally useful for the non-perturbative study of 
supersymmetry in other settings and extensions.

\acknowledgments{Helpful discussions and earlier
collaborations with Jens Braun, Holger Gies, Moshe Moshe, Tobias Hellwig, Axel
Maas and Edouard Marchais are gratefully acknowledged. This work has been
supported by the DFG under GRK 1523 and grant Wi 777/11, 
and by the Science and Technology Facilities Council (STFC) 
under grant number ST/J000477/1.}

\appendix

\section{Conventions} 
\label{sec:Conventions} 

Relevant
 symmetry relations and Fierz identities for Majorana spinors are  
 $\bar{\Psi}\chi = \bar{\chi}\Psi$, $\bar{\Psi}\gamma^{\mu} \chi = - 
\bar{\chi}\gamma^{\mu}\Psi$ and $\theta_k \bar{\theta}_l = 
- \frac{1}{2} (\bar{\alpha}\alpha)\mathds{1}_{kl}$.
One of the main features of the action is its invariance under
supersymmetry transformations. The latter are characterized by  the
supersymmetry variations $\delta_{\epsilon}\Phi^{i}$, generated by 
$\mathcal{N}=1$ fermionic generator  $\mathcal{Q}$.
We have
\begin{align}
&\delta_{\epsilon}\Phi^{i}(x) = i \bar{\epsilon}_k\mathcal{Q}_k\Phi^{i}(x)
\quad\quad \mbox{with} \notag\\
 \mathcal{Q}_k &= -i\partial_{\bar{\theta}_k} -
 \gamma^{\mu}_{kl}\theta_l\partial_{\mu}, \;\,\bar{\mathcal{Q}}_k =
 -i\partial_{\theta_k} - \bar{\theta}_l\gamma^{\mu}_{lk}\partial_{\mu}.
\label{eq:variation}
 \end{align} 
Thus,  (\ref{eq:variation}) leads to the
supersymmetry variations
\begin{equation}
\delta \phi^{i} = \bar{\epsilon}\psi^{i}, 
  \delta\psi^{i} = (F^{i} + i\partial\!\!\!/\phi^{i})\epsilon \;\,
 \mbox{and}\;\, \delta F^{i} = i\bar{\epsilon}\partial\!\!\!/\psi^{i}
\label{eq:variationcomp}
\end{equation}
of the component fields. The anticommuting sector of the superalgebra 
is given by the anticommutator of two supercharges
\begin{equation}
 \{\mathcal{Q}_k, \bar{\mathcal{Q}}_l\} = 2 i \gamma^{\mu}_{kl} \partial_{\mu}.
 \label{eq:anticommutator}
 \end{equation}
 The derivation of the supersymmetric flow equation is given in appendix B of
 \cite{Litim:2011bf}.

\bibliographystyle{unsrt}       
\bibliography{References}

\begin{thebibliography}{10}

\bibitem{Litim:2011bf}
Daniel~F. Litim, Marianne~C. Mastaler, Franziska Synatschke-Czerwonka, and
  Andreas Wipf.
\newblock {Critical behavior of supersymmetric O(N) models in the large-N
  limit}.
\newblock {\em Phys.Rev.}, D84:125009, 2011.

\bibitem{ZinnJustin}
Jean Zinn-Justin.
\newblock {\em {Quantum Field Theory and Critical Phenomena}}.
\newblock Oxford University Press, third edition, 1996.

\bibitem{Bardeen:1983rv}
William~A. Bardeen, Moshe Moshe, and Myron Bander.
\newblock {Spontaneous Breaking of Scale Invariance and the Ultraviolet Fixed
  Point in O($N$)-Symmetric $(\phi^{6}_3)$ Theory}.
\newblock {\em Phys. Rev. Lett.}, 52:1188, 1984.

\bibitem{David:1984we}
Francois David, David~A. Kessler, and Herbert Neuberger.
\newblock {The Bardeen-Moshe-Bander Fixed Point and the Ultraviolet Triviality
  of $(\Phi^2)^3_3$}.
\newblock {\em Phys.Rev.Lett.}, 53:2071, 1984.

\bibitem{David:1985zz}
Francois David, David~A. Kessler, and Herbert Neuberger.
\newblock {A Study of $(\Phi^2)^3_3$ at $N=\infty$}.
\newblock {\em Nucl.Phys.}, B257:695--728, 1985.

\bibitem{Bardeen:1984dx}
William~A. Bardeen, Kiyoshi Higashijima, and Moshe Moshe.
\newblock {Spontaneous Breaking of Scale Invariance in a Supersymmetric Model}.
\newblock {\em Nucl. Phys.}, B250:437, 1985.

\bibitem{Moshe:2003xn}
Moshe Moshe and Jean Zinn-Justin.
\newblock {Quantum field theory in the large N limit: A review}.
\newblock {\em Phys. Rept.}, 385:69--228, 2003.

\bibitem{Dawson:2005uw}
John~F. Dawson, Bogdan Mihaila, Per Berglund, and Fred Cooper.
\newblock {Supersymmetric approximations to the 3D supersymmetric O(N) model}.
\newblock {\em Phys. Rev.}, D73:016007, 2006.

\bibitem{Suzuki:1985uk}
Tsuneo Suzuki.
\newblock {Three-dimensional $O(N)$ model with Fermi and scalar fields}.
\newblock {\em Phys. Rev.}, D32:1017, 1985.

\bibitem{Suzuki:1985pw}
Tsuneo Suzuki and Hisashi Yamamoto.
\newblock {A nontrivial Ultraviolet fixed point and stability of
  three-dimensional $O(N)$ model with fermions and bosons}.
\newblock {\em Prog. Theor. Phys.}, 75:126, 1986.

\bibitem{Gudmundsdottir:1984yk}
Ragnheidur Gudmundsdottir and Gunnar Rydnell.
\newblock {On a supersymmetric version of $(\phi^2)^3_3$ theory}.
\newblock {\em Nucl. Phys.}, B254:593, 1985.

\bibitem{Feinberg:2005nx}
J.~Feinberg, M.~Moshe, Michael Smolkin, and J.~Zinn-Justin.
\newblock {Spontaneous breaking of scale invariance and supersymmetric models
  at finite temperature}.
\newblock {\em Int. J. Mod. Phys.}, A20:4475--4483, 2005.

\bibitem{Matsubara:1987iz}
Yoshimi Matsubara, Tsuneo Suzuki, Hisashi Yamamoto, and Ichiro Yotsuyanagi.
\newblock {On a phase with spontaneously broken scale invariance in
  three-dimensional $O(N)$ models}.
\newblock {\em Prog. Theor. Phys.}, 78:760, 1987.

\bibitem{Wetterich:1992yh}
Christof Wetterich.
\newblock {Exact evolution equation for the effective potential}.
\newblock {\em Phys. Lett.}, B301:90--94, 1993.

\bibitem{Berges:2000ew}
Jurgen Berges, Nikolaos Tetradis, and Christof Wetterich.
\newblock {Non-perturbative renormalization flow in quantum field theory and
  statistical physics}.
\newblock {\em Phys. Rept.}, 363:223--386, 2002.
\newblock hep-ph/0005122.

\bibitem{Litim:2000ci}
Daniel~F. Litim.
\newblock {Optimization of the exact renormalization group}.
\newblock {\em Phys.Lett.}, B486:92--99, 2000.

\bibitem{Litim:2001fd}
Daniel~F. Litim.
\newblock {Mind the gap}.
\newblock {\em Int.J.Mod.Phys.}, A16:2081--2088, 2001.

\bibitem{Litim:2001up}
Daniel~F. Litim.
\newblock {Optimized renormalization group flows}.
\newblock {\em Phys.Rev.}, D64:105007, 2001.

\bibitem{Tetradis:1992xd}
N.~Tetradis and C.~Wetterich.
\newblock {The high temperature phase transition for phi**4 theories}.
\newblock {\em Nucl.Phys.}, B398:659--696, 1993.

\bibitem{Tetradis:1995br}
N.~Tetradis and D.~F. Litim.
\newblock {Analytical Solutions of Exact Renormalization Group Equations}.
\newblock {\em Nucl. Phys.}, B464:492--511, 1996.

\bibitem{Bagnuls:2000ae}
C.~Bagnuls and C.~Bervillier.
\newblock {Exact renormalization group equations. An Introductory review}.
\newblock {\em Phys.Rept.}, 348:91, 2001.

\bibitem{Litim:2002cf}
Daniel~F. Litim.
\newblock {Critical exponents from optimized renormalization group flows}.
\newblock {\em Nucl.Phys.}, B631:128--158, 2002.

\bibitem{Litim:2003kf}
Daniel~F. Litim and Lautaro Vergara.
\newblock {Subleading critical exponents from the renormalization group}.
\newblock {\em Phys.Lett.}, B581:263--269, 2004.

\bibitem{Canet:2003qd}
Leonie Canet, Bertrand Delamotte, Dominique Mouhanna, and Julien Vidal.
\newblock {Nonperturbative renormalization group approach to the Ising model: A
  Derivative expansion at order partial**4}.
\newblock {\em Phys.Rev.}, B68:064421, 2003.

\bibitem{Bervillier:2007rc}
Claude Bervillier, Andreas Juttner, and Daniel~F. Litim.
\newblock {High-accuracy scaling exponents in the local potential
  approximation}.
\newblock {\em Nucl.Phys.}, B783:213--226, 2007.

\bibitem{Benitez:2009xg}
F.~Benitez, J.-P. Blaizot, H.~Chate, B.~Delamotte, R.~Mendez-Galain, et~al.
\newblock {Solutions of renormalization group flow equations with full momentum
  dependence}.
\newblock {\em Phys.Rev.}, E80:030103, 2009.

\bibitem{Litim:2010tt}
Daniel~F. Litim and Dario Zappala.
\newblock {Ising exponents from the functional renormalisation group}.
\newblock {\em Phys.Rev.}, D83:085009, 2011.

\bibitem{Vian:1998kv}
F.~Vian.
\newblock {Supersymmetric gauge theories in the exact renormalization group
  approach}.
\newblock 1998.
\newblock hep-th/9811055.

\bibitem{Bonini:1998ec}
M.~Bonini and F.~Vian.
\newblock {Wilson renormalization group for supersymmetric gauge theories and
  gauge anomalies}.
\newblock {\em Nucl. Phys.}, B532:473--497, 1998.
\newblock hep-th/9802196.

\bibitem{Synatschke:2008pv}
Franziska Synatschke, Georg Bergner, Holger Gies, and Andreas Wipf.
\newblock {Flow Equation for Supersymmetric Quantum Mechanics}.
\newblock {\em JHEP}, 03:028, 2009.

\bibitem{Gies:2009az}
Holger Gies, Franziska Synatschke, and Andreas Wipf.
\newblock {Supersymmetry breaking as a quantum phase transition}.
\newblock {\em Phys. Rev.}, D80:101701, 2009.

\bibitem{Synatschke:2009nm}
Franziska Synatschke, Holger Gies, and Andreas Wipf.
\newblock {Phase Diagram and Fixed-Point Structure of two dimensional N=1
  Wess-Zumino Models}.
\newblock {\em Phys. Rev.}, D80:085007, 2009.

\bibitem{Synatschke:2010jn}
Franziska Synatschke-Czerwonka, Thomas Fischbacher, and Georg Bergner.
\newblock {The two dimensional N=(2,2) Wess-Zumino Model in the Functional
  Renormalization Group Approach}.
\newblock {\em Phys. Rev.}, D82:085003, 2010.

\bibitem{Synatschke:2010ub}
Franziska Synatschke, Jens Braun, and Andreas Wipf.
\newblock {N=1 Wess Zumino Model in d=3 at zero and finite temperature}.
\newblock {\em Phys. Rev.}, D81:125001, 2010.

\bibitem{Synatschke:2009da}
Franziska Synatschke, Holger Gies, and Andreas Wipf.
\newblock {The Phase Diagram for Wess-Zumino Models}.
\newblock {\em AIP Conf. Proc.}, 1200:1097--1100, 2010.

\bibitem{Falkenberg:1998bg}
Sven Falkenberg and Bodo Geyer.
\newblock {Effective average action in N = 1 super-Yang-Mills theory}.
\newblock {\em Phys. Rev.}, D58:085004, 1998.
\newblock hep-th/9802113.

\bibitem{Rosten:2008ih}
Oliver~J. Rosten.
\newblock {On the Renormalization of Theories of a Scalar Chiral Superfield}.
\newblock {\em JHEP}, 03:004, 2010.
\newblock arXiv:0808.2150 [hep-th].

\bibitem{Sonoda:2009df}
Hidenori Sonoda and Kayhan Ulker.
\newblock {An elementary proof of the non-renormalization theorem for the
  Wess-Zumino model}.
\newblock {\em Prog. Theor. Phys.}, 123:989--1002, 2010.
\newblock arXiv:0909.2976 [hep-th].

\bibitem{Sonoda:2008dz}
Hidenori Sonoda and Kayhan Ulker.
\newblock {Construction of a Wilson action for the Wess-Zumino model}.
\newblock {\em Prog. Theor. Phys.}, 120:197--230, 2008.
\newblock arXiv:0804.1072 [hep-th].

\bibitem{Tissier:2011zz}
Matthieu Tissier and Gilles Tarjus.
\newblock {Supersymmetry and Its Spontaneous Breaking in the Random Field Ising
  Model}.
\newblock {\em Phys.Rev.Lett.}, 107:041601, 2011.

\bibitem{Tissier:2011mu}
Matthieu Tissier and Gilles Tarjus.
\newblock {Nonperturbative Functional Renormalization Group for Random Field
  Models. III: Superfield formalism and ground-state dominance}.
\newblock {\em Phys.Rev.}, B85:104202, 2012.

\bibitem{Tissier:2011mv}
Matthieu Tissier and Gilles Tarjus.
\newblock {Nonperturbative Functional Renormalization Group for Random Field
  Models. IV: Supersymmetry and its spontaneous breaking}.
\newblock {\em Phys.Rev.}, B85:104203, 2012.

\bibitem{Karsch:1987uj}
Frithjof Karsch and Hildegard Meyer-Ortmanns.
\newblock {Phase structure of $O(N)$ symmetric $\phi^6_3$ models at small and
  intermediate $N$}.
\newblock {\em Phys.Lett.}, B193:489, 1987.

\end{thebibliography}

\end{document}